\documentclass[a4paper,11pt]{article}
\pdfoutput=1
\usepackage{jheppub}

\usepackage[utf8]{inputenc}
\usepackage{tikz}
\usepackage{tcolorbox}
\usepackage{xcolor}

\begin{document}

\vspace*{-2.5cm}
\begin{center}
\centerline{\large EUROPEAN ORGANIZATION FOR NUCLEAR RESEARCH (CERN)}
\end{center}

\hrule

\vskip 1.5cm
\begin{center}
    {\Huge SHADOWS} 
\end{center}

\vskip 0.5cm
\begin{center}
{\LARGE \underline{S}earch for \underline{H}idden \underline{A}nd \underline{D}ark \underline{O}bjects \underline{W}ith the \underline{S}PS}
\end{center}
\vskip 5mm

\begin{center}
{\LARGE \it Expression of Interest}
\end{center}

\vskip 5mm
\begin{center}
{\large 
W.~Baldini$^{(1)}$,
A.~Balla$^{(2)}$,
J.~Bernhard$^{(3)}$,
A.~Calcaterra$^{(2)}$,
V.~Cafaro$^{(4)}$,
A.~Ceccucci$^{(3)}$,
V.~Cicero$^{(4)}$,
P.~Ciambrone$^{(2)}$,
H. Danielsson$^{(3)}$,
G.~D’Alessandro$^{(3)}$,
G.~Felici$^{(2)}$,
L.~Gatignon$^{(5)}$,
A.~Gerbershagen$^{(3)}$,
V.~Giordano$^{(4)}$,
G.~Lanfranchi$^{(2)}$,
A.~Montanari$^{(4)}$,
A.~Paoloni$^{(2)}$,
G.~Papalino$^{(2)}$,
T.~Rovelli$^{(4)}$,
A.~Saputi$^{(2)}$,
S.~Schuchmann$^{(6)}$,
F.~Stummer$^{(7)}$,
N.~Tosi$^{(4)}$
}

\vskip 5mm
{\small \it
$^{(1)}$ INFN, Sezione di Ferrara, Ferrara, Italy \\
$^{(2)}$ INFN, Laboratori Nazionali di Frascati, Frascati (Rome), Italy, \\
$^{(3)}$ CERN \\
$^{(4)}$ INFN, Sezione di Bologna, Bologna, Italy \\
$^{(5)}$ University of Lancaster, Lancaster, UK \\
$^{(6)}$ University of Mainz, Germany \\
$^{(7)}$ University of Vienna, Austria \\
}

\end{center}

\vskip 1.5cm

\begin{center}
    {\Large \bf Executive Summary}
\end{center}

\vskip 0.3cm
We propose a new beam-dump experiment, SHADOWS, to search for a large variety of feebly-interacting particles possibly produced in the interactions of a 400~GeV proton beam  with a high-Z material dump.
SHADOWS will use the 400~GeV primary proton beam extracted from the CERN SPS currently serving the NA62 experiment in the CERN North area and will take data off-axis 
when the P42 beam line is operated in beam-dump mode.
SHADOWS can accumulate up to a $\sim 2 \cdot 10^{19}$ protons on target per year  and expand the exploration for a large variety of FIPs well beyond the state-of-the-art in the mass range of MeV-GeV in a parameter space that is allowed by cosmological and astrophysical observations.
So far the strongest bounds on the interaction strength of new feebly-interacting light particles with Standard Model particles exist up to the kaon mass; above this threshold the bounds weaken significantly. SHADOWS can do an important step into this still poorly explored territory and has the potential to discover them if they have a mass between the kaon and the beauty mass. If no signal is found, SHADOWS will push the limits on their couplings with SM particles between one and four orders of magnitude in the same mass range, depending on the model and scenario.

\clearpage
\noindent
\setcounter{tocdepth}{3}
\tableofcontents

\clearpage

\section{Introduction}
\label{sec:introduction}

\vskip 1mm
Feebly-interacting particles (FIPs) with masses below the electroweak (EW) scale and possibly belonging to a rich dark sector represent a complementary approach with respect to the traditional Beyond the Standard Model (SM) physics explored at the LHC. They can be the answer to many open questions in modern particle physics: the baryon asymmetry of the Universe, the nature of dark matter (DM), the origin of the neutrino masses and oscillations, the cosmological inflation, the strong $CP$ problem, the hierarchy of scales, and the cosmological constant~\cite{Lanfranchi:2020crw}. 

\vskip 1mm
Prominent examples of FIPs are: i) {\it the axions}, that may be responsible for the conservation of CP symmetry in strong interactions, and constitute a fraction or entirety of DM and its extensions {\it axion-like particles (ALPs)}, that can be mediators between DM and SM particles;  ii) {\it DM candidates and related mediators with thermal origin} and masses in the MeV-GeV range that can arise from new gauge groups added to the SM;  iii) {\it heavy neutrinos} with masses below the EW scale that could account for the origin of neutrino masses and the matter anti-matter asymmetry via a leptogenesis mechanism based on neutrino oscillations;  iv) {\it Light scalar particles mixed with the Higgs boson} that might provide some novel ways of addressing the hierarchy problem (e.g. via the so-called relaxion mechanism), or could have been responsible for cosmological inflation.

\vskip 1mm
The interest for FIPs in the broader particle and astroparticle communities is continuously growing, and a multitude of initiatives has emerged in the recent years. The profound interest in this emerging field is reported in many important reports and documents from the community at large: the Cosmic~\cite{Feng:2014uja} and Intensity~\cite{Hewett:2014qja} Frontier Reports of the 2013 Snowmass Process, the Dark Sector community report~\cite{Alexander:2016aln}
, the Cosmic Vision report~\cite{Battaglieri:2017aum}, the LHC Long-Lived Particle community white paper~\cite{Alimena:2019zri}, the Physics Beyond Colliders (PBC) BSM report~\cite{Beacham:2019nyx}, the White Paper on new opportunities for next generation neutrino experiments~\cite{Arguelles:2019xgp}, the Briefing Book of the European Strategy for Particle Physics~\cite{EuropeanStrategyforParticlePhysicsPreparatoryGroup:2019qin}, and the Report from the FIPs2020 workshop~\cite{Agrawal:2021dbo}.  

The recent recommendations of the European Strategy for Particle Physics (ESPP) Update~\cite{ESPP} include the physics of feebly-interacting particles among the {\it essential scientific activities} for particle physics to be pursued in the next decade:
{\it  "The quest for dark matter and the exploration of flavour and fundamental symmetries are crucial components of the search for new physics. This search can be done in many ways, for example through precision measurements of flavour physics and electric or magnetic dipole moments, and {searches for axions,} \underline{ dark sector candidates and feebly interacting particles}."}.

\vskip 2mm
{SHADOWS aims to be a main player in the searches for FIPs in the range of the familiar matter (MeV-GeV) by exploiting the potential of the existing CERN accelerator complex.}

\vskip 1mm
In the range of MeV-GeV, the strongest bounds on the interaction strength of new light particles with SM particles exist up to the kaon mass; above this mass the bounds weaken significantly. SHADOWS can do an important step into this still poorly explored territory and has the potential to discover them if they have a mass between the kaon and the beauty mass. If no signal is found, SHADOWS will push the limits on their couplings with SM particles between one and four orders of magnitude in the same mass range, depending on the model and scenario, opening new directions in model building.

This paper is organized as follows.
Section~\ref{sec:shadows_at_SPS} explains why SHADOWS should be placed in the ECN3/TCC8 area currently hosting the NA62 experiment~\cite{NA62:2017rwk} served by a primary proton beam line (P42/K12) from the CERN SPS.
Sections~\ref{sec:beamline} and Section~\ref{sec:ECN3} introduce the characteristics of the P42/K12 proton beam line and the ECN3/TCC8 experimental complex, respectively. The main requirements for the detector are described in Section~\ref{sec:detector} and a description of the main detector components is provided along with suitable detector technologies. The physics reach of the experiment for one and five integrated years of data taking is shown in Section~\ref{sec:physics} following the benchmarks proposed by the Physics Beyond Colliders BSM working group~\cite{Beacham:2019nyx}. In Section~\ref{sec:background} the most important beam-induced backgrounds are examined and a solution found for their reduction. A tentative schedule of the experiment is sketched in Section~\ref{sec:schedule}. Finally Section~\ref{sec:conclusions} draws the conclusions.

\section{SHADOWS at the CERN SPS}
\label{sec:shadows_at_SPS}
Feebly-interacting particles can be originated from the decay of beauty, charm and strange hadrons and from photons produced in the interaction of protons with a target. Their couplings to SM particles are very suppressed leading to expected production rates of $10^{-10}$ or less. Since the charm and beauty cross-sections steeply increase with the energy~\cite{Lourenco:2006vw, ZEUS:2013fws}, a high-intensity, high-energy (but slowly extracted) proton beam is required to improve over the current results~\cite{Agrawal:2021dbo}. 

\vskip 2mm
To date the world best beam line to produce high intensity fluxes of beauty and charm hadrons through the interactions of protons on a high-$Z$ dump is the 400~GeV P42/K12 primary proton beam line extracted from the CERN SPS and serving the NA62 experiment hosted in the TCC8/ECN3 experimental area.
FIPs emerging from the decays of charm and beauty hadrons have a non negligible transverse momentum and therefore can be detected by an experiment placed {\it off-axis} with respect to the direction of the impinging beam on the dump. This feature opens up the possibility of installing SHADOWS in the ECN3/TTC8 area {\color{black} and taking data when the P42 line is operated in beam-dump mode~\cite{NA62_Addendum}}.
The SHADOWS project has to be compatible with future plans for kaon physics in the ECN3/TCC8 area currently investigated by the NA62 collaboration.

\vskip 2mm
The smallness of the couplings implies that FIPs are also very long-lived (up to ~0.1 sec) compared to the bulk of the SM particles. Therefore the decays to SM particles can be optimally detected  using an experiment with tens of meter long decay volume and with tracking and particle identification capabilities. 
The long ($\sim~200$~m) ECN3/TTC8 experimental hall allows such an experiment to be installed. 

\vskip 2mm
SHADOWS main goal is to detect FIPs produced in charm and beauty decays and decaying to visible SM final states. The striking experimental signature will be two (or more) tracks and/or photons originating from the same point in the decay volume and nothing else. The experimental challenge is to disentangle the signal from backgrounds that mimic the same signature.

\section{The K12/P42 Beam line}
\label{sec:beamline}

\begin{figure}[h]
\begin{center}
\includegraphics[width=0.8\textwidth]{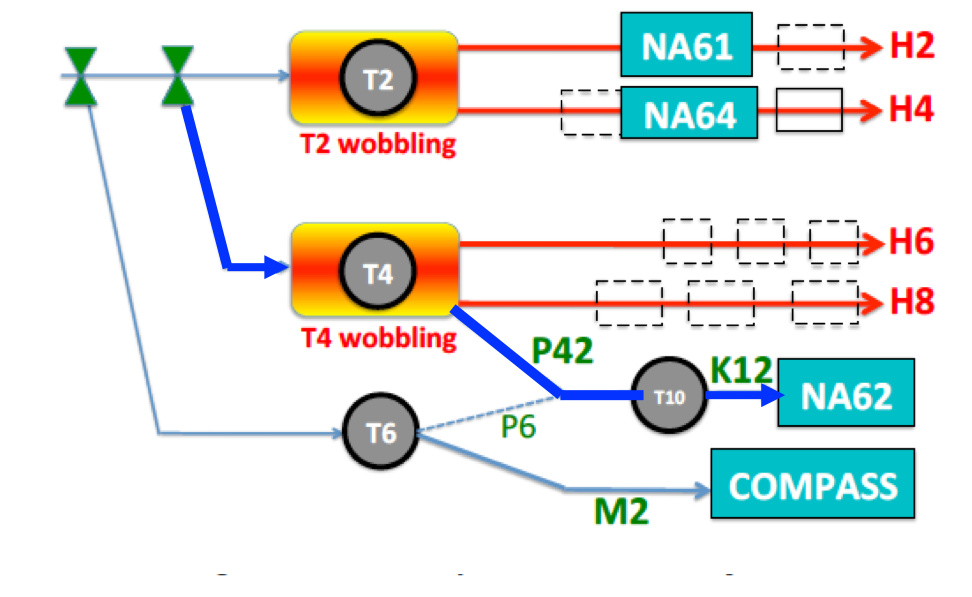}
\end{center}
\caption{\small A schematic layout of the North Area complex. The blue arrows show the path of the protons towards the T10 target in ECN3. Figure taken from Ref.~\cite{Montbarbon:2020lcf}. }
\label{fig:NA_lines}
\end{figure}

Figure~\ref{fig:NA_lines} shows the schematic layout of the extracted beam lines from the SPS in the CERN North Area
complex~\cite{NA_note}. Primary protons extracted from the SPS impinge on three different targets, T2, T4 and T6.
The T2 target produces the H2 and H4 beam lines which are normally operated as versatile secondary or tertiary beams but may occasionally be configured as attenuated primary beams.
The T6 target produces the M2 beam for the COMPASS experiment in the EHN2 hall. 

The P42 line is served by primary slowly-extracted protons from the SPS that impinge on the T4 target. Secondaries produced in the T4 target provide the secondary or tertiary H4 and H6 beams to test beam areas. Non-interacting protons are transported to the T10 target located almost 900~m downstream of T4.

The K12 beam line is derived from a 400~GeV primary proton beam impinging on a 400-mm long, 2-mm diameter cylindrical Beryllium (Be) target that is used to produce a secondary positively charged hadron beam of 75 GeV/c momentum.
In normal (kaon) operation, only 40\% of the protons interact with the Be target and the remaining 60\% is dumped onto the NA62 dump collimators (TAXes) that act as a hadron stopper $\sim$~23~m downstream of the target.  The TAXes are made of blocks of Cu-Fe for a total $\sim$ 21$~\lambda_I$ and a length of 3.2~m.

\begin{figure}[h]
\begin{center}
\includegraphics[width=\textwidth]{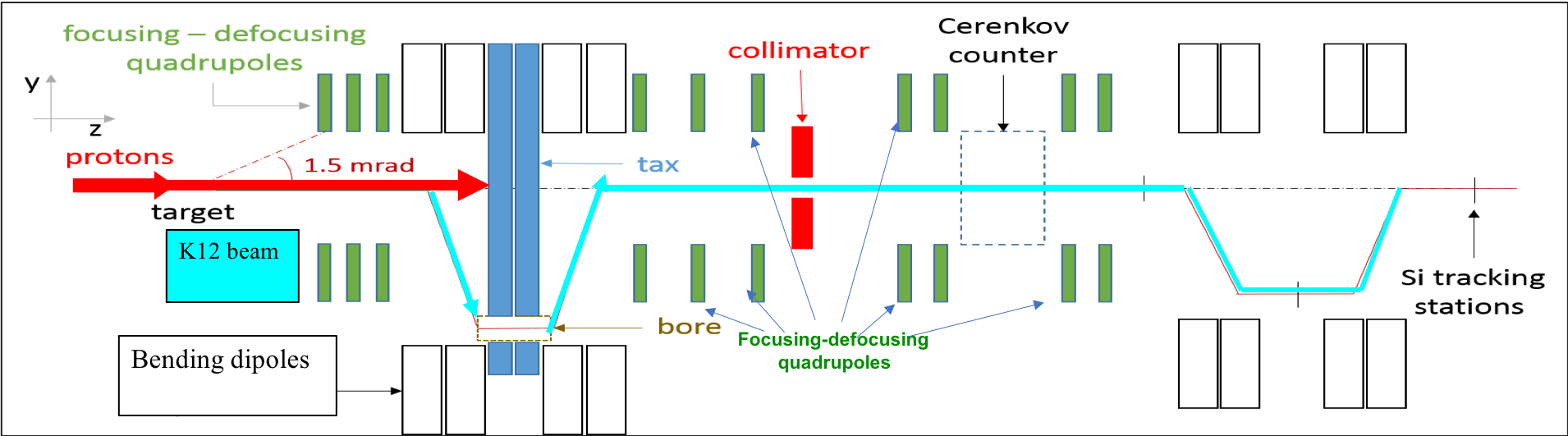}
\end{center}
\caption{\small The P42/K12 beam line serving NA62. Primary 400 GeV protons from the P42 beam line impinge on the T10 target and create a mixed (pion, kaon, proton) beam (K12 line, light blue arrows) serving the NA62 experiment. If the T10 target is lifted, the primary proton beam (red arrow) impinges directly on the TAX collimators that are used as a dump.}
\label{fig:K12}
\end{figure}

\vskip 2mm
The K12 beam line can be operated also in {\it beam-dump mode}. In beam-dump mode the T10 target is lifted and the proton beam is fully dumped onto the K12 TAXes. This is illustrated by the red arrow in Figure~\ref{fig:K12}. 

The switching between kaon-mode to dump-mode is a quick and fully reversible operation: it has been already performed in the 2017-2018 run and about $3 \cdot 10^{16}$ protons-on-target (pot) successfully recorded by NA62 in dump-mode.
NA62 aims to collect additional o$(10^{18})$ pot in dump mode during the current run (2021-2024)~\cite{NA62_Addendum}. 

\vskip 2mm
About $2 \cdot 10^{18}$~pot are delivered by the P42 line to NA62 every year. This assumes about 200 days of SPS up time per year, a beam intensity of $3.3\cdot 10^{12}$ protons per 4.8 sec long spill and about 3000 spills delivered per day\footnote{This is a typical number averaged over the past few years. It takes into account the SPS super-cycle structure, the LHC filling, and the SPS efficiency.}.

\begin{figure}[h]
\begin{center}
\includegraphics[width=0.6\textwidth]{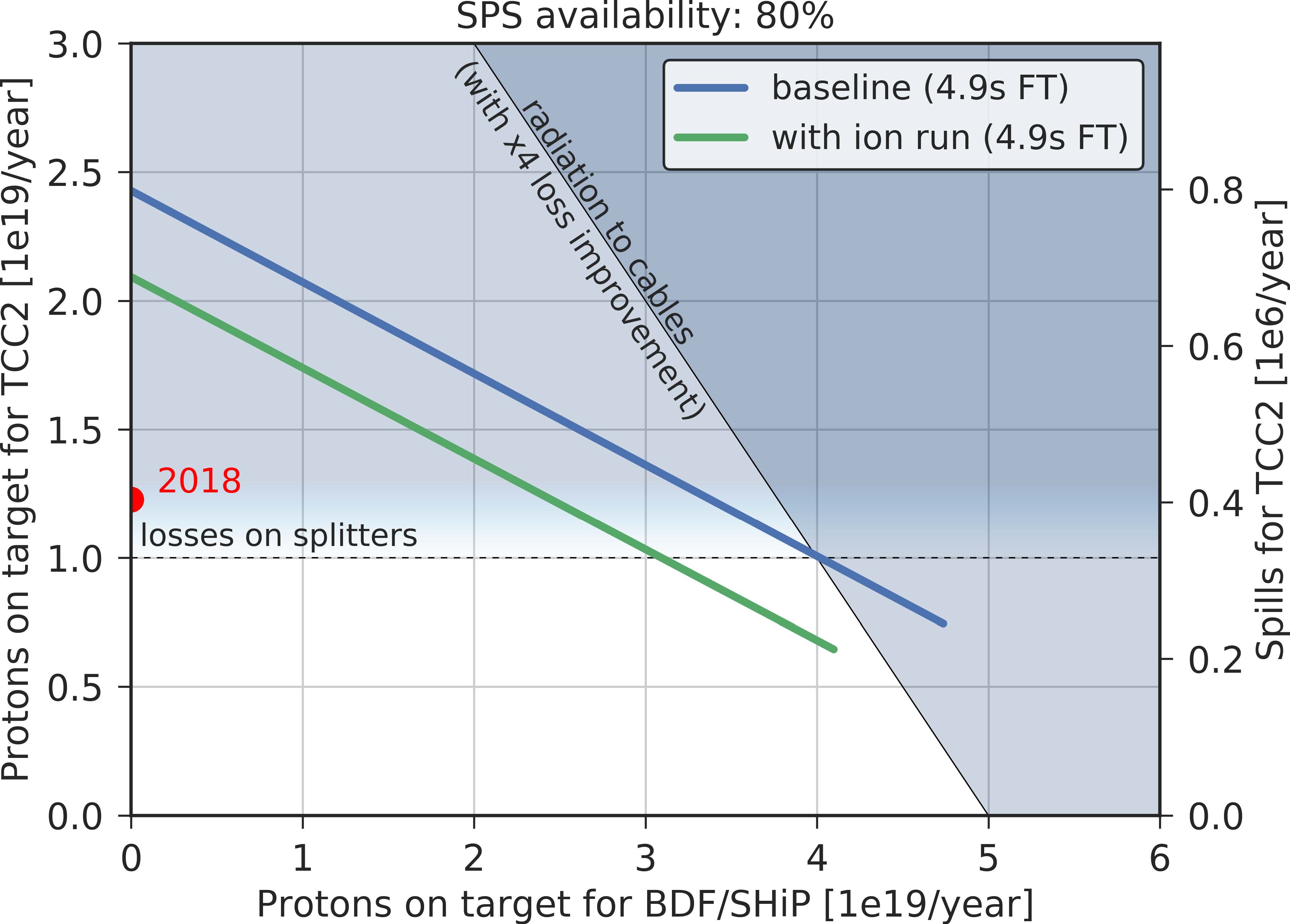}
\end{center}
\caption{\small Intensity limitations (shaded areas) from SPS slow extraction and operationally achieved protons on target in 2018 (red dot).  The secondary $y$ axis shows the number of spills. From Ref.~\cite{CERN-ACC-NOTE-2018-0082}.}
\label{fig:BeamIntensity}
\end{figure}

With some upgrade of the target and dump complex, the intensity of the P42 primary proton beam can be increased up to a factor 6-7 and still being compatible with the operation of all the other existing projects/experiments in the North Area. This is illustrated in Figure~\ref{fig:BeamIntensity} where up to $2-2.5\cdot 10^{19}$ protons can be delivered to the North Area experiments every year assuming that no protons are delivered to the Beam Dump Facility (BDF). The current plan is that the BDF can be considered for approval only in 2027 after the next update of the European Strategy for Particle Physics.

\vskip 2mm
Recent studies have shown that an intensity up to $8 \cdot 10^{12}$ protons per spill would be compatible with the current NA62 target and dump complex and with the radiation protection classification of the TCC8/ECN3 area.  For higher beam intensities, instead, a new design of the TAXes is required with optimised block materials and cooling. Higher beam intensities are currently requested 
by the KLEVER project~\cite{KLEVERProject:2019aks} that requires an increase in the primary proton beam intensity up to a factor 6-7 for generating a neutral kaon beam suitable for the measurement of the branching fraction of the rare decay $K_L \to \pi^0 \nu \overline{\nu}$ with 20\%~accuracy.

\vskip 2mm
Studies are ongoing within the BE-EA CERN department to assess the thermal load on the target and TAXes for a beam up to $\times 7$ the nominal NA62 intensity and related radiation protection issues.

\vskip 2mm
In evaluating SHADOWS sensitivity for FIPs, we assume that the P42 primary proton beam is dumped on TAXes for 200 days per year for a maximum of five integrated years of data taking. 
Two scenarios are considered: 

\begin{itemize}
    \item[-] {\bf scenario n.1:} one integrated year (200 days) of data taking at $\times 4$ the nominal NA62 beam intensity, corresponding to $N_{\rm pot} \sim 10^{19}$;
    \item[-] {\bf scenario n.2:} five integrated years of data taking at $\times4$ the nominal NA62 beam intensity (or three integrated years at the KLEVER nominal intensity) for a total number $N_{\rm pot} \sim 5 \times 10^{19}$.
\end{itemize}

In scenario n.1 (n.2), about $2 (10) \cdot 10^{16}$ $D-$mesons and $\sim 1 (5) \cdot 10^{12}$ $b-$hadrons are produced. This large amount of charm and beauty hadrons opens up the possibility to search for feebly-interacting particles emerging from their decays in the dump, as heavy axions, ALPs, dark scalars, and heavy neutral leptons (HNLs).

\clearpage
\section{The ECN3/TTC8 experimental complex}
\label{sec:ECN3}

\begin{figure}[h]
\begin{center}
\includegraphics[width=0.8\textwidth]{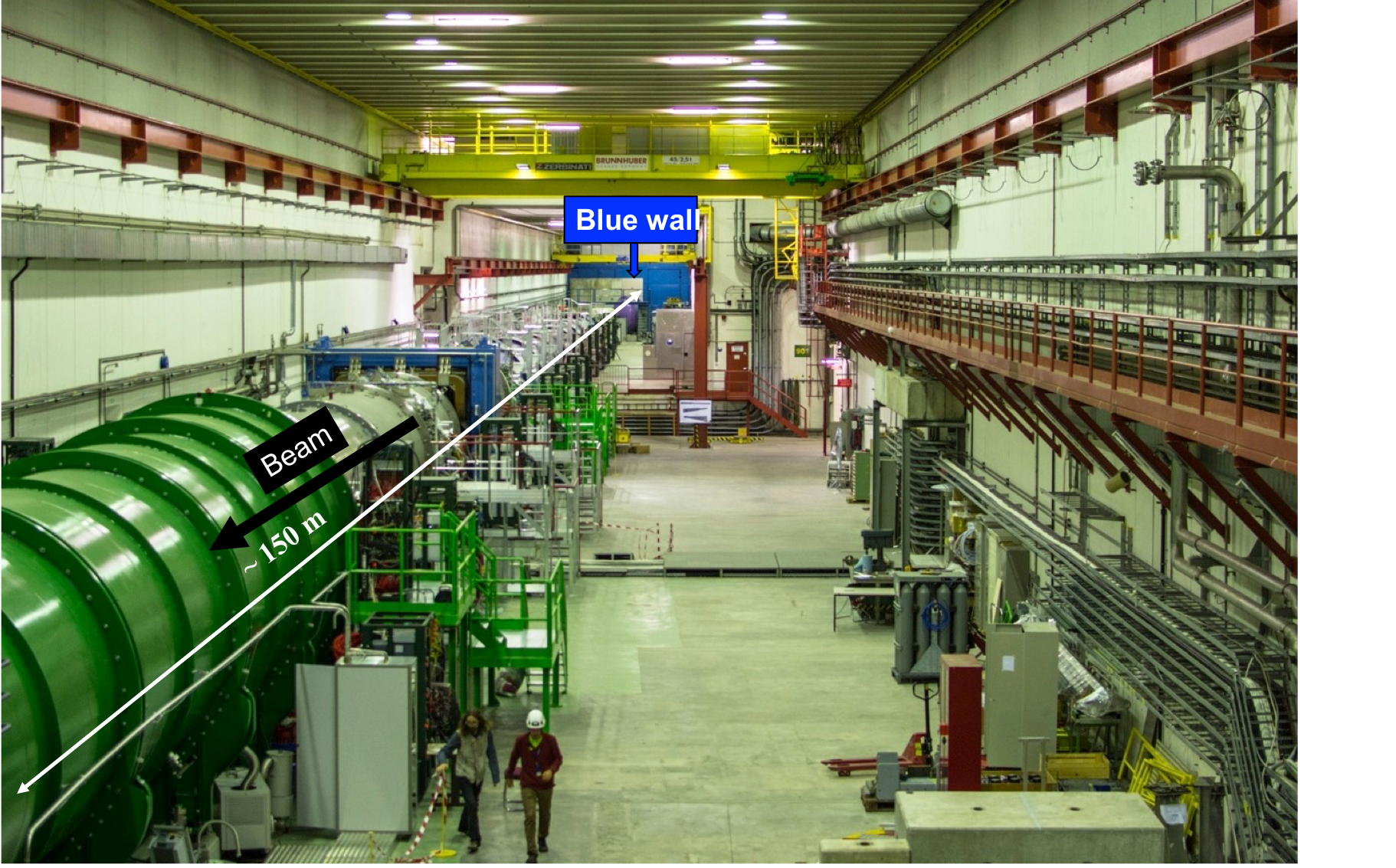}
\end{center}
\caption{\small View of the ECN3 hall hosting the NA62 experiment.}
\label{fig:NA62picture}
\end{figure}

The ECN3/TTC8 experimental hall is placed 15~m underground in the CERN North Area and hosts the NA62 experiment. This is shown in  Figure~\ref{fig:NA62picture}. The blue wall at the very end of the picture defines the separation between the experimental hall and the {\it target area}, a supervised zone in the TCC8 tunnel hosting the target and the TAX collimators.   

\begin{figure}[h]
\begin{center}
\includegraphics[width=\textwidth]{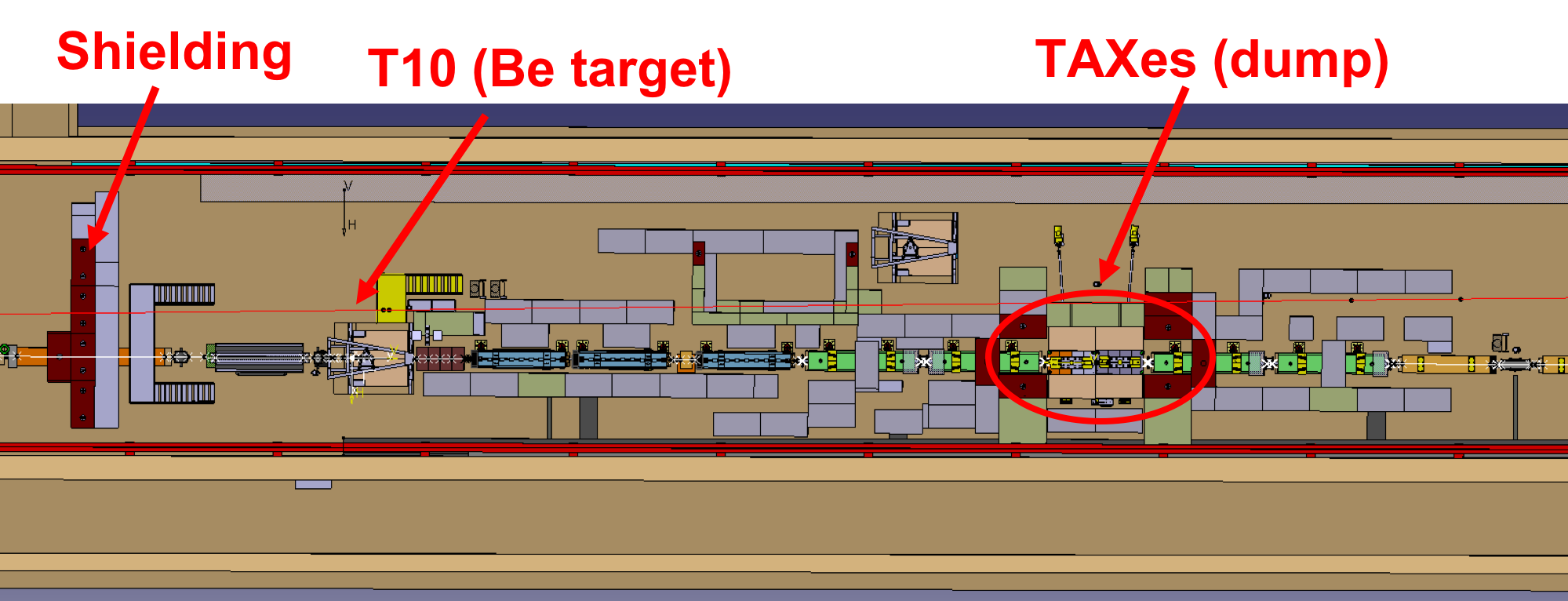}
\end{center}
\caption{\small Schematics of the target/dump area, with the T10 Be target and the TAXes complex in the TCC8 tunnel. The proton beam direction goes from left to right.}
\label{fig:target_complex}
\end{figure}

\begin{figure}[h]
\begin{center}
\includegraphics[width=0.6\textwidth]{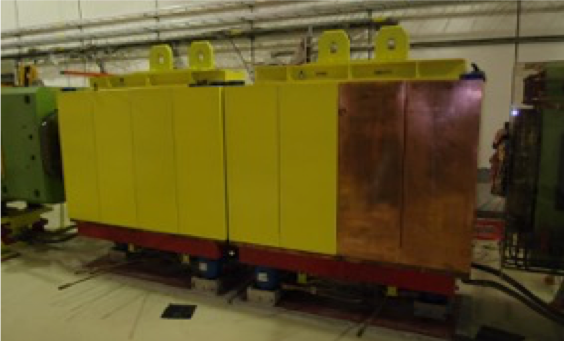}
\end{center}
\caption{\small Picture of the 3.2~m long, water cooled, Fe-Cu based TAXes.}
\label{fig:TAXes}
\end{figure}

Figure~\ref{fig:target_complex} shows a layout of the target area in the TCC8 tunnel, with the T10 target and the TAXes complex 23~m downstream of the T10 target. A picture of the 3.2~m long water-cooled Cu-Fe based TAXes is shown in Figure~\ref{fig:TAXes}.

\begin{figure}[h]
\begin{center}
\includegraphics[width=\textwidth]{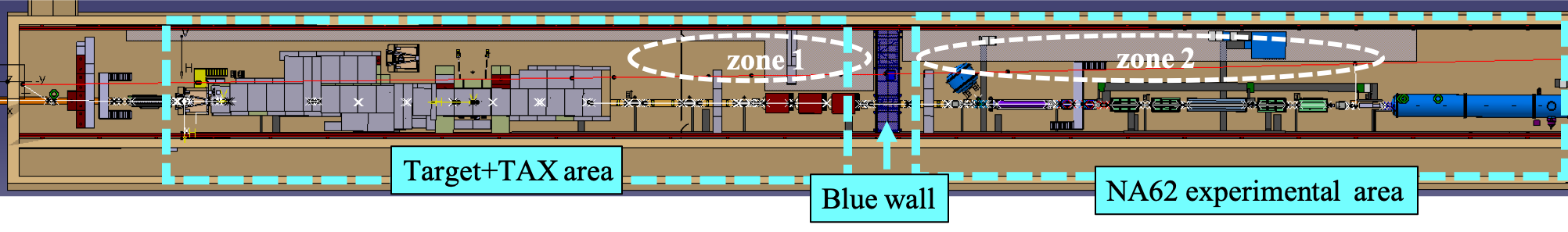}
\end{center}
\caption{\small Schematics of the TCC8 tunnel comprising the target area and a fraction of the NA62 zone. The SHADOWS experiment could be placed in the zones marked as {\it zone 1} and {\it zone 2}. The beam runs from left to right.}
\label{fig:shadows_place}
\end{figure}

\begin{figure}[h]
\begin{center}
\includegraphics[width=\textwidth]{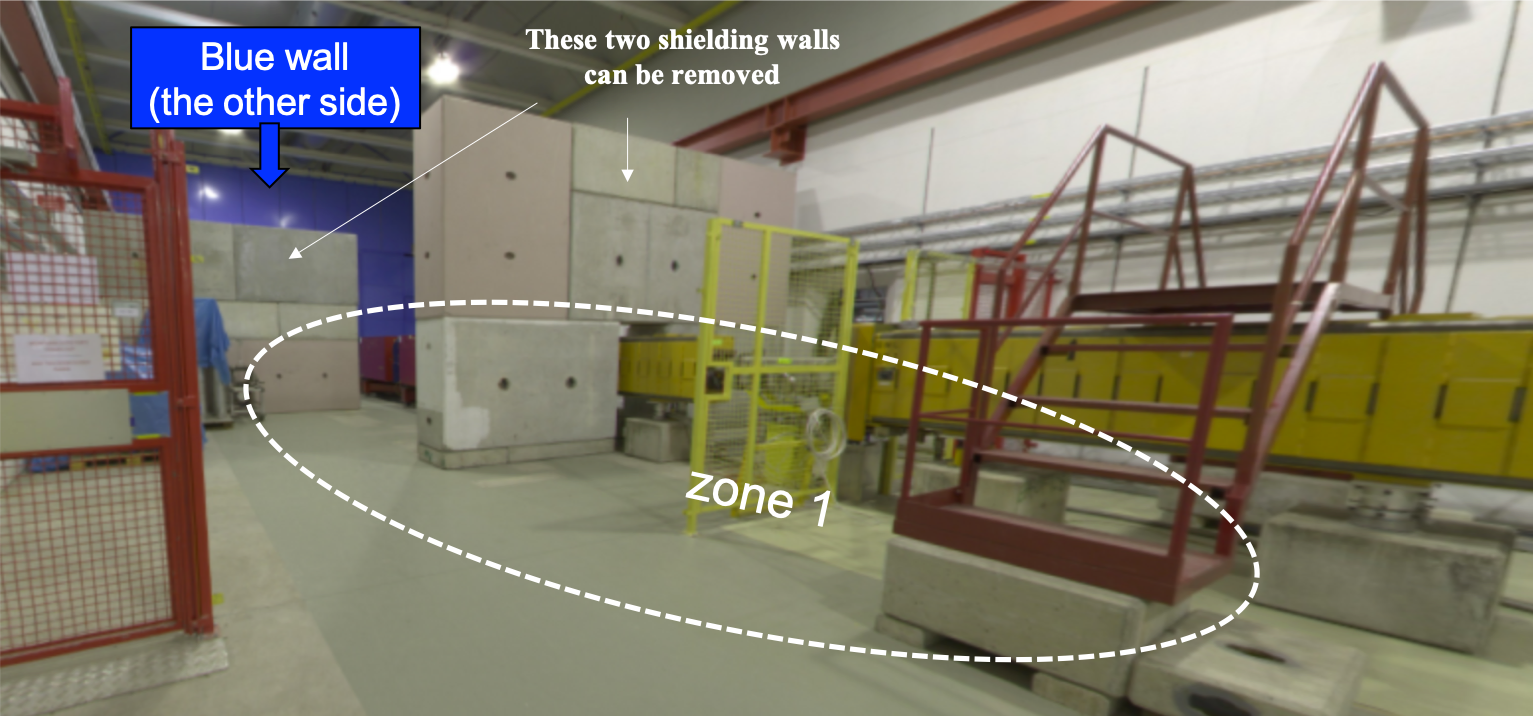}
\end{center}
\caption{\small Picture of the zone adjacent to the beam line in the target area ({\it zone 1)} . The two concrete shielding walls visible in the picture can be moved closer to the dump to free the space for SHADOWS.}
\label{fig:shadows_zone1}
\end{figure}

\begin{figure}[h]
\begin{center}
\includegraphics[width=\textwidth]{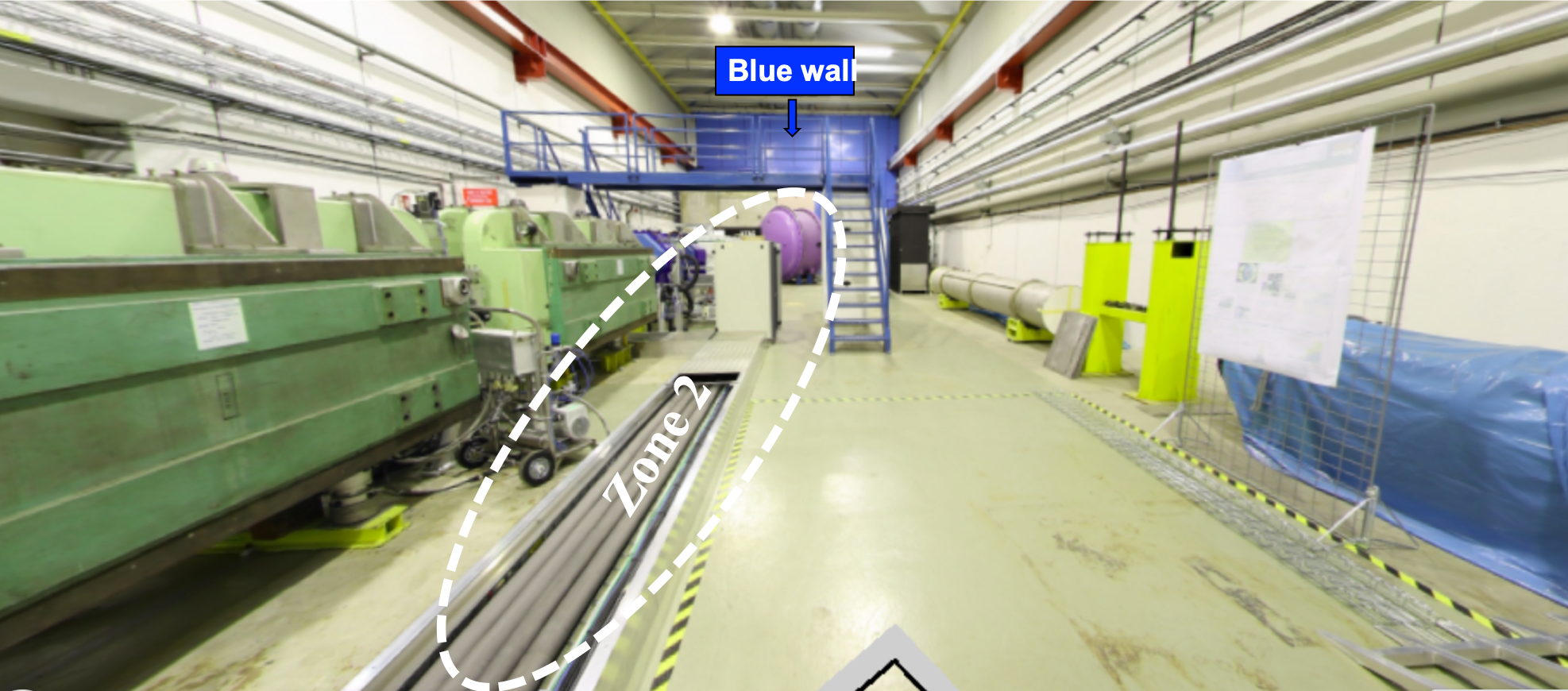}
\end{center}
\caption{\small Picture of the zone adjacent to the beam line in the NA62 experimental hall ({\it zone 2}).}
\label{fig:shadows_zone2}
\end{figure}

\vskip 2mm
A larger view of the target area including the beginning of the NA62 experimental area is shown in Figure~\ref{fig:shadows_place}.
The two labels {\it zone 1} and {\it zone 2} show two possible zones suitable to host the SHADOWS experiment. {\it Zone 1}  could be immediately available with minor modifications provided no show-stoppers arise from radiation protection issues. {\it Zone 2} would be available with non-negligible (but likely addressable) changes in the area setup. A picture of the space available in zone 1 and zone 2 is shown in Figure~\ref{fig:shadows_zone1} and ~\ref{fig:shadows_zone2}, respectively. The two concrete walls shown in zone 1 could be moved upstream to free space for the detector. The blue wall also should be moved  closer to the target and TAX area to allow the access to the SHADOWS detector area.

\vskip 2mm
{\it The SHADOWS experiment in its first phase aims to install a first spectrometer in zone 1, leaving to an optional second phase the installation of a second spectrometer in zone 2, as detailed in Section~\ref{sec:detector} }.

\vskip 2mm
After the EoI stage, a detailed integration/engineering design will be needed to address important integration issues related to zone 1, namely: i)  the access to the magnets of the beam line; ii) the radiation protection constraints in particular upstream of the blue wall.
On a longer term, for the second phase of the experiment after the long-shutdown LS4, other important issues would need to be addressed for zone 2 (eg: accessibility to other magnets of the K12 beam line).

\clearpage
\section{The detector concept}
\label{sec:detector}

\begin{figure}[h]
\begin{center}
\includegraphics[width=0.48\textwidth]{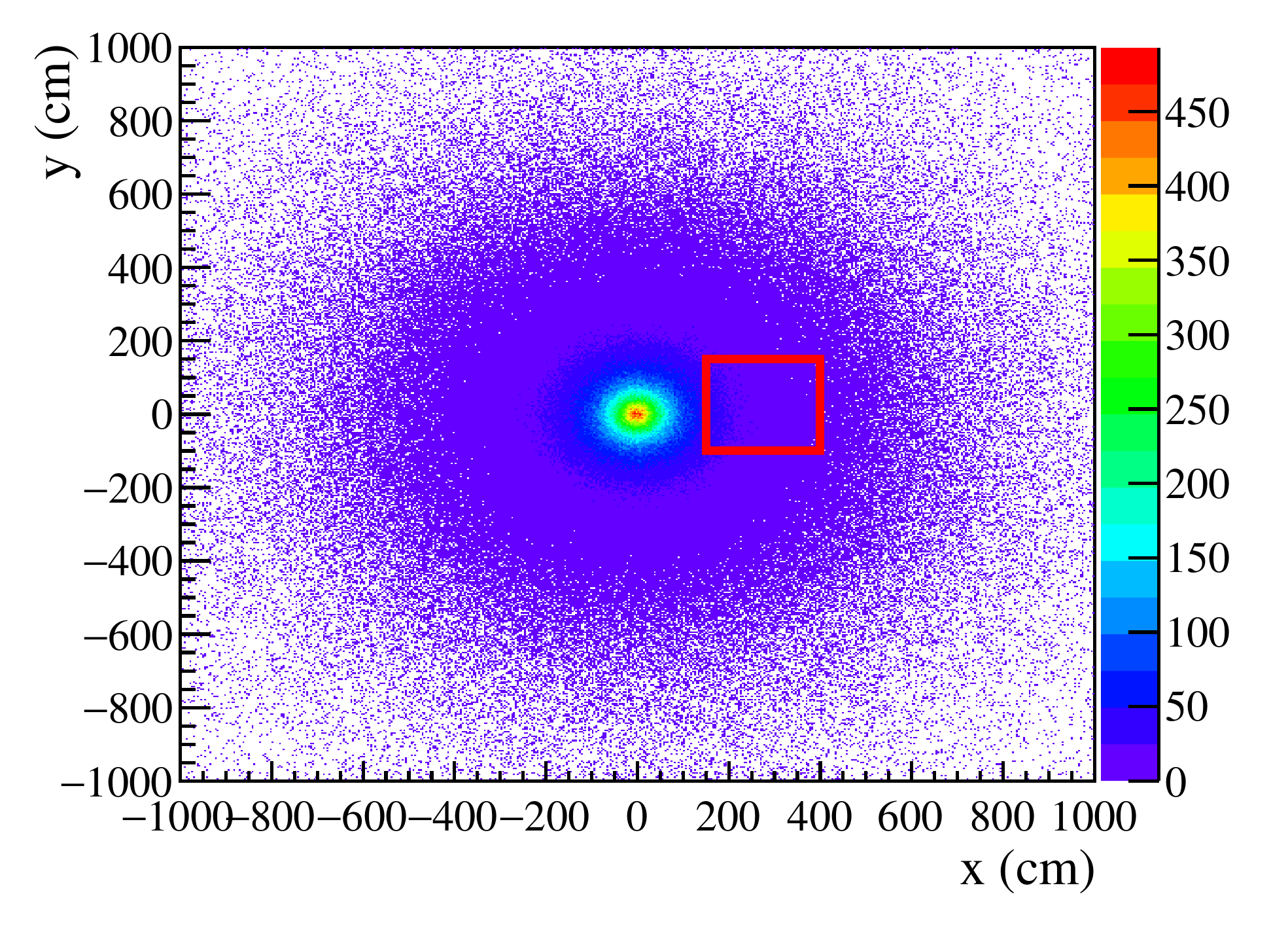}
\includegraphics[width=0.48\textwidth]{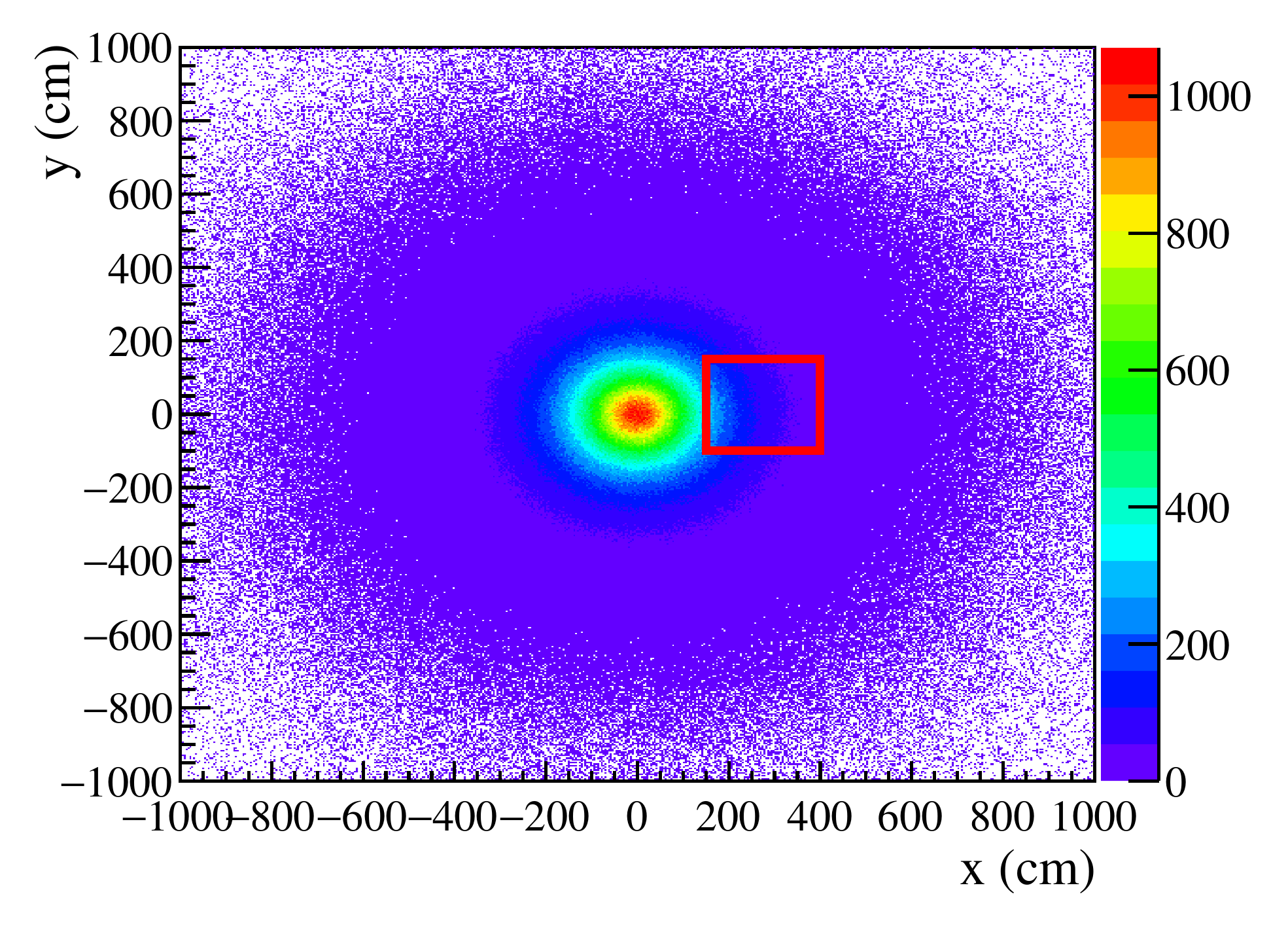}
\end{center}
\caption{\small Illumination at the first SHADOWS tracking layer ($\sim 50$~m from the dump) of the $\pi^+ \mu^-$ pair from the decay of a HNL of mass 1.5 GeV from charm (left) and from beauty (right) hadron decays.}
\label{fig:illumination}
\end{figure}

The detector requirements are defined by the characteristics of FIPs produced in the interactions of 400 GeV proton beam with a dump. At these energies, FIPs with masses above the kaon mass are mostly produced in the decays of charmed and beauty hadrons and in proton bremsstrahlung and/or Primakoff effect occurring in the dump.
The smallness of the couplings implies that FIPs are naturally long-lived, eg. at the SPS energies they can fly on average from a few meters to several tens or hundreds of km, depending on the model and scenario. On the other side the SPS centre-of-mass energy ($\sqrt{s} \sim 23$~GeV) implies that the hadrons are produced with a relatively small boost and therefore FIPs emerging from their decays have a large polar angle. Figure~\ref{fig:illumination} shows the illumination of the $\pi^+ \mu^-$ pair from the decay of a HNL of mass 1.5~GeV in a scoring plane placed at $\sim 50$~m from the dump. The HNL is assumed to be created in the decays of charmed hadrons (Figure~\ref{fig:illumination}, left) and beauty hadrons (Figure~\ref{fig:illumination}, right) produced in the interactions of the 400~GeV proton beam with the dump. The pions and muons illuminate the plane up to several meters away from the beam axis, validating the concept of an off-axis detector.

\vskip 2mm
The distance of the detector with respect to the impinging point of the proton beam onto the dump is a compromise between the maximisation of FIP flux in acceptance (that requires short distances) and the maximisation of the probability that the FIP decays before reaching the detector (that requires long distances).
The optimal distance varies as a function of the FIP model and benchmark: for example, HNLs with very long lifetimes
(hundreds of km at the SPS energy) require long decay volumes in order to increase the probability of their decay before reaching the spectrometer. Light dark scalars, instead, have a lifetime of order of meters for a large majority of the available parameter space for a mass in the GeV range, and therefore require relatively short decay volumes.

\vskip 2mm
There is another consideration to be taken into account.
On the one hand,  the detector should be placed on-axis with respect to the beam in order to maximise the acceptance for FIPs. On the other hand, an on-axis detector is hindered by the large flux of muon and neutrino backgrounds induced by the interactions of the protons with the dump. A detector placed off-axis is less affected by these backgrounds that are mostly produced in the forward direction and can profit of the non negligible polar angle of a FIP coming from the decay of a hadron containing a heavy quark (charm or beauty).
The transverse distance with respect to the beam axis is a compromise between the loss of signal acceptance and the background reduction.

\vskip 2mm
The interplay between these diverse requirements defines the driving principles for the design of the detector. {\it Ultimately, an optimal detector for FIPs at a proton beam dump is made of a series of spectrometers, each preceded by a decay volume whose length is a function of the spectrometer transverse dimensions. If the series of spectrometers is placed off-axis the loss in acceptance is mitigated by a large background reduction with respect to an on-axis setup and by the fact that the detector can be placed much closer to the FIP production point.}

\vskip 2mm
This proposal describes the implementation of the first  of a series of spectrometers that could be placed off-axis in the ECN3/TTC8 area. The first spectrometer could be placed in zone 1. However physics sensitivities are computed also for the case of two spectrometers, placed one after the other in zone 1 and zone 2, respectively, to show in perspective the physics potential of the SHADOWS proposal. Going beyond two spectrometers is not considered in the present study, but it is a compelling option that will be studied in the future.

\begin{figure}[h]
\begin{center}
\includegraphics[width=0.8\textwidth]{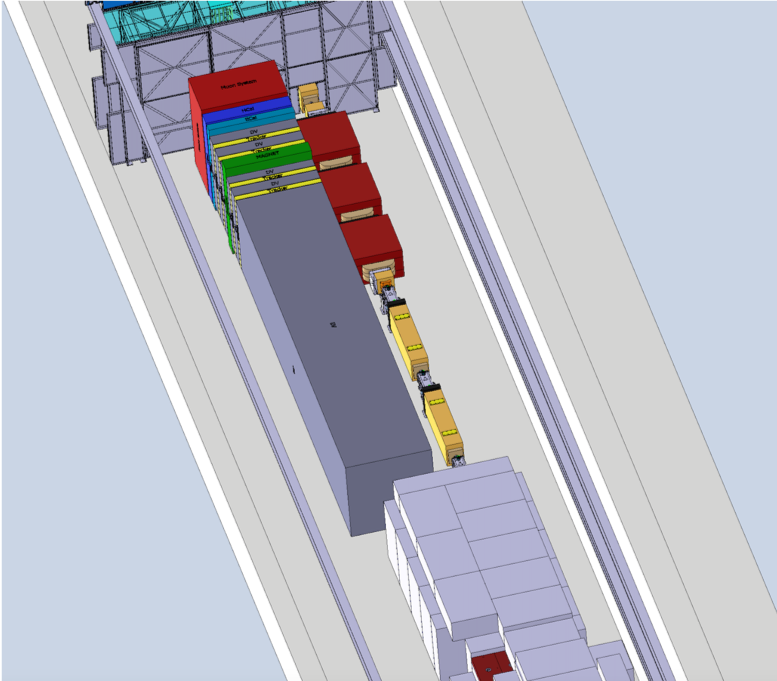}
\includegraphics[width=0.8\textwidth]{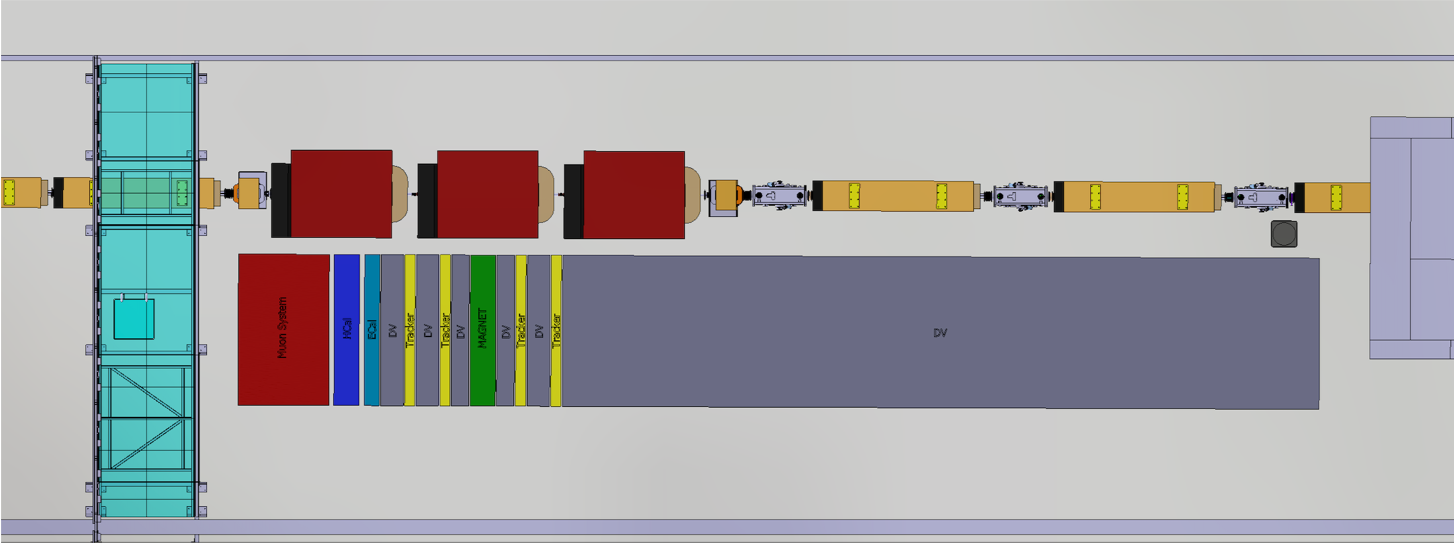}
\end{center}
\caption{\small 3D view of the SHADOWS detector located in zone 1 adiacent to the beam line and close to the TAX shielded zone. Top plot: 3D view. Bottom plot: top view.}
\label{fig:shadows_layout}
\end{figure}

\vskip 2mm
The dimension of the first spectrometer is driven by the dimensions of the available space in zone 1 and by the constraints coming from safety rules and accessibility.
The baseline SHADOWS detector in zone 1 consists of a decay volume and a spectrometer. The decay cvolume is 20~m long with transverse dimensions of 2.5$\times$2.5 m$^2$, placed $\sim 1.5$~m  off-axis with respect to the beam direction, and starting 10~m downstream of the dump. The  spectrometer of about the same transverse dimensions and a length of about 7-8~m follows the decay volume.
Figure~\ref{fig:shadows_layout} show the 3D view (top) and the top view (bottom) of the SHADOWS decay volume and spectrometer in zone 1.

\begin{figure}[h]
\begin{center}
\includegraphics[width=\textwidth]{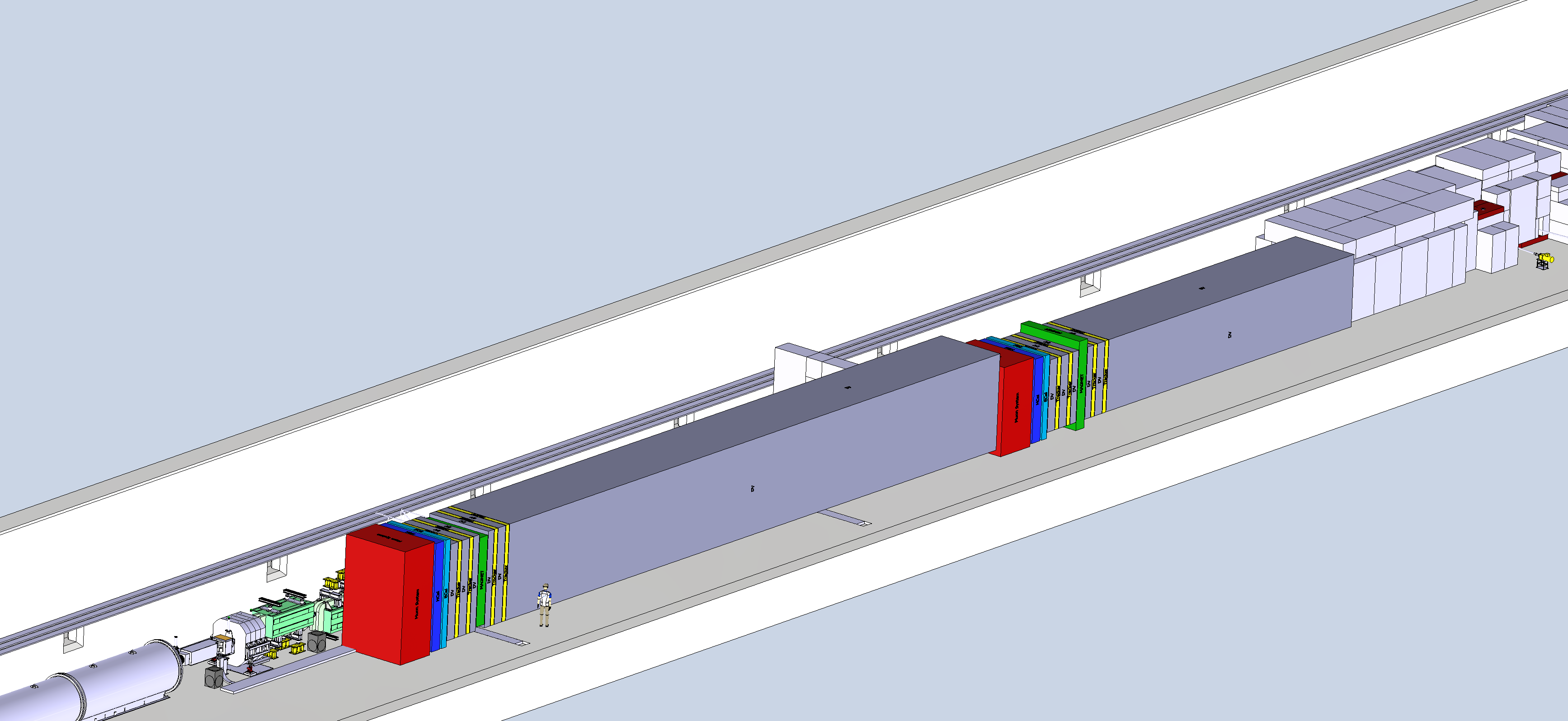}
\includegraphics[width=\textwidth]{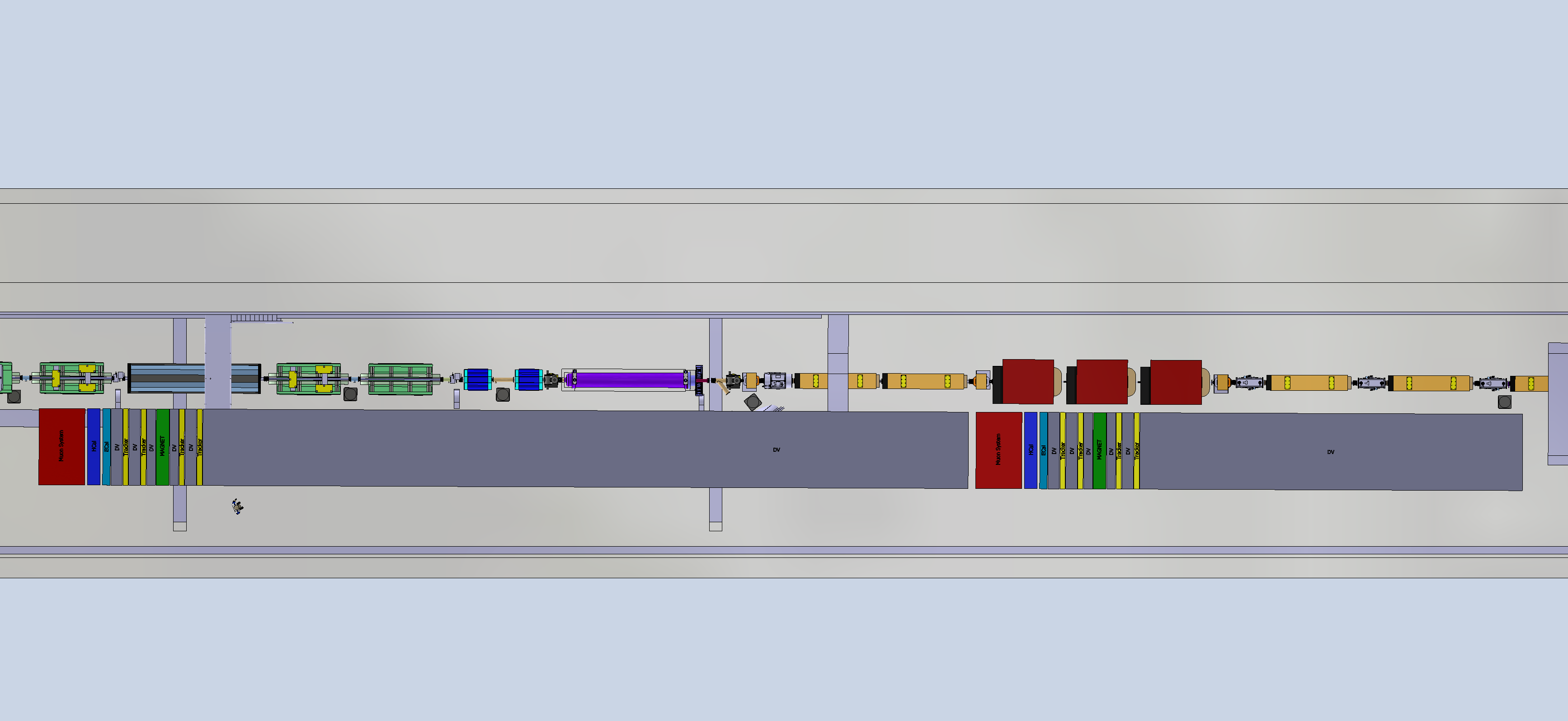}
\end{center}
\caption{\small 3D view (top) and top view (bottom) of the SHADOWS detectors located both in zone 1 and zone 2 adjacent to the beam line in the TCC8 area. The end point of the two spectrometers is defined by the beginning of the decay volume of the NA62 experiment, clearly visible after the second spectrometer.}
\label{fig:shadows2}
\end{figure}

\begin{figure}[h]
\begin{center}
\includegraphics[width=\textwidth]{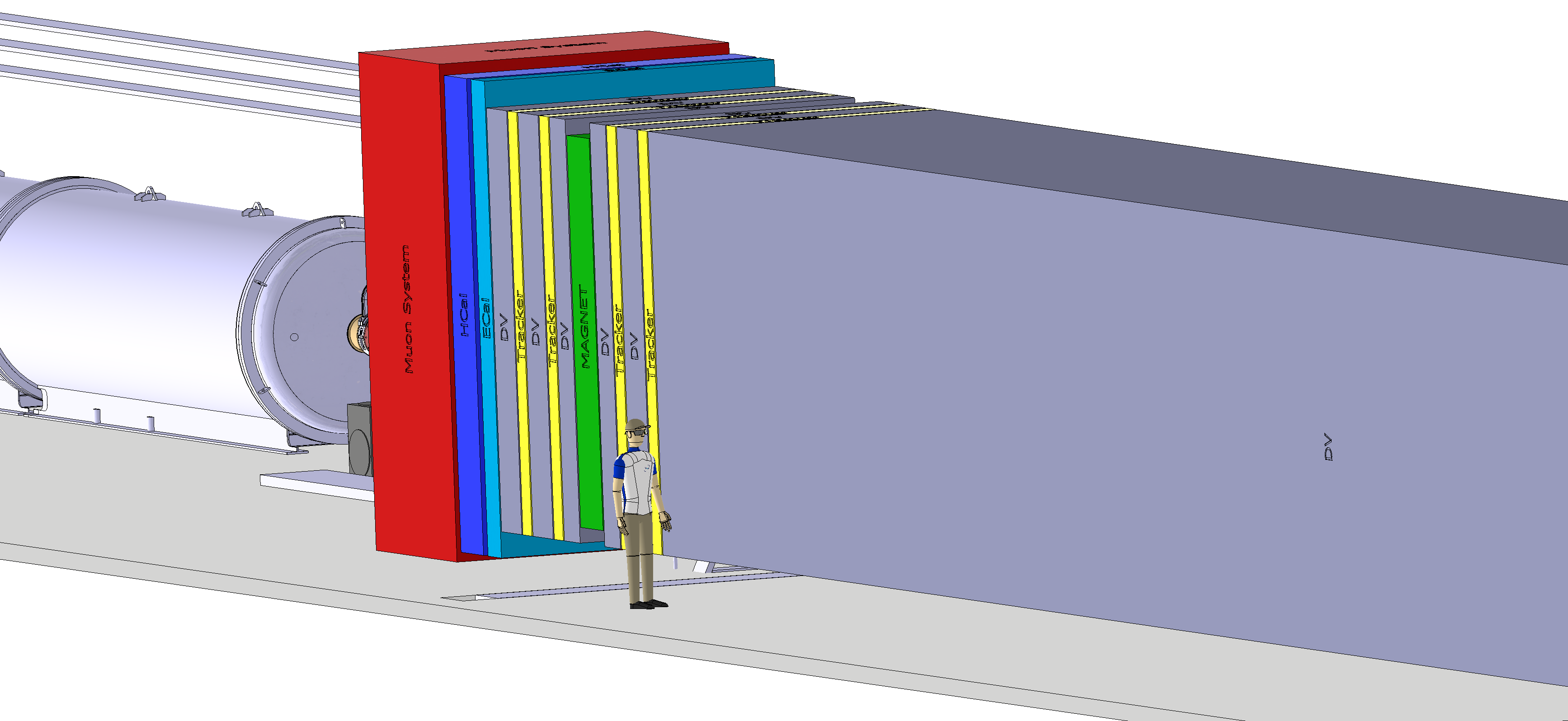}
\end{center}
\caption{\small Detail of the second SHADOWS spectrometer. The beginning of the decay tank of the NA62 experiment (Blue tube) is clearly visible after the spectrometer.}
\label{fig:shadows2_detail}
\end{figure}

\vskip 2mm
About $\sim$40~m of space could be available in zone 2, before the beginning of the decay volume of the NA62 experiment. This space could be used in the future to build another detector,
with a 30~m long, 3~m wide and 4~m high decay volume followed by a spectrometer with the same transverse dimensions. A possible layout with the two spectrometers is shown in Figure~\ref{fig:shadows2} and a zoom of the second spectrometer is shown in Figure~\ref{fig:shadows2_detail}. In the following Sections we will discuss the physics reach mostly for one spectrometer, but we include the study of the performance that can be obtained also with a second spectrometer for FIPs with very long lifetimes, as for example Heavy Neutral Leptons, in order to provide a projection of the full SHADOWS potential. 

\clearpage
\section{Physics reach}
\label{sec:physics}
The SHADOWS detector, being off-axis, is sensitive to FIPs emerging from the dump with a sizeable polar angle. This can occur in the decay of mesons/baryons containing heavy quarks (charm and beauty) produced in the interactions of 400~GeV protons with the $Cu$-based dump. The probability of producing a $b \overline{b}$ and $c \overline{c}$ pair  is given by $\chi_{c \overline{c}} = \sigma_{c \overline{c}} / \sigma_{pp} \sim 3.9 \cdot 10^{-3}$ and $\chi_{b \overline{b}} = \sigma_{b \overline{b}} / \sigma_{pp}  \sim 2.7 \cdot 10^{-7}$\cite{Lourenco:2006vw, Hans-Thomas}, which include also secondary production. The kinematic distributions of charmed and beauty mesons are obtained with Pythia 6 by simulating the interactions of 400 GeV protons with a $Cu$-based dump following the study in Ref.~\cite{Hans-Thomas}.

In the following Sections we will present SHADOWS sensitivity for light dark scalars, ALPs with fermion couplings, and Heavy Neutral Leptons, as they are representatives benchmarks for FIPs emerging from charm and beauty decays. More benchmarks are certainly possible and will be considered in the future. In evaluating the sensitivity for the benchmark channels we closely follow the prescriptions of the PBC-BSM working group~\cite{Beacham:2019nyx}. 

\subsection{Sensitivity to light dark scalars}
\label{ssec:scalars}
A light scalar particle mixing with the Higgs with angle $\theta$ can be a mediator between DM and SM particles.  The overall Langrangian is of the form:
\begin{equation}
\label{scalar}
{\cal L}_{\rm scalar} = {\cal L}_{\rm SM} + {\cal L}_{\rm DS} - (\mu S+ \lambda S^2)H^\dagger H.
\end{equation}

where $ {\cal L}_{\rm SM}$ is the SM lagrangian, 
${\cal L}_{\rm DS} $ is a overall {\it dark sector} lagrangian that can include a light dark scalar among several other states, and the remaining terms represent the interactions of a light dark scalar with the Higgs boson.
The minimal scenario assumes for simplicity that $\lambda = 0$ and all production and decay processes of the dark scalars are controlled by the same parameter $\mu$.
At low energy, the Higgs field can be substituted for $H = (v + h)/ \sqrt{2}$, where v = 246 GeV is
the electroweak vacuum expectation value and $h$ is the field corresponding to the physical 125~GeV Higgs
boson. The nonzero $\mu$ leads to the mixing of $h$ and $S$ states. In the limit of small mixing it can be written as follows:
\[
\theta = \frac{ \mu v }{ m^2_h - m^2_S }.
\]
 
For $m_h >> m_S$, $\theta \sim \mu$ and the parameter space for this model is $(\theta, m_S)$.
A more general approach consists in having both $\lambda$ and $\sin \theta$ being different from zero:
in this case, the parameter space is $\{\lambda, \theta, m_S\}$, and $\lambda$ is assumed to dominate
the production via {\em e.g.} $h\to SS$, $B \to K^{(*)}SS$, $B^0 \to SS$ etc. In this study we consider only the minimal scenario which gives the dominant contribution at the SPS energy.

At the SPS energy the dominant production process for a light dark scalar is via decay of secondary mesons containing a $s$ or $b$ quark, while the production via charmed mesons is CKM suppressed.  Since we are interested in decays emerging with a large polar angle we concentrate to the case of production via $b$ decays. In that case the inclusive production probability for a light dark scalar is:

\begin{equation}
P_{\rm scalar} \sim 2 \chi_{b \overline{b}} \times {\rm BR}(b \to S s) \sim 13 \, \theta^2 \, \chi_{b \overline{b}} \left ( 1 - {m^2_S \over m^2_b} \right )^2.
\label{eq:scalar_prod}
\end{equation}

\begin{figure}[h]
\begin{center}
\includegraphics[width=0.48\textwidth]{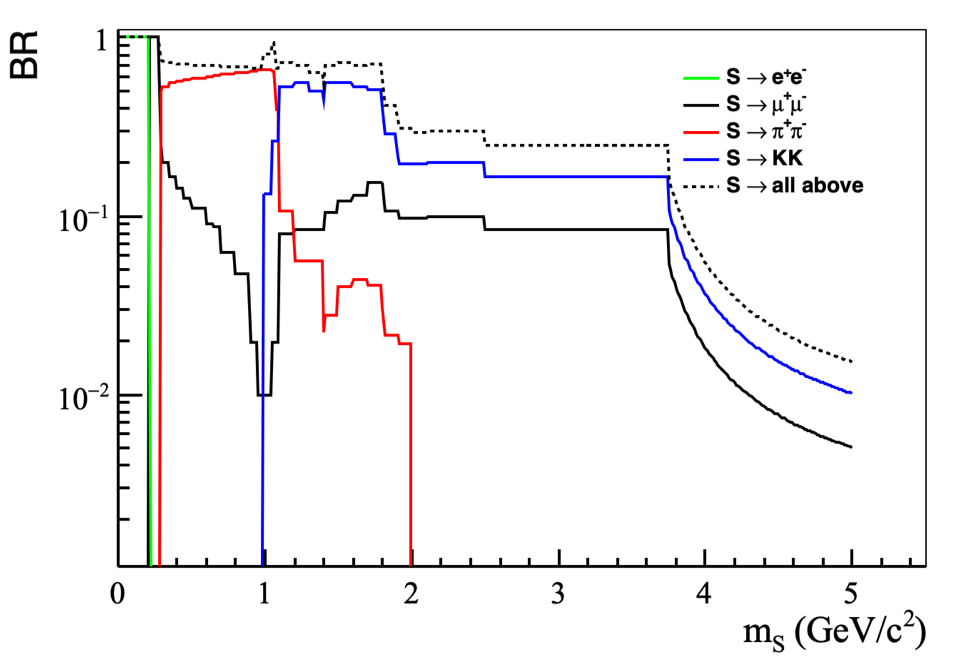}
\includegraphics[width=0.48\textwidth]{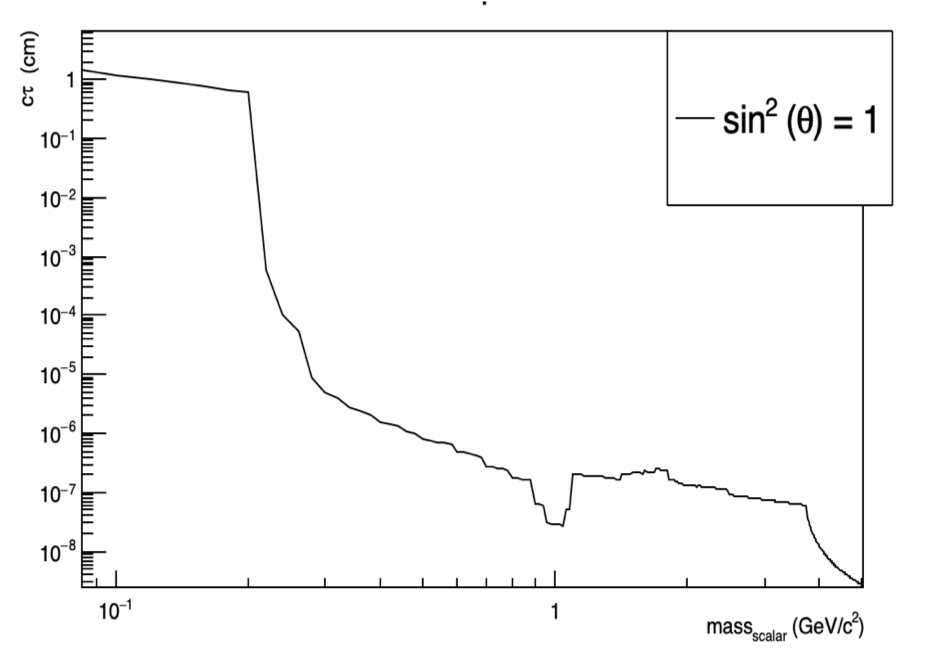}
\end{center}
\caption{\small Scalar branching fractions in two charged track final states (left) and scalar lifetime (for $\sin \theta =1$) (right) as a function of the scalar mass.}
\label{fig:scalar_br}
\end{figure}

The main visible decay channels of the light dark scalar are photons, leptons and hadrons.
Below the $\pi \pi$ threshold, the scalar decays into photons, electrons and muons. We follow Ref.~\cite{Winkler:2018qyg} for the evaluation of the partial decay widths that can be computed perturbatively. As example of perturbative partial decay width is the one for leptonic decays: 

\begin{equation}
    \Gamma (S \to \ell^+ \ell^-) = { \sin^2 \theta G_F m_S \over 4 \sqrt{2} \pi } m^2_{\ell} \beta^3_{\ell}
\end{equation}

where $\ell = (e,\mu,\tau$), $G_F$ is the Fermi constant, and $\beta_{\ell} = \sqrt{1- 4 m^2_{\ell}/m^2_S}$. Hadronic decays are instead more involved due to the presence of strong interactions in the final state, namely in the proximity of the $f_0 (980)$ resonance, where non-perturbative computations are needed. There is a vast literature about the evaluation of the hadronic widths near the $\sim 1$~GeV threshold (see for example Ref.~\cite{Winkler:2018qyg}). In this work we follow the PBC recommendations and use the computation done by Donoghue et al. ~\cite{Donoghue:1990xh} for the partial decay widths for $\pi \pi$ and $ KK$ final states around the 1~GeV threshold. The scalar branching fractions for scalar decays into two charged track final states ($e^+ e^-, \mu^+ \mu^-, \pi^+ \pi^-, K^+ K^-$) and the scalar lifetime (computed for $\sin \theta = 1$) are shown in Figure~\ref{fig:scalar_br} (left and right, respectively) as a function of the scalar mass.

\vskip 2mm
In this study we consider only final states with two charged tracks, neglecting for the time being, multi pion and $\tau \tau$ final states. On this respect the sensitivity evaluation should be considered conservative. 

\vskip 2mm
The number of events of $S \to X^+ X^-$ ($X = e,\mu, \pi, K$) reconstructed in the SHADOWS detector is given by:

\begin{equation}
 N_{\rm obs} = N_{\rm pot} \times P_{\rm scalar} \times \sum_{X=ee,\mu\mu,\pi\pi,KK} BR(S \to f)  \times {\cal A}(f) \times \varepsilon (f)
\label{eq:scalar}
\end{equation}

where $N_{\rm pot}$ is the number of protons on dump, $P_{\rm scalar}$ is given by Eq.~\ref{eq:scalar_prod},   ${\cal A} (f)$ is the acceptance for a light scalar particle of a given mass $m_S$ and coupling parameter $\sin^2 \theta$ decaying into a final state $f = X^+ X^-$.

The acceptance is given by the product of the probability that the light scalar particle decays inside the fiducial volume and the probability  ${\cal P}(X^+ X^-)$ that the two charged  tracks in the final state are reconstructible in the magnetic spectrometer. 
The efficiency $\varepsilon (f)$ is the product of the trigger, reconstruction and selection efficiencies for the final state. We consider $\varepsilon (f)=1$ in the present study, knowing that for reasonable values of $\varepsilon(f)$ the results could change by a factor of two at most.

\vskip 2mm
The position of the decay vertex of the scalar is required to be inside the decay volume. The decay volume starts 10~m after the production point of the $b$-hadrons into the dump. The decay products of the scalar are required to be within the acceptance of the magnet, placed 2.5~m after the end of the fiducial volume with an aperture of $2.5\times 2.5$~m$^2$.
We assume that all the decay products reaching the magnet aperture will hit the muon detector placed at the end of the SHADOWS spectrometer, about 5~m away from the magnet. This is a reasonable assumption given the fact that the spectrometer is very compact.

\begin{figure}[h]
\begin{center}
\includegraphics[width=0.8\textwidth]{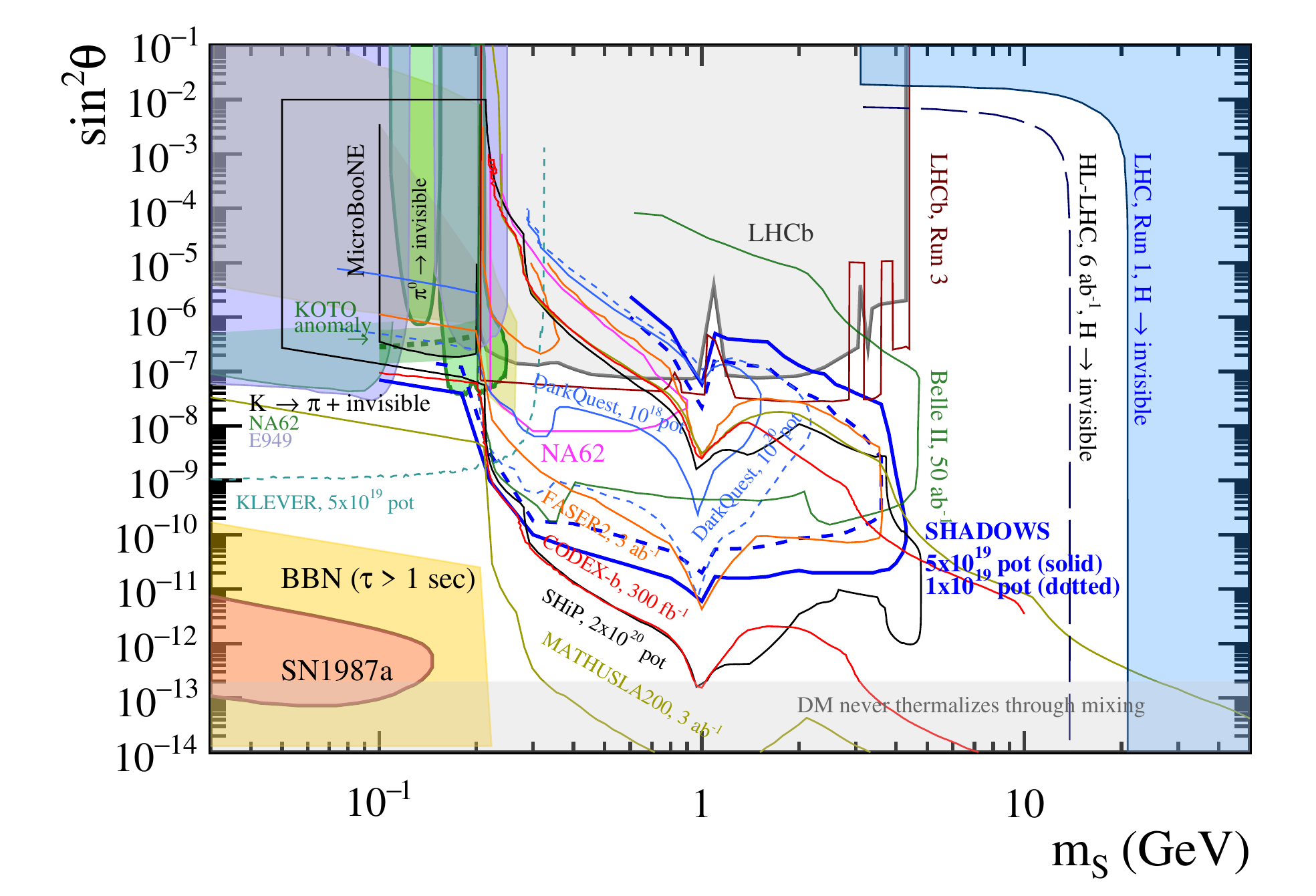}
\end{center}
\caption{\small Sensitivity of SHADOWS experiment for light feebly-interacting scalars for $N_{pot} = 10^{19}$ (scenario 1) and $N_{pot} = 5 \cdot 10^{19}$ (scenario 2).
SHADOWS with $N_{pot} = 10^{19}$ is competitive with FASER2
with 3~ab$^{-1}$ and with $N_{pot} = 5 \times 10^{19}$ covers a sizeable fraction of the parameter space that could be possibly explored by SHiP (with $N_{pot} = 2 \times 10^{20}$), MATHUSLA and CODEX-b (with 3~ab$^{-1}$) below the $b-$mass.}
\label{fig:shadows_scalar}
\end{figure}

\vskip 2mm
Figure~\ref{fig:shadows_scalar} shows the sensitivity of SHADOWS first spectrometer for $N_{pot} = 10^{19}$ (scenario 1) and $N_{pot} = 5 \times 10^{19}$ (scenario 2) along with the current bounds and future projections for other experiments/proposals.
SHADOWS with $N_{pot} = 10^{19}$ is competitive with FASER2
with 3~ab$^{-1}$ and with $N_{pot} = 5 \times 10^{19}$ covers a sizeable fraction of the parameter space that could be possibly explored by SHiP (with $N_{pot} = 2 \times 10^{20}$), MATHUSLA (with 3~ab$^{-1}$) and CODEX-b (with 3~ab$^{-1}$) below the $b-$mass. It is worth noting that SHADOWS could cover this parameter space much earlier than FASER2, CODEX-b, and SHiP.  

\subsection{Sensitivity to ALPs with fermion couplings}
\label{ssec:pseudo-scalars}

Light pseudo-scalars with fermion couplings can be produced in meson decays, most notably in $K \to \pi a$ and $B \to K^{*} a$,
as $D$ decays are suppressed by CKM.

As for the scalar case, in order to avoid uncertainties due to the form factors, in the present study
we consider the inclusive decays $b \to X_s a$, as documented in Ref~\cite{Dolan:2014ska}:

\begin{equation}
P (b \to X_s a) = { 1 \over 8 \pi } \left( { m_b^2 - m_a^2 \over m_b^3 } \right )^2 |h^S_{sb}|^2
\label{eq:pseudo-inclusive}
\end{equation}

where $X_s$ can be any strange meson, $a$ the light pseudo-scalar, $b$ the $b-$quark. Here $|h_{sb}^S|^2 = (h_{sb}^R + h_{sb}^L)/2$ contains the ALP fermionic coupling to quarks 
$g_Y = v /f_a$  
with $v$ is the Higgs vev and $f_a$ is the scale of spontaneous symmetry breaking. Here we report the explicit expression of $h_{sb}^R$  ($h_{sb}^L $ is analogous to $h_{sb}^R$ with $m_b \to m_s$):

\begin{equation}
 h_{sb}^R = - {\alpha g_Y m_b m^2_t \over  4 \sqrt{2} \pi m^2_W \sin(\theta_W)^2 v}
 V_{tb} V^*_{ts} \log \left ({\Lambda^2 \over m^2_t} \right ).
\end{equation}

In this expression $m_b$, $m_t$, $m_W$ are the masses of the bottom, top quarks and $W$, respectively, $\theta_W$ is the Weinberg angle, and $V_{tb} V^*_{ts}$ are the CKM matrix values.

\vskip 2mm
It is important to note that the production rate of an ALP depends on a cut-off scale $\Lambda$ that cures the divergency in the FCNC diagram. This divergency is absent in the scalar case. This means that numerical computations depend on the specific chosen UV completion. Here we follow what proposed in \cite{Dolan:2014ska} and use $\Lambda= $ 1 TeV. Also the analytic formulae for the branching fractions are taken from the same reference.

Below $2\cdot m_{\mu}$ the ALPs decay rate is dominated by the electron and photon mode, above $2\cdot m_{\mu}$ by the muon mode and above $2 \cdot m_{\tau}$ by the $\tau$ modes.
The branching fraction in hadrons is instead negligible and open only above the $3 \pi$ threshold as a consequence of the
pseudoscalar nature of the ALP that cannot decay in two pions.
In the current study we consider only the $e^+ e^-$ and $\mu^+ \mu^-$ final states which are the dominant ones (apart a small window between 100 and 200 MeV which is dominated by $\gamma \gamma$ final state) below $\sim 3$~GeV.

\begin{figure}[h]
\begin{center}
\includegraphics[width=0.8\textwidth]{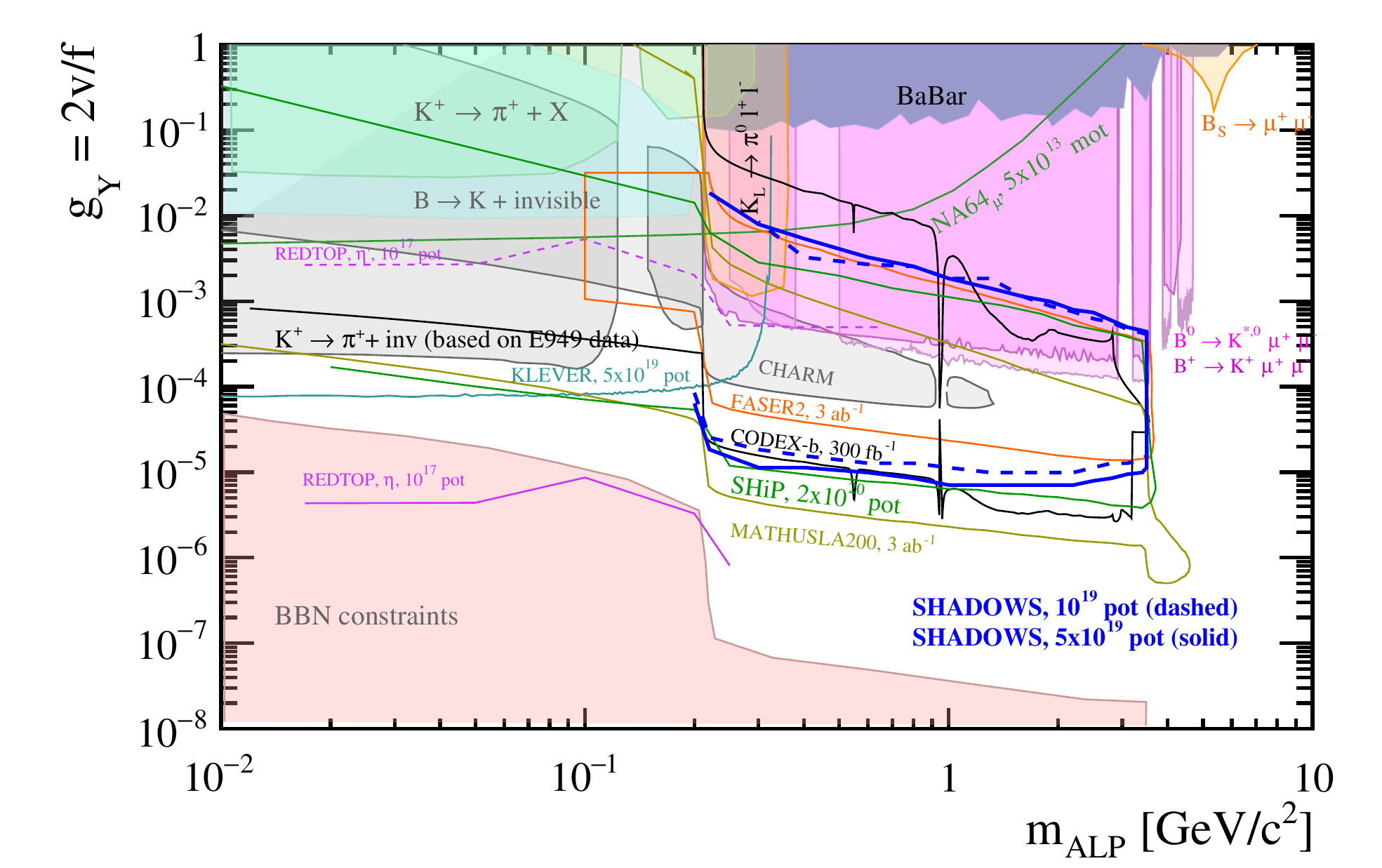}
\end{center}
\caption{\small Sensitivity of SHADOWS experiment for light feebly-interacting axion-like particles with fermion couplings for $N_{pot} = 10^{19}$ (scenario 1) and $N_{pot} = 5 \times 10^{19}$ (scenario 2). SHADOWS is better than FASER2 with $3$~ab$^{-1}$, equivalent to CODEX-b (with 3 ab$^{-1}$) and SHiP (with $2\cdot 10^{20}$~pot), and covers a sizeable part of the parameter space below the $b$-mass that could be possibly covered by MATHUSLA.}
\label{fig:shadows_alps}
\end{figure}

\vskip 2mm
As for the light scalar case, the number of events of $S \to X^+ X^-$ ($X = e,\mu$) reconstructed in the SHADOWS detector is given by:

\begin{equation}
 N_{\rm obs} = N_{\rm pot} \times P (b \to X_s a) \times \sum_{X=ee,\mu\mu} BR(a \to f)  \times {\cal A}(f) \times \varepsilon (f)
\label{eq:pseudoscalar}
\end{equation}

where we use the same notations as in Eq.~\ref{eq:scalar} and $P (b \to X_s a)$ is defined in Eq.~\ref{eq:pseudo-inclusive}.
Figure~\ref{fig:shadows_alps} shows the SHADOWS sensitivity with one spectrometer for $N_{\rm pot} \sim 10^{19}$ (scenario 1) and $5 \cdot 10^{19}$ (scenario 2) compared against projections from other experiments~\cite{Agrawal:2021dbo}. 
In both scenarios SHADOWS is better than FASER2 with $3$~ab$^{-1}$, equivalent to CODEX-b (with 3 ab$^{-1}$) and SHiP (with $2\cdot 10^{20}$~pot), and covers a sizeable part of the parameter space below the $b$-mass that could be possibly covered by MATHUSLA. It is worth noting also in this case that SHADOWS could cover this parameter space much earlier than FASER2, CODEX-b, and SHiP.

\clearpage
\subsection{Sensitivity to Heavy Neutral Leptons}
\label{ssec:hnls}
The neutrino portal extension of the SM is very motivated by the fact that it can be tightly related with the neutrino mass generation mechanism. The neutrino portal operates with one or several dark fermions $N$, that can be also called {\it heavy neutral leptons} or HNLs.  The general form of the Lagrangian that includes the neutrino portal is:

\begin{equation}
\label{neutrino}
{\cal L}_{\rm vector} = {\cal L}_{\rm SM} + {\cal L}_{\rm DS} + \sum F_{\alpha I}(\bar L_\alpha H) N_I
\end{equation}

where ${\cal L}_{\rm SM}$ is the SM Lagrangian, ${\cal L}_{\rm DS}$ is the Lagrangian describing a dark sector that includes HNL as one of the possible states, and the remaining term describe the interactions of HNL with the SM. Here
 $H$ is the Higgs field and the summation goes over the flavour of lepton doublets $L_{\alpha}$, and the number of available HNLs, $N_I$.
The $F_{\alpha I}$ are the corresponding Yukawa couplings.
The dark sector Lagrangian $\mathcal{L}_{\rm DS}$ should include the mass terms for HNLs, that can be both Majorana or Dirac type. For a more extended review, see Ref.~\cite{Gorbunov:2007ak,Alekhin:2015byh}.

Setting the Higgs field to its vacuum expectation value (vev), and diagonalizing mass terms for neutral fermions, one arrives at $\nu_i-N_J$ mixing, that is usually parametrized by a matrix called $U$.
Therefore, in order to obtain interactions of HNLs, inside the SM interaction terms one can replace $\nu_\alpha \to \sum_I U_{\alpha I} N_I$. In the minimal HNL models, both the production and decay of an HNL are controlled by the elements of matrix $U$. 

\vskip 2mm 
Following the PBC prescriptions as in Ref.~\cite{Beacham:2019nyx}, we evaluate SHADOWS sensitivity to HNLs assuming the single-flavor dominance, eg: we assume only one HNL state N that mixes with electron or muon or tau neutrinos, separately. Under this assumption all HNL production and decay probability can be determined as a function of the parameters ($m_N, |U^2_{e}|$, ($m_N, |U^2_{\mu}|$) and ($m_N, |U^2_{\tau}|)$, depending on the chosen flavor.
 The single flavor dominance approximation is currently being reconsidered in the PBC and multiple HNL states with comparable couplings to different flavours are now introduced to make HNL physics compatible with the parameters describing mixing and oscillations of the active neutrinos~\cite{Agrawal:2021dbo}. The determination of the SHADOWS sensitivity in this updated scenarios will be performed in the coming months following the PBC activities.

\begin{figure}[h]
\begin{center}
\includegraphics[width=0.7\textwidth]{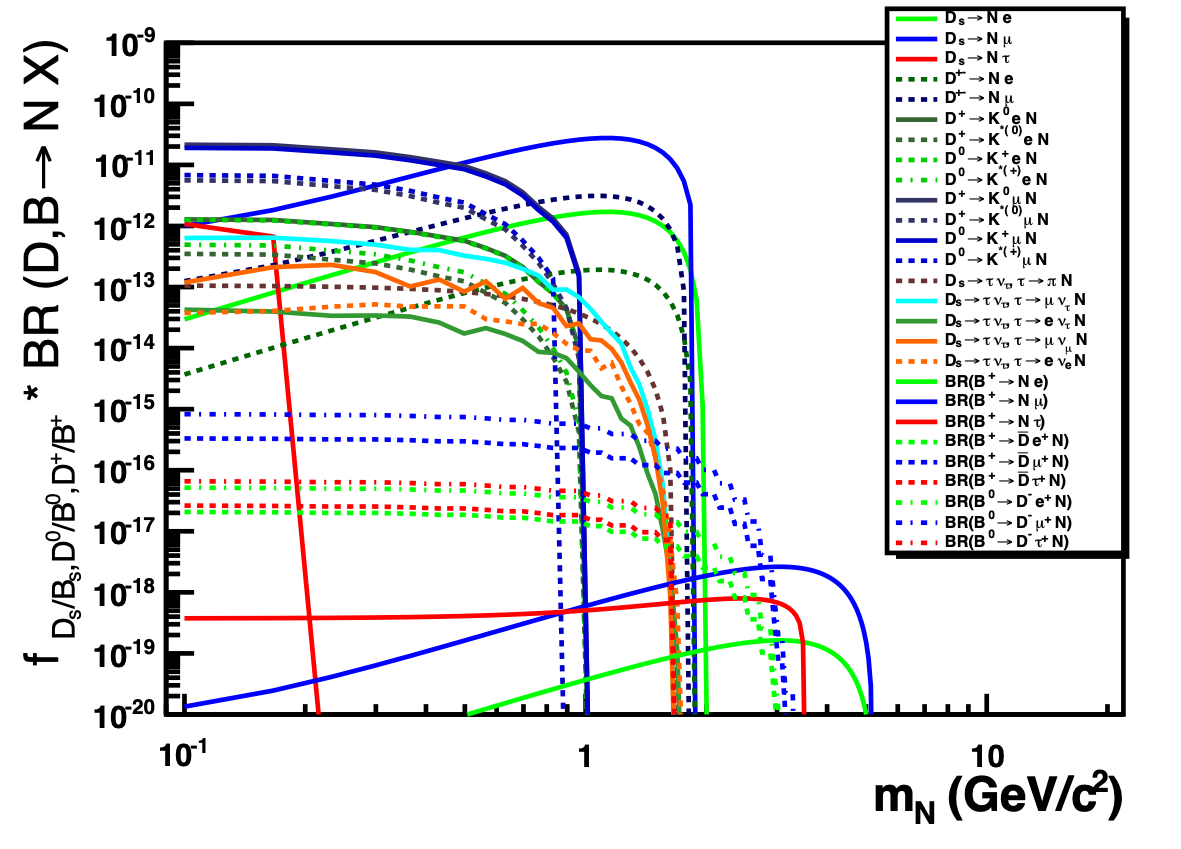}
\includegraphics[width=0.7\textwidth]{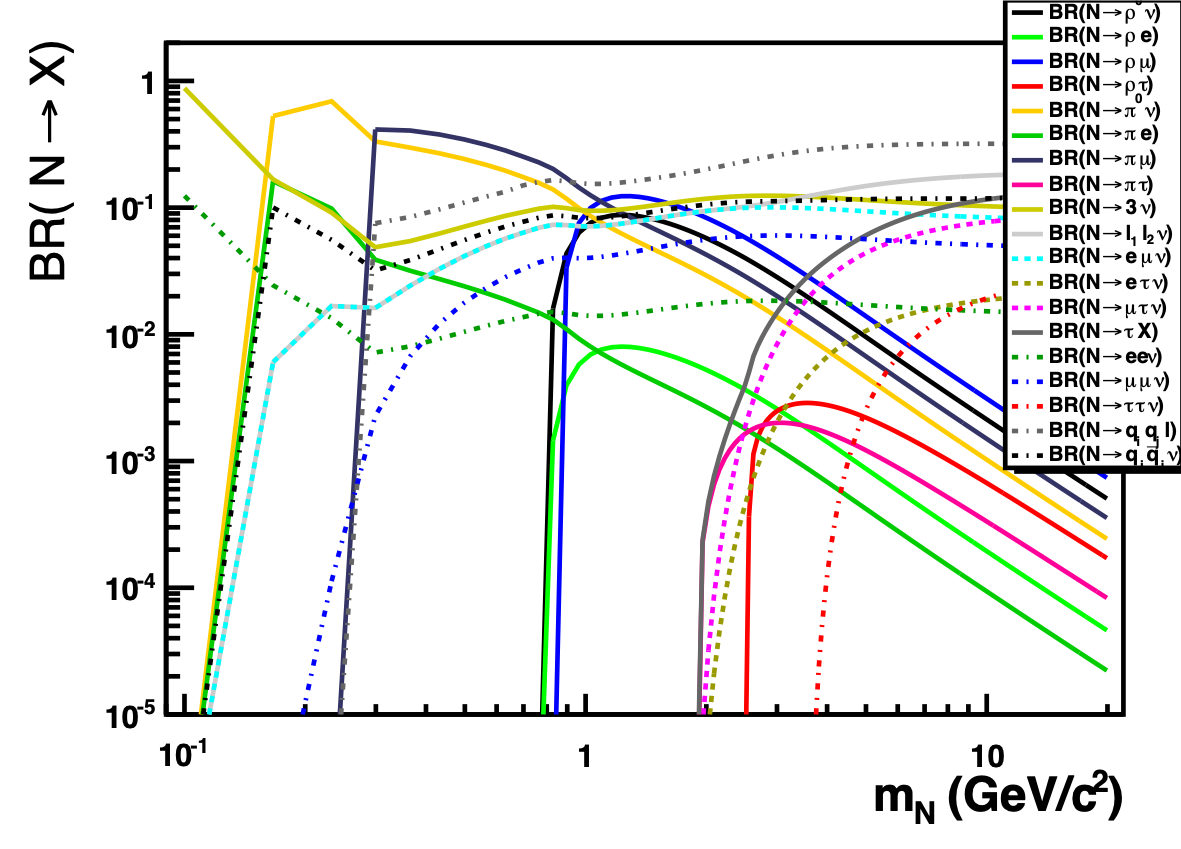}
\end{center}
\caption{\small Production modes (top) and branching fractions (bottom) of HNL decaying in several final states as a function of the HNL mass for a specific choice of the ratios of the HNL couplings.} 
\label{fig:hnl_br}
\end{figure}

\begin{figure}[h]
\begin{center}
\includegraphics[width=0.7\textwidth]{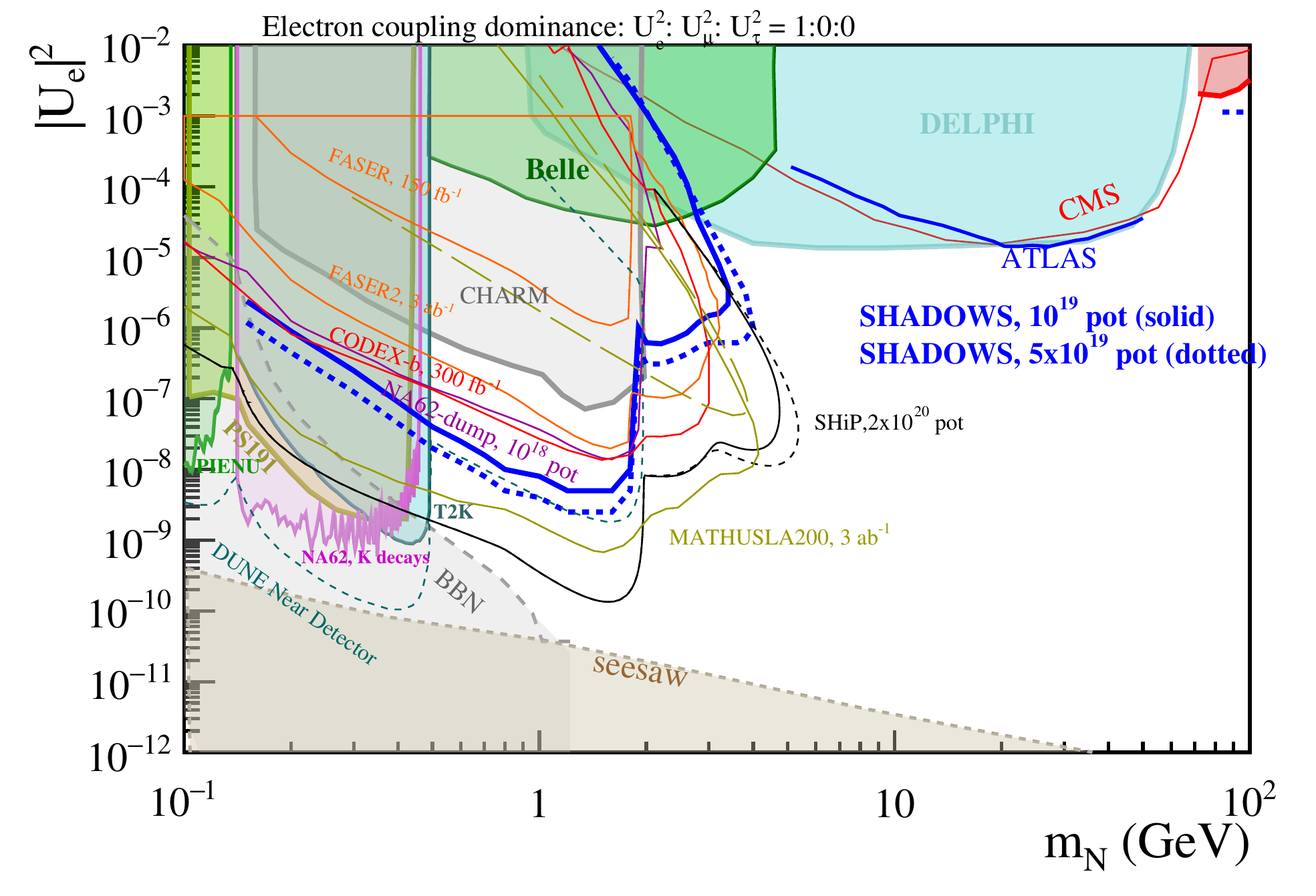}
\includegraphics[width=0.7\textwidth]{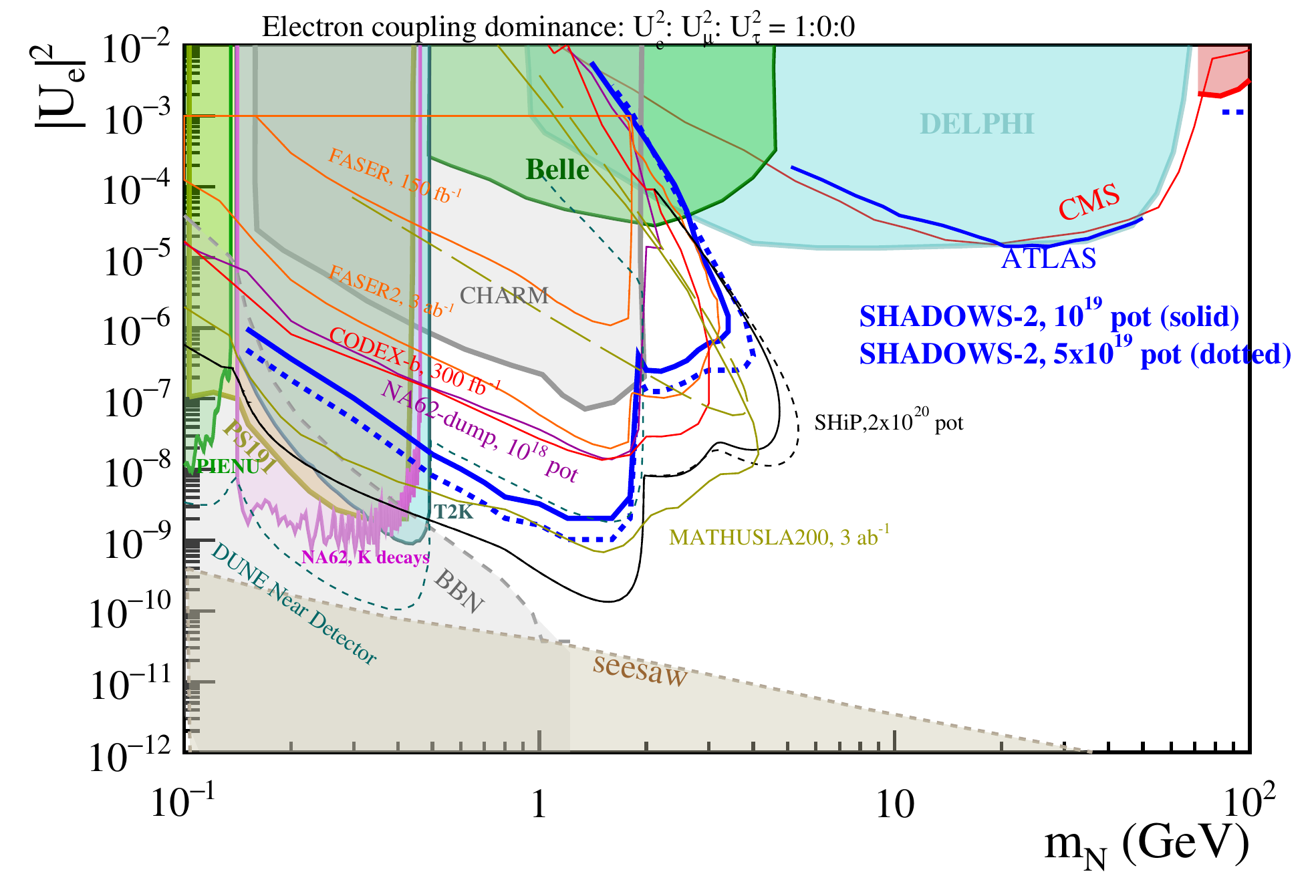}
\end{center}
\caption{\small Top: Sensitivity of SHADOWS experiment with one spectrometer for light feebly-interacting heavy neutral leptons coupled to the first lepton generation. Bottom: same as above but with two spectrometers.
Below the charm threshold, SHADOWS with one spectrometer is better than any existing or proposed experiment in same mass range apart MATHUSLA and SHiP. Below the charm threshold SHADOWS with one (two) spectrometer(s) is competitive with MATHUSLA and only one (1.5) order of magnitude above the SHiP sensitivity. Above the charm threshold and below the b-mass, SHADOWS with one (two) spectrometer(s) allows to improve the current bounds by 2.0 (2.5) orders of magnitude and is better of any experiment in the same timescale.} 
\label{fig:shadows_HNL_electron}
\end{figure}

\begin{figure}[h]
\begin{center}
\includegraphics[width=0.7\textwidth]{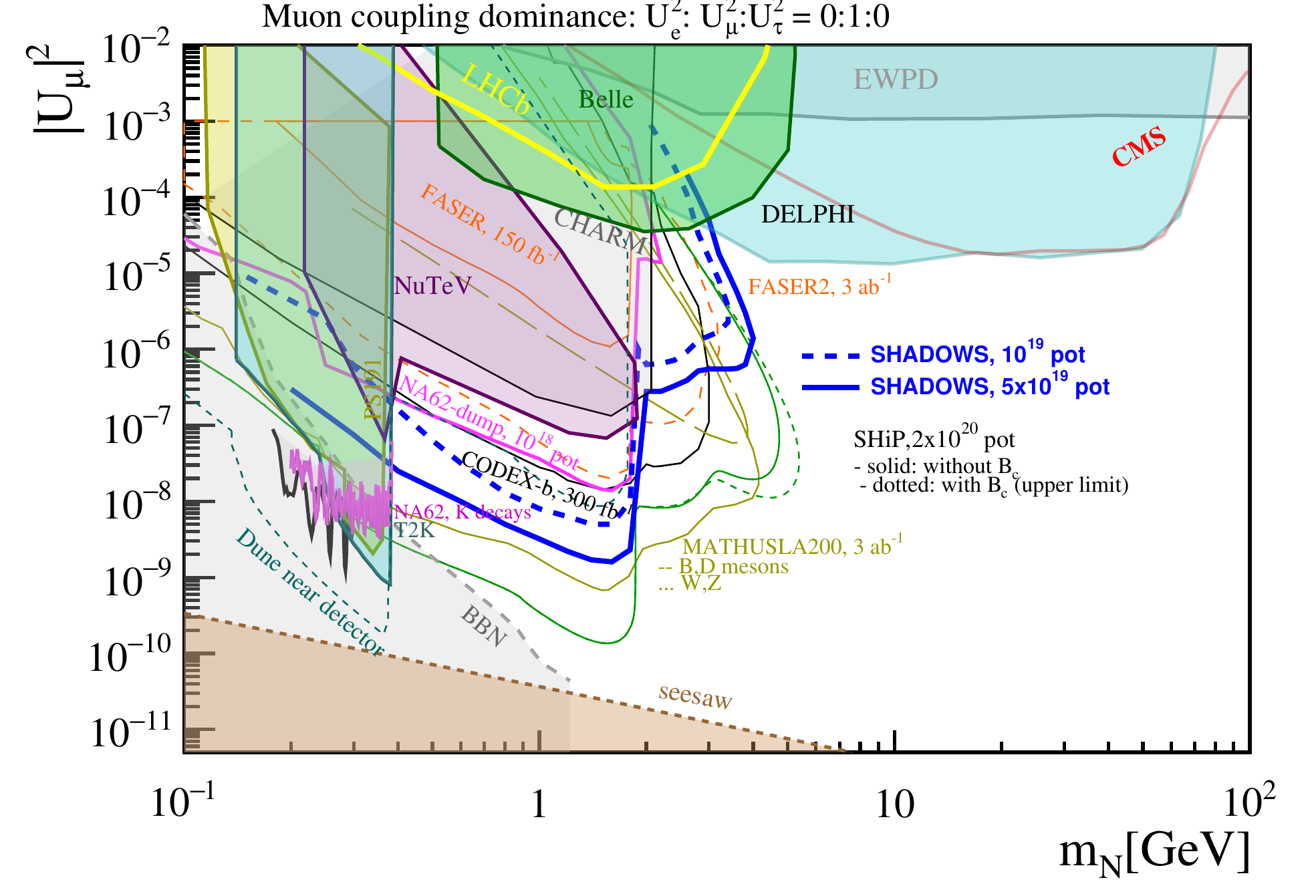}
\includegraphics[width=0.7\textwidth]{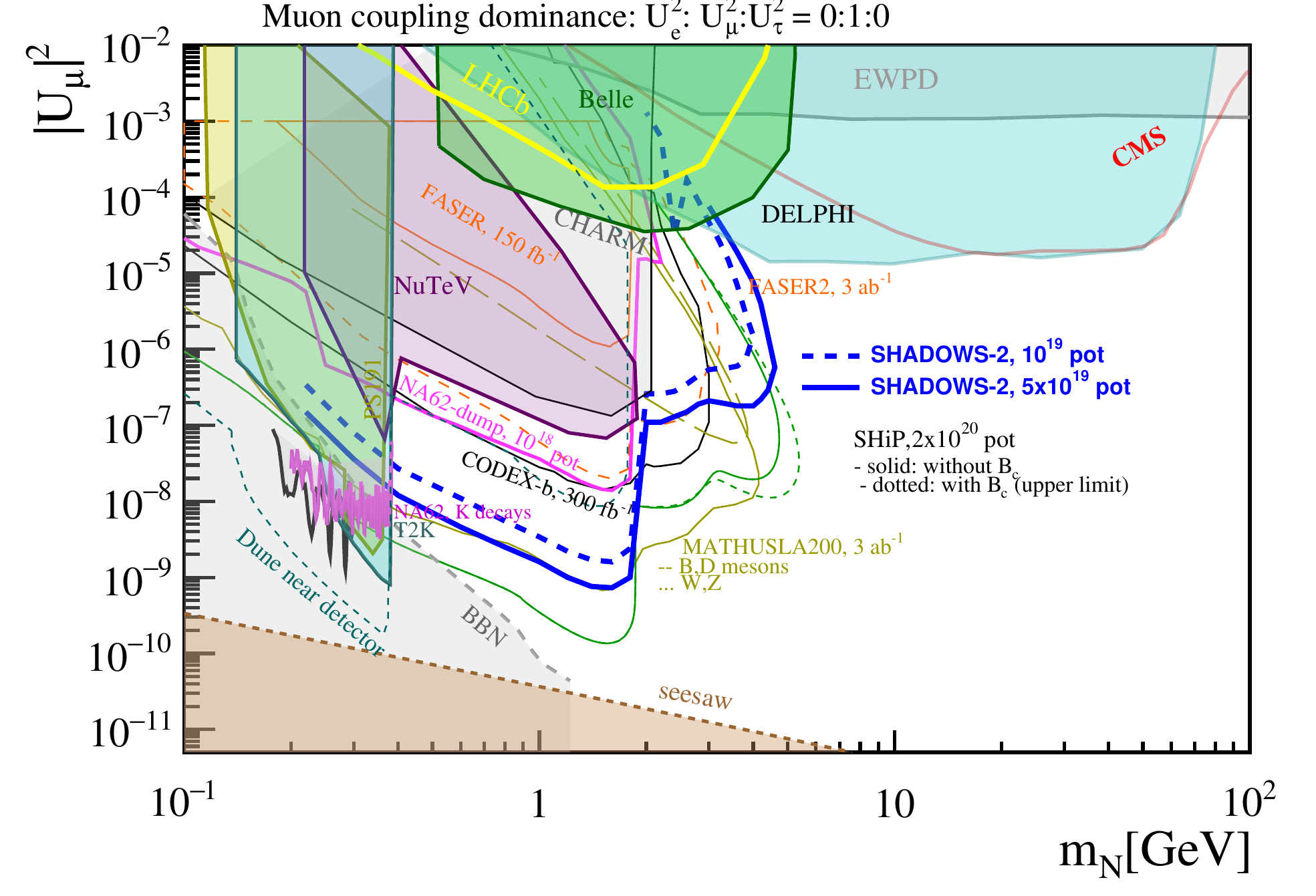}
\end{center}
\caption{\small Top: Sensitivity of SHADOWS experiment with one spectrometer for light feebly-interacting heavy neutral leptons coupled to the second lepton generation. Bottom: same as above but with two spectrometers. Below the charm threshold SHADOWS with one (two) spectrometer(s) is competitive with MATHUSLA and only one (1.5) order of magnitude above the SHiP sensitivity. Above the charm threshold and below the b-mass, SHADOWS with one (two) spectrometer(s) allows to improve the current bounds by 2.0 (2.5) orders of magnitude and is better of any experiment in the same timescale.} 
\label{fig:shadows_HNL_muon}
\end{figure}

\begin{figure}[h]
\begin{center}
\includegraphics[width=0.7\textwidth]{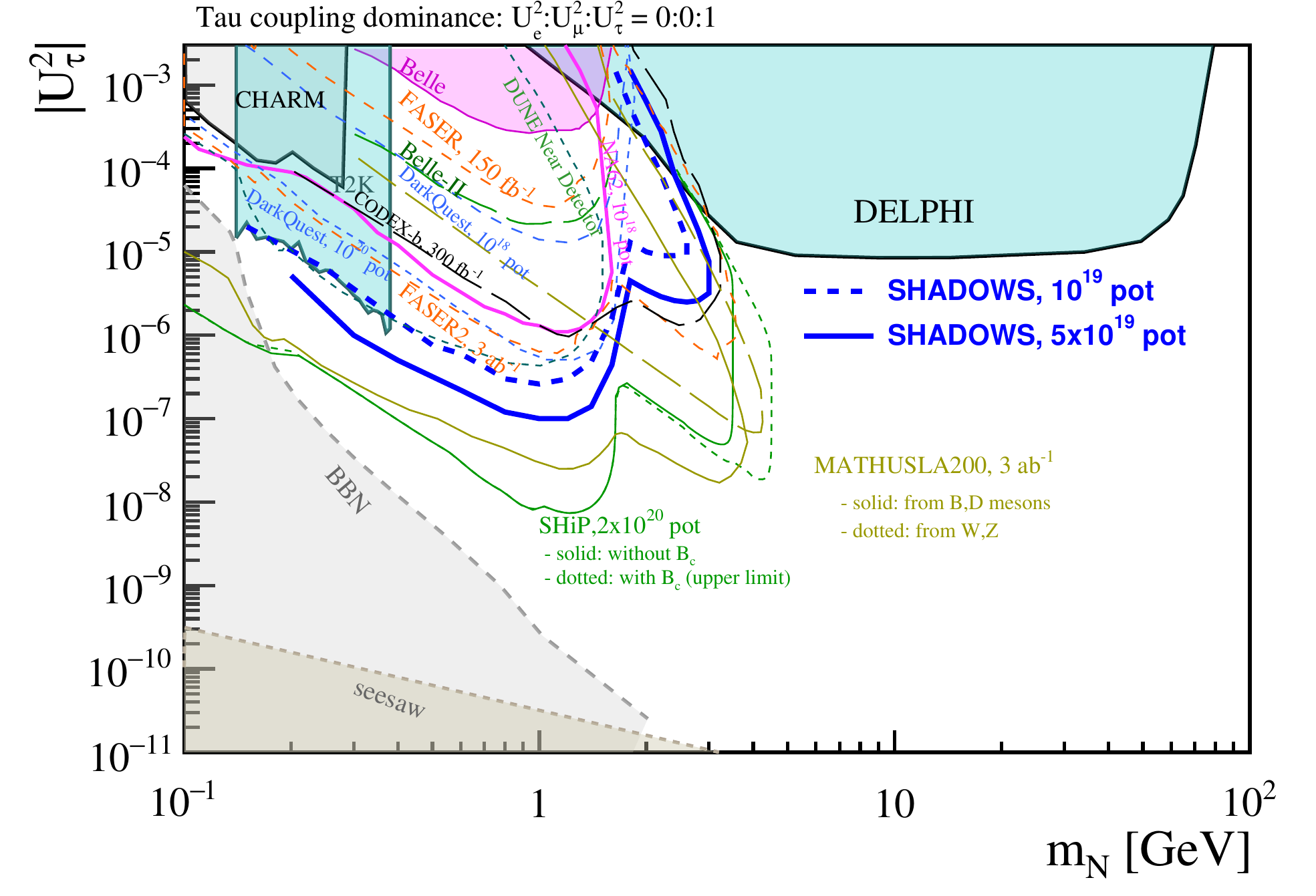}
\includegraphics[width=0.7\textwidth]{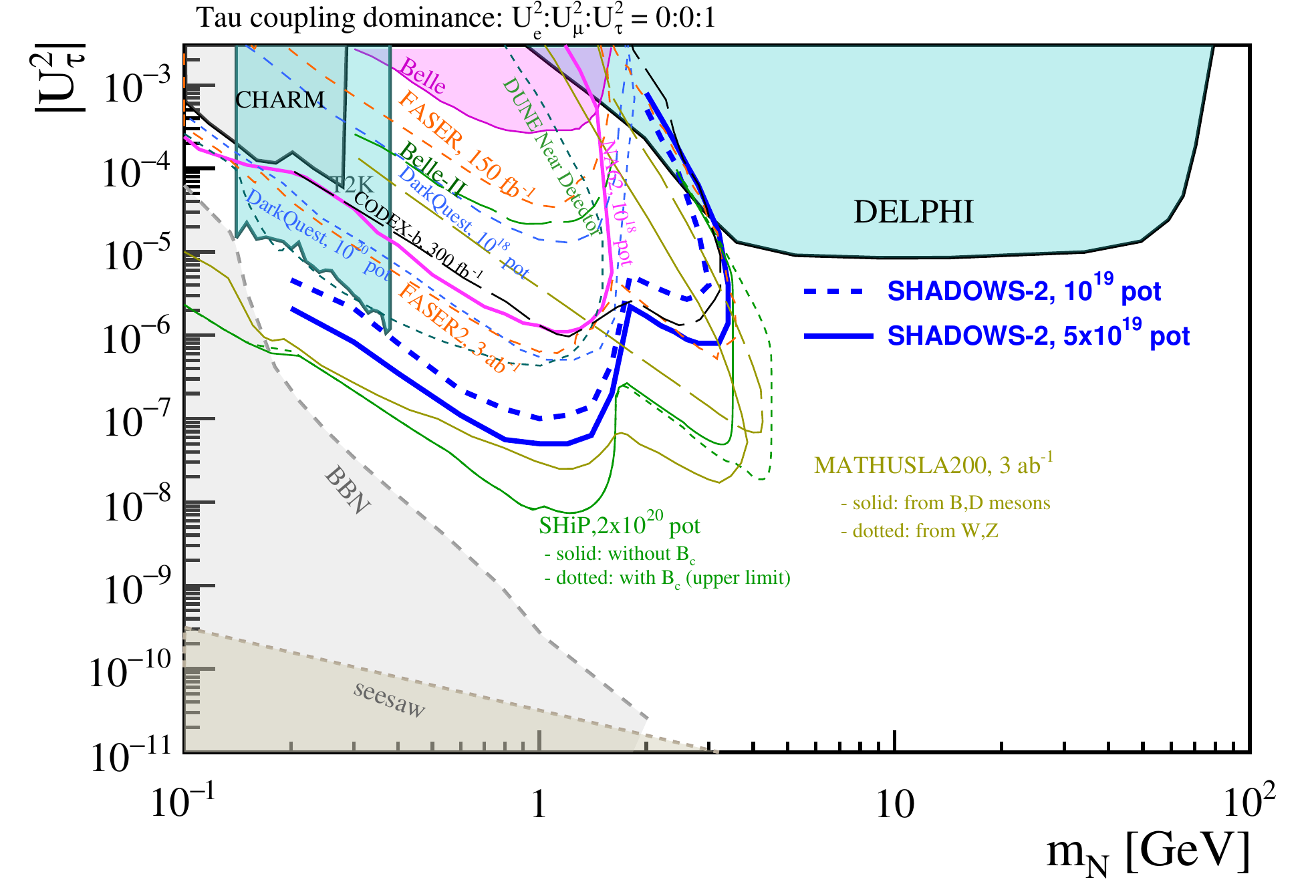}
\end{center}
\caption{\small Top: Sensitivity of SHADOWS experiment with one spectrometer for light feebly-interacting heavy neutral leptons coupled to the third lepton generation. Bottom: same as above but with two spectrometers. Below the charm threshold SHADOWS with two spectrometers is competitive with MATHUSLA and only one order of magnitude above the SHiP sensitivity. Above the charm threshold and below the b-mass, SHADOWS with one (two) spectrometer(s) allows to improve the current bounds by 2.0 (2.5) orders of magnitude and is better of any experiment in the same timescale.} 
\label{fig:shadows_HNL_tau}
\end{figure}

\vskip 2mm
Heavy neutral leptons at the SPS energy arise from the decays of charmed and beauty mesons. The production rates and possible final states are taken from  Ref.~\cite{Gorbunov:2007ak}.
An example of HNL branching fractions as a function of the HNL mass for a specific choice of the ratios of the couplings is shown in Figure~\ref{fig:hnl_br}. 
Figures~\ref{fig:shadows_HNL_electron}, \ref{fig:shadows_HNL_muon}, and \ref{fig:shadows_HNL_tau} show SHADOWS sensitivity for the coupling to the first, second and third lepton generation for the two considered scenarios, $10^{19}$~pot and $5 \cdot 10^{19}$~pot. The top and bottom plots correspond to SHADOWS with one and two spectrometers, respectively. For the HNL case the addition of the second spectrometer is equivalent to an increase of the proton intensity of a factor $\sim (5-6)$. 

\vskip 2mm
Below the charm threshold for all considered scenarios (electron, muon, and tau coupling) SHADOWS with one spectrometer is better than any existing or proposed experiment in same mass range apart MATHUSLA and SHiP. SHADOWS with two spectrometers is competitive with MATHUSLA and only one order of magnitude above the SHiP sensitivity. Above the charm threshold and below the b-mass, SHADOWS with one (two) spectrometer(s) allows to improve the current bounds by 2.0 (2.5) orders of magnitude in all the considered scenarios and is better of any experiment in the same timescale.

\clearpage
\section{Backgrounds}
\label{sec:background}
The proton interactions with the dump, along with feebly-interacting particles, give rise to a copious direct production of short-lived resonances, and pions and kaons. While the TAX length ($\sim 22 \, \lambda_{\rm I}$) is sufficient to absorb the hadrons and the electromagnetic radiation produced in the proton interactions, the decays of pions, kaons and short-lived resonances result in a large flux of muons and neutrinos. Muons and neutrinos from the dump are the two major sources of background for FIP searches.

\vskip 2mm
For an off-axis detector, however, the background originated by neutrino interactions with the air or the detector material are fully negligible as the neutrinos emerging from the decays of pions and kaons are intrinsically produced in the forward direction. The most relevant backgrounds for SHADOWS are those originated by the beam-induced muon flux.

\vskip 2mm
In order to study the rate, composition, and evolution of the beam and related backgrounds,
a Geant4-based simulation of the P42/K12 beam line including the target and TAX area and all the magnetic elements placed along the beam  has been developed by the NA62 collaboration and the BE-EA-LE team. This simulation is based on the package {\it Beam Delivery Simulation or BDSim}~\cite{Nevay:2018zhp}. The products of the interactions of the 400~GeV proton beam with the TAX material are simulated, transported across the magnetic elements of the K12 beam line, and their kinematics and composition stored for further processing. 
The results of the simulation have been cross-checked with data when the experiment is operated in kaon-mode and the agreement between data and simulation has been found to be within a factor of two. 

\begin{figure}[h]
\begin{center}
\includegraphics[width=0.49\textwidth]{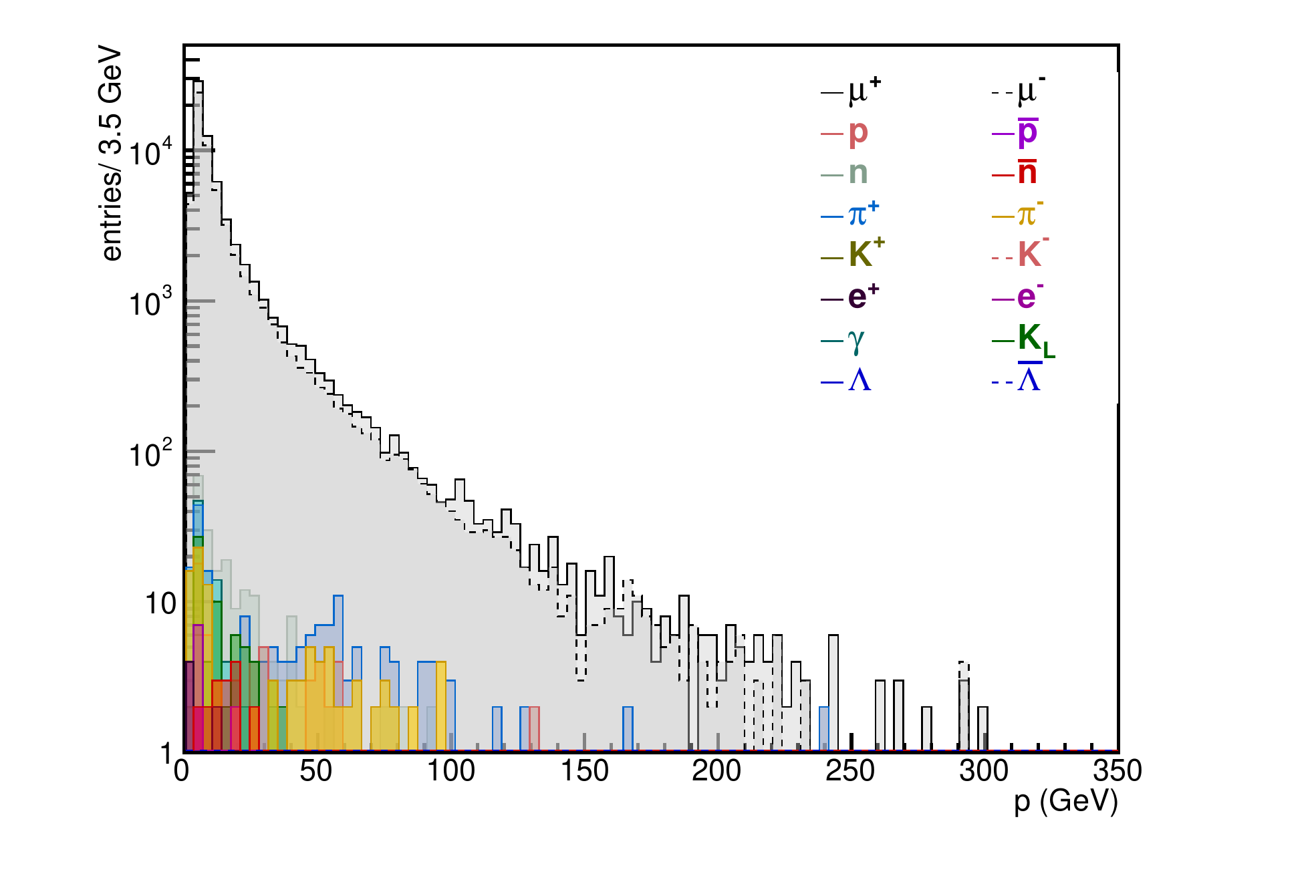}
\includegraphics[width=0.49\textwidth]{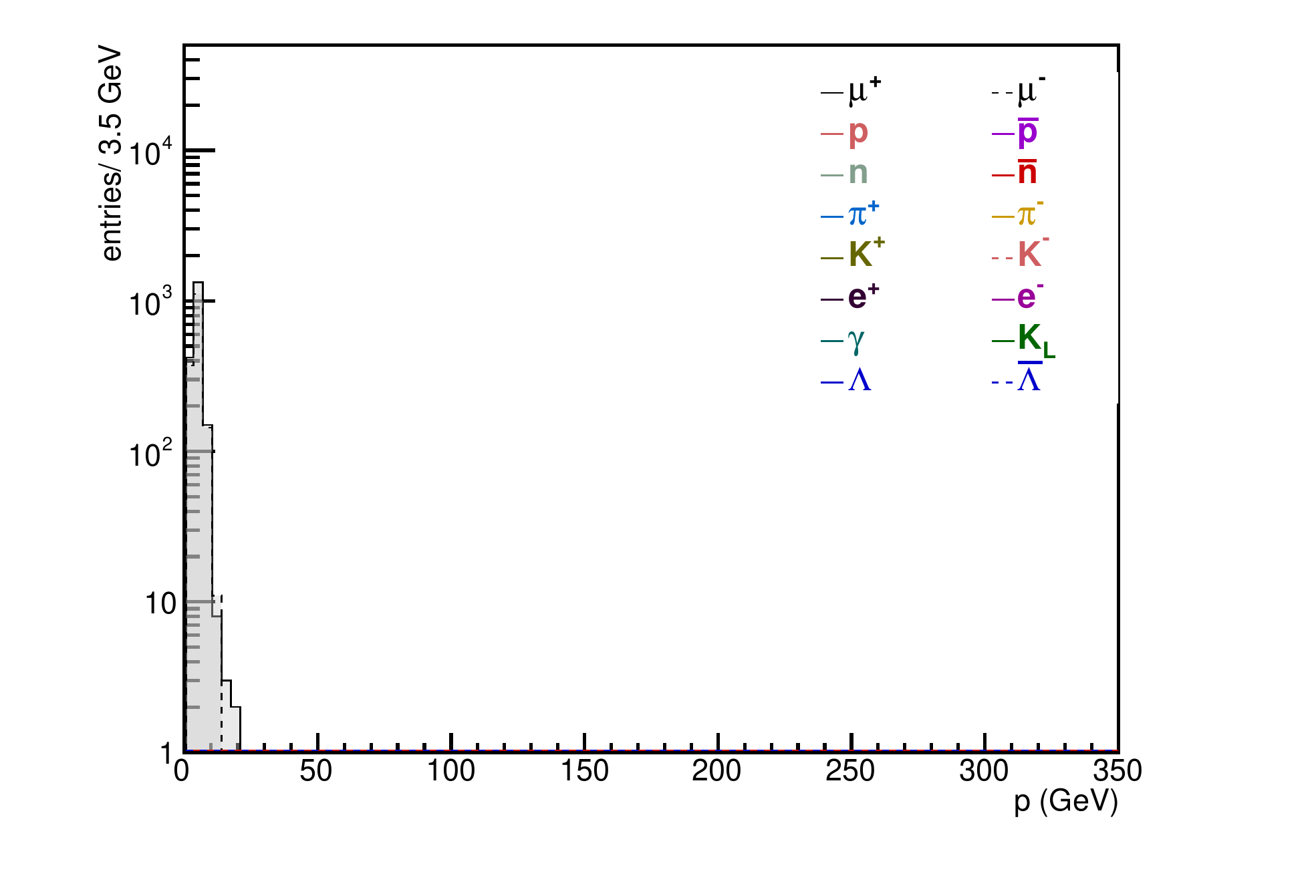}
\end{center}
\caption{\small Left: Simulated momentum distribution of particles emerging from TAXes scored immediately upstream of the beginning of the decay vessel of the SHADOWS detector.
Right: Same as left plot but requiring that the particles are in SHADOWS acceptance ($1.5<x<4.0$~m and $2.5<y<-1$~m).}
\label{fig:muon_p}
\end{figure}

\vskip 2mm
Figure~\ref{fig:muon_p} shows the simulated momentum distribution of the particles produced in the dump and scored immediately upstream of the beginning of the decay vessel ($z=34.4$~m with respect to the target) for: i) any $x,y$ (left plot); ii) $1.5 < x < 4.0$~m, $-1 < y < 1.5$~m (right plot) which define the transverse dimensions of the decay vessel. As expected, most of the background is produced on-axis and misses the SHADOWS acceptance. The only relevant background emerging with a non-negligible $p_T$ due to multiple scattering in the TAX material are low  momentum ($< 20$~GeV) muons (see Figure~\ref{fig:muon_p}, right).

\begin{figure}[h]
\begin{center}
\includegraphics[width=0.8\textwidth]{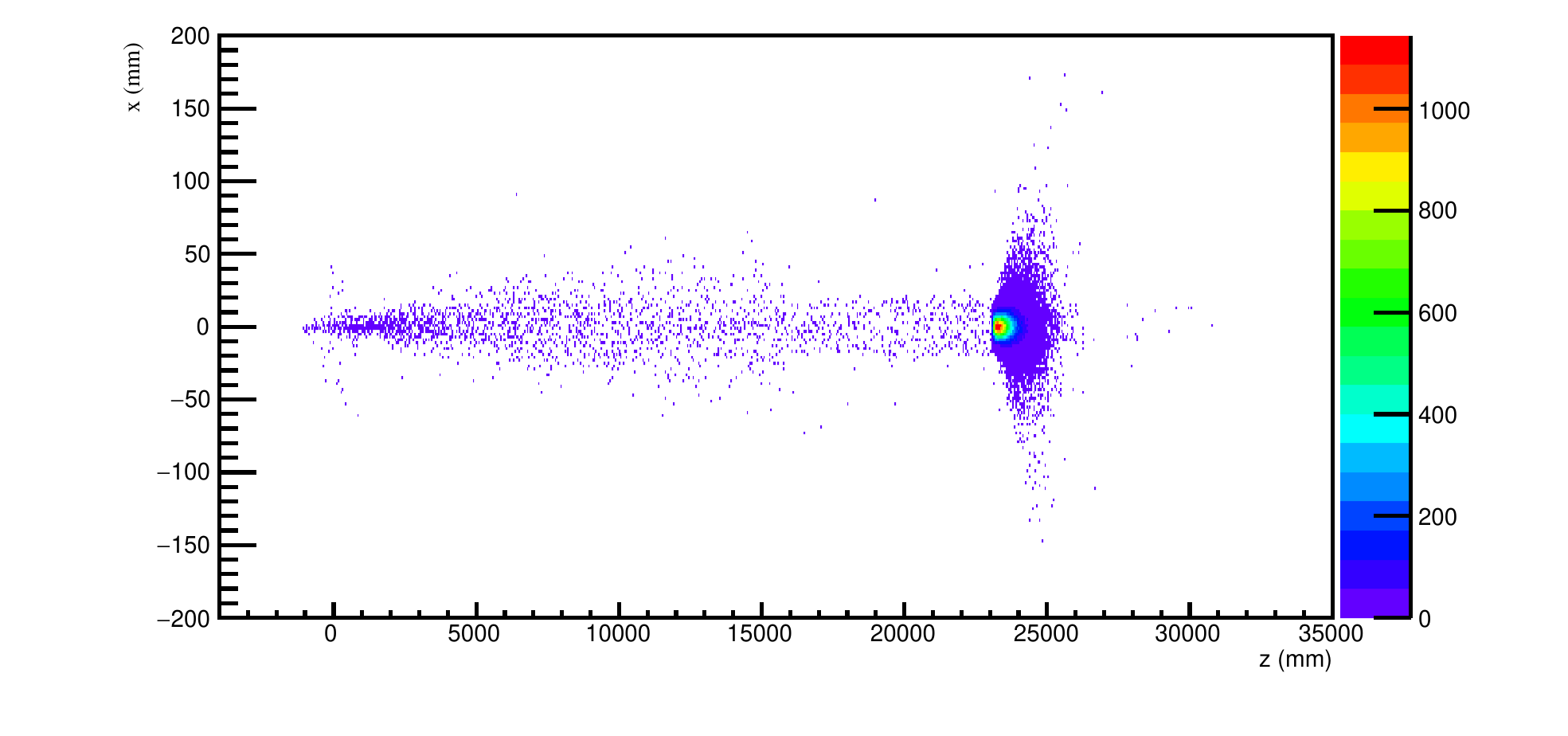}
\includegraphics[width=0.8\textwidth]{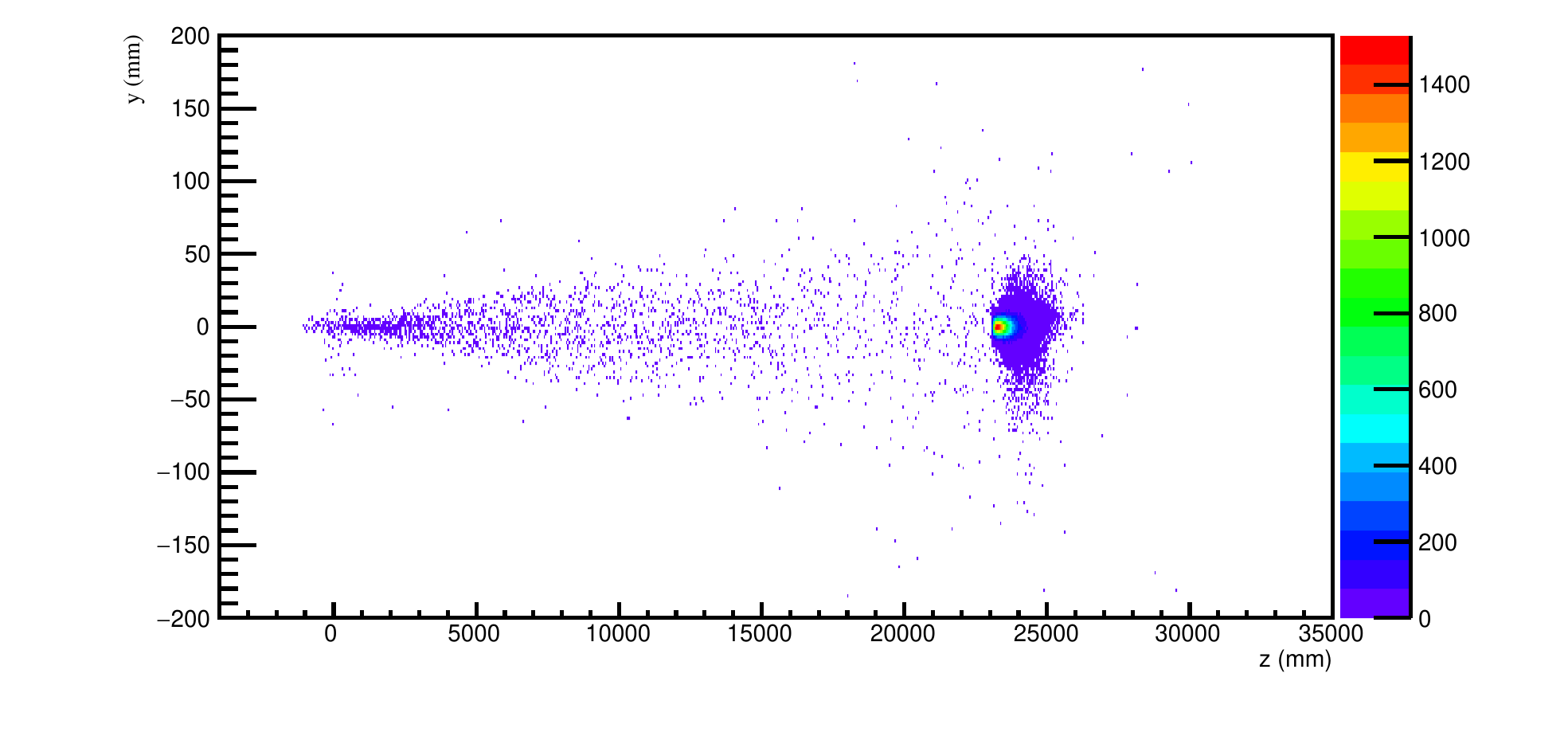}
\includegraphics[width=0.8\textwidth]{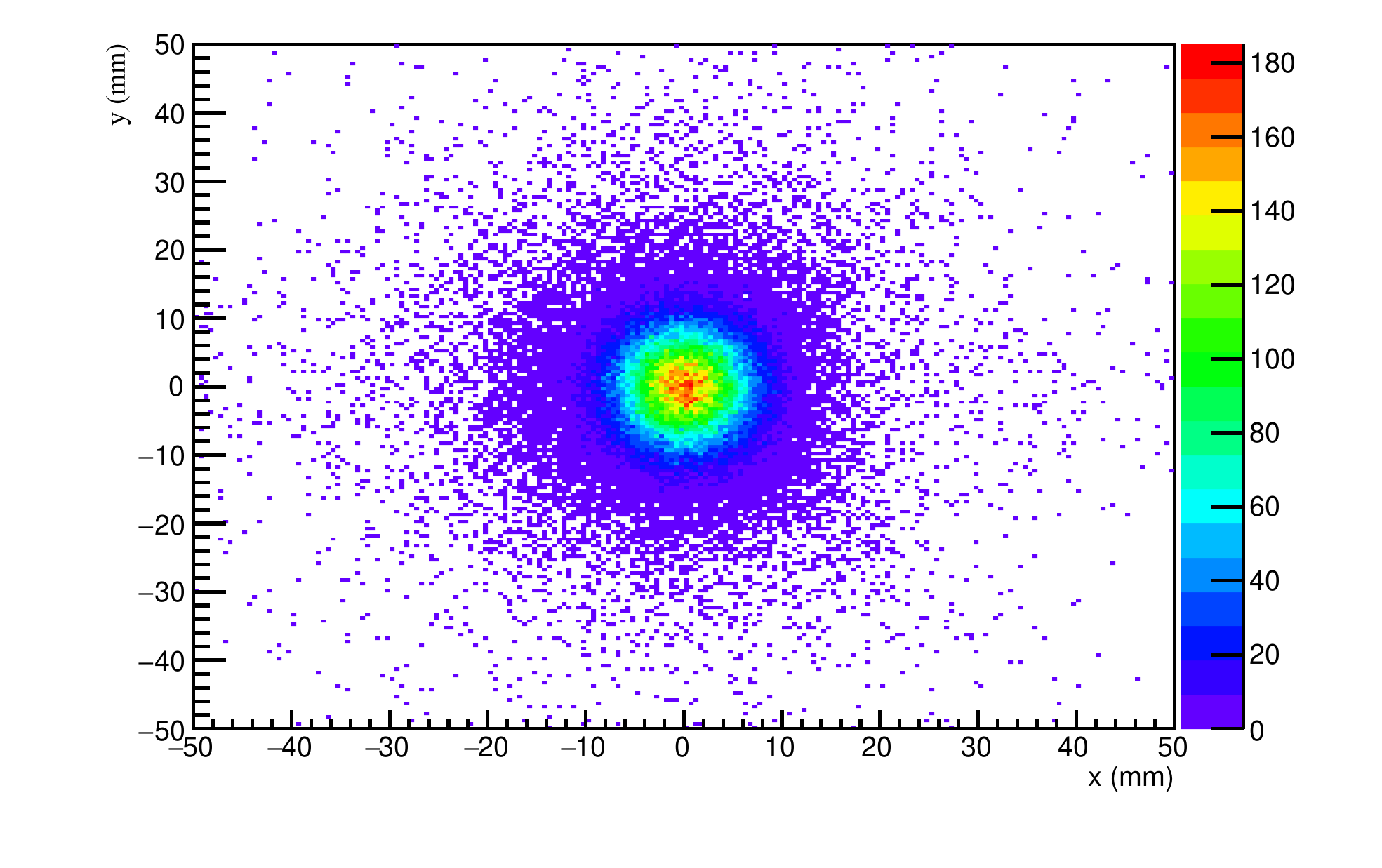}
\end{center}
\caption{\small Origin point of the muons emerging from TAXes. Top: $(x,z)$ view. Center: $(y,z)$ view. Bottom: $(y,x)$ view. The T10 target is placed at z=0, the TAXes start at z=23~m. Most of the muons are originated in the decay of pions and kaons produced by the (primary and secondary) inelastic interactions of the protons in the first interaction lengths of the dump. A small fraction is originated by kaons and pions produced in the proton interactions with the target box, as shown by the long tail in the top and middle plot. The target box will be removed during SHADOWS operation.}
\label{fig:muon_init}
\end{figure}

\vskip 2mm
The origin point of the muons emerging from the dump is shown in Figure~\ref{fig:muon_init}, in the three possible views. Most of the muons, as expected, are originated in the decay of pions and kaons produced by the (primary and secondary) inelastic interactions of the protons within the first few interaction lengths of the dump. This origin point is contained within a radius of a few cm with respect to the beam impinging point. A small fraction is due to the decays of kaons and pions produced in the proton interactions with the target box, corresponding to the long tail starting at z=0~m. 

\vskip 2mm

\clearpage
Muons originated in the dump can produce either combinatorial background and secondary tracks and $V^0$ through  inelastic interactions in the last TAX interaction lengths and in the decay vessel material. Both backgrounds are described below. 

\begin{itemize}
    \item [-] {\it Muon inelastic interactions:} muons emerging from the TAXes can interact inelastically with the material upstrean of or forming the decay volume. These interactions can generate particles, including $V^0$ ($K_{\rm S}, K_{\rm L}$ and $\Lambda$), that enter the decay volume and mimic signal events. 
    
    \item[-] {\it Combinatorial muon background:} A dangerous source of background comes from random combinations of opposite charged muon tracks entering the decay vessel and mimicking signal events with dimuon final state by forming a fake vertex in the fiducial volume. 
    
\end{itemize}

\vskip 2mm
As detailed in Ref.~\cite{NA_note},
the front end of the K12 beam consists of a strong quadrupole triplet followed by a dogleg, called ‘achromat’ of four identical dipoles. The first two dipoles provide a vertical offset of 110~mm and the last pair brings back the beam to the straight axis.
Once back on the original axis, the beam is collimated and the momentum definition refined in another collimator, COLL3.
After COLL3, a series of three 2~m long dipoles with the gap filled with soft iron, except for a 40~mm diameter field-free hole for the beam passage, sweeps away muons (both $\mu^+$ and
$\mu^-$) produced in the region around the T10 target and TAX dumps.

\vskip 2mm
When operated in beam dump mode the T10 target head is lifted up and the proton beam is dumped on the TAXEs. The configuration of the beam line magnets used for studying the background hitting the SHADOWS acceptance is the standard NA62 magnet configuration with the only difference 
putting the second pair of dipoles at the same polarity with a nominal field strength of approximately -1.67~T in the iron (and perhaps switching off the first pair of dipoles). This configuration was already studied in the Physics Beyond Colliders Conventional Beam Working Group for the operation of NA62 in dump mode~\cite{CERN-PBC-REPORT-2018-002}.

\begin{figure}[h]
\begin{center}
\includegraphics[width=0.7\textwidth]{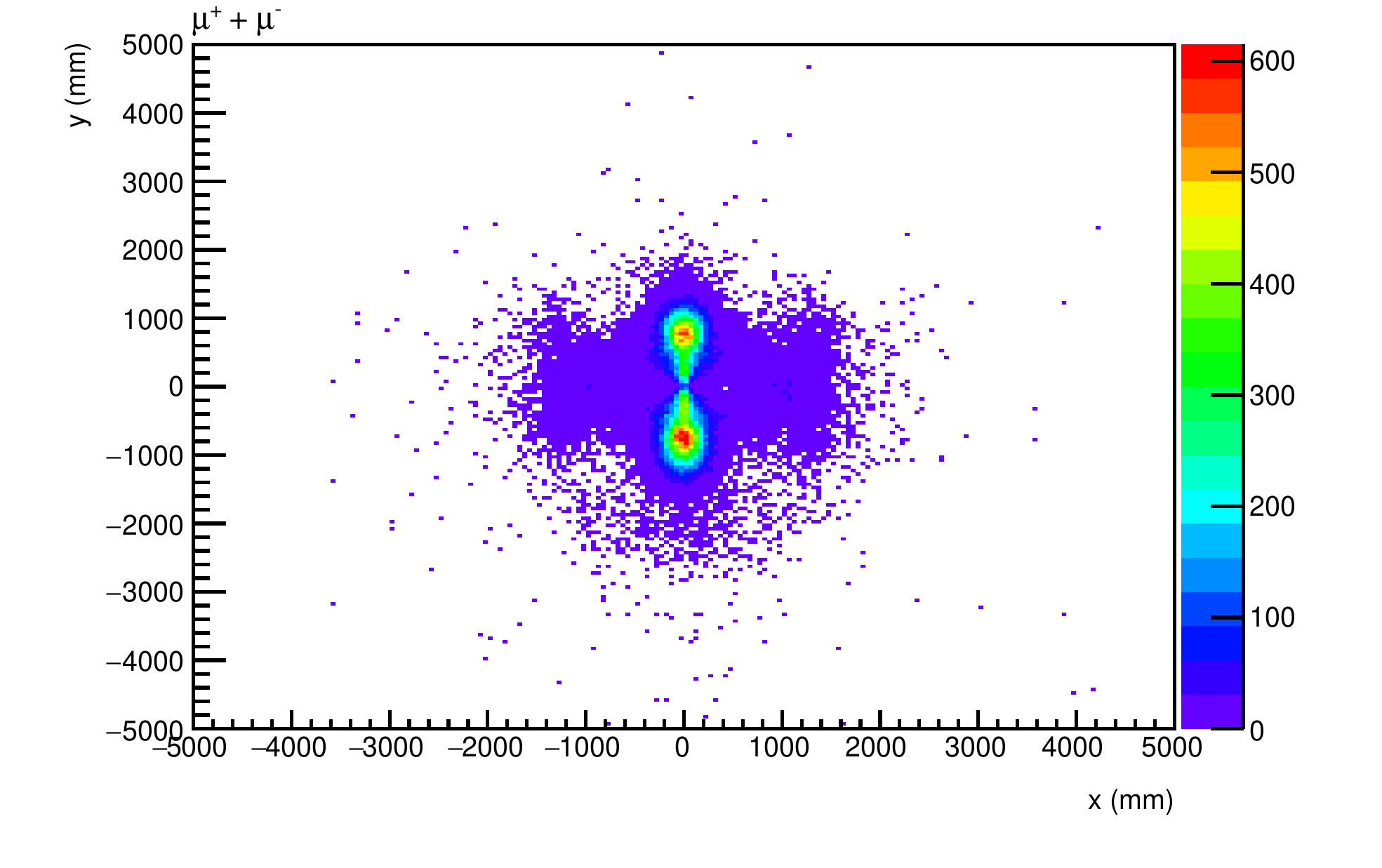}
\includegraphics[width=0.48\textwidth]{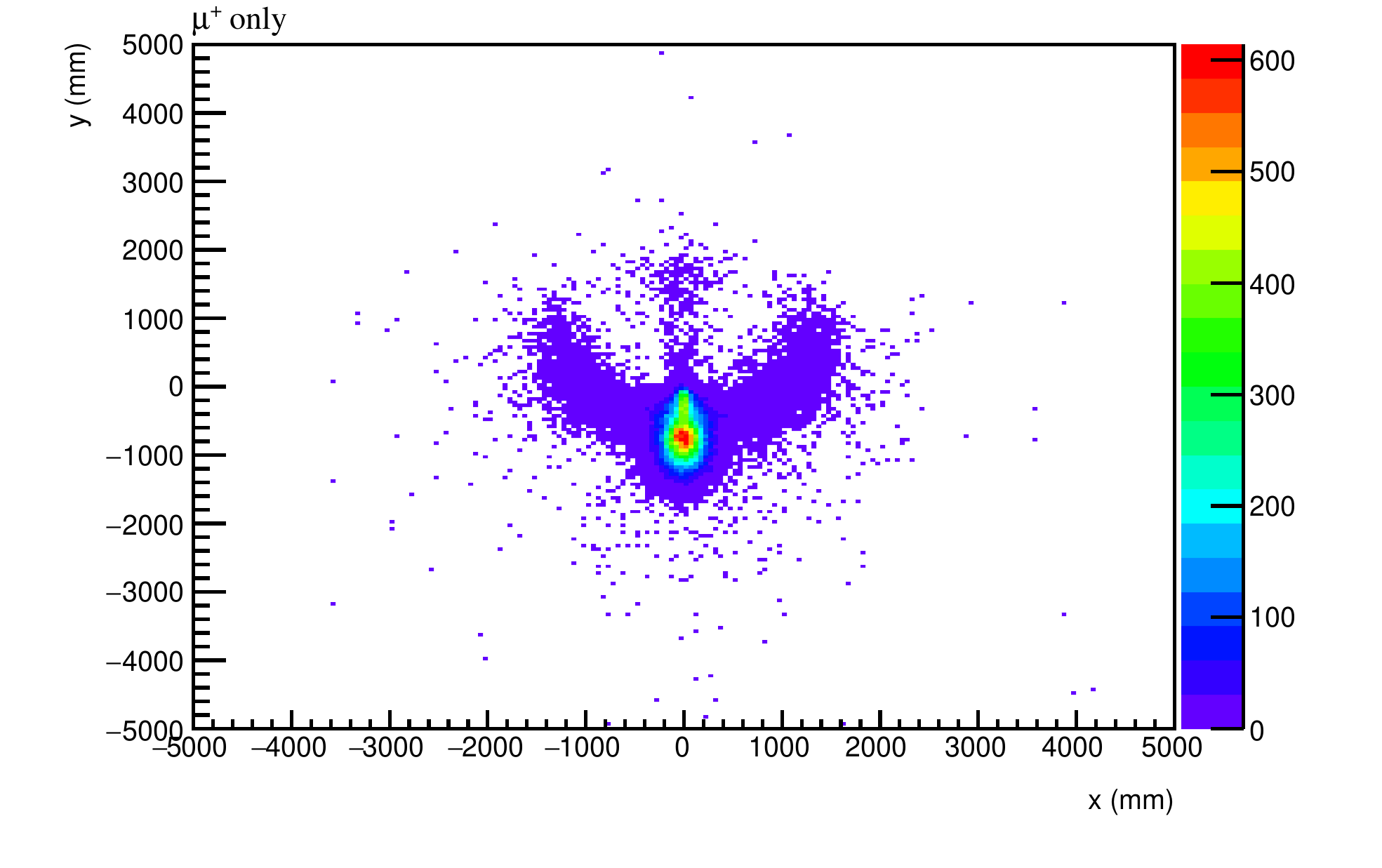}
\includegraphics[width=0.48\textwidth]{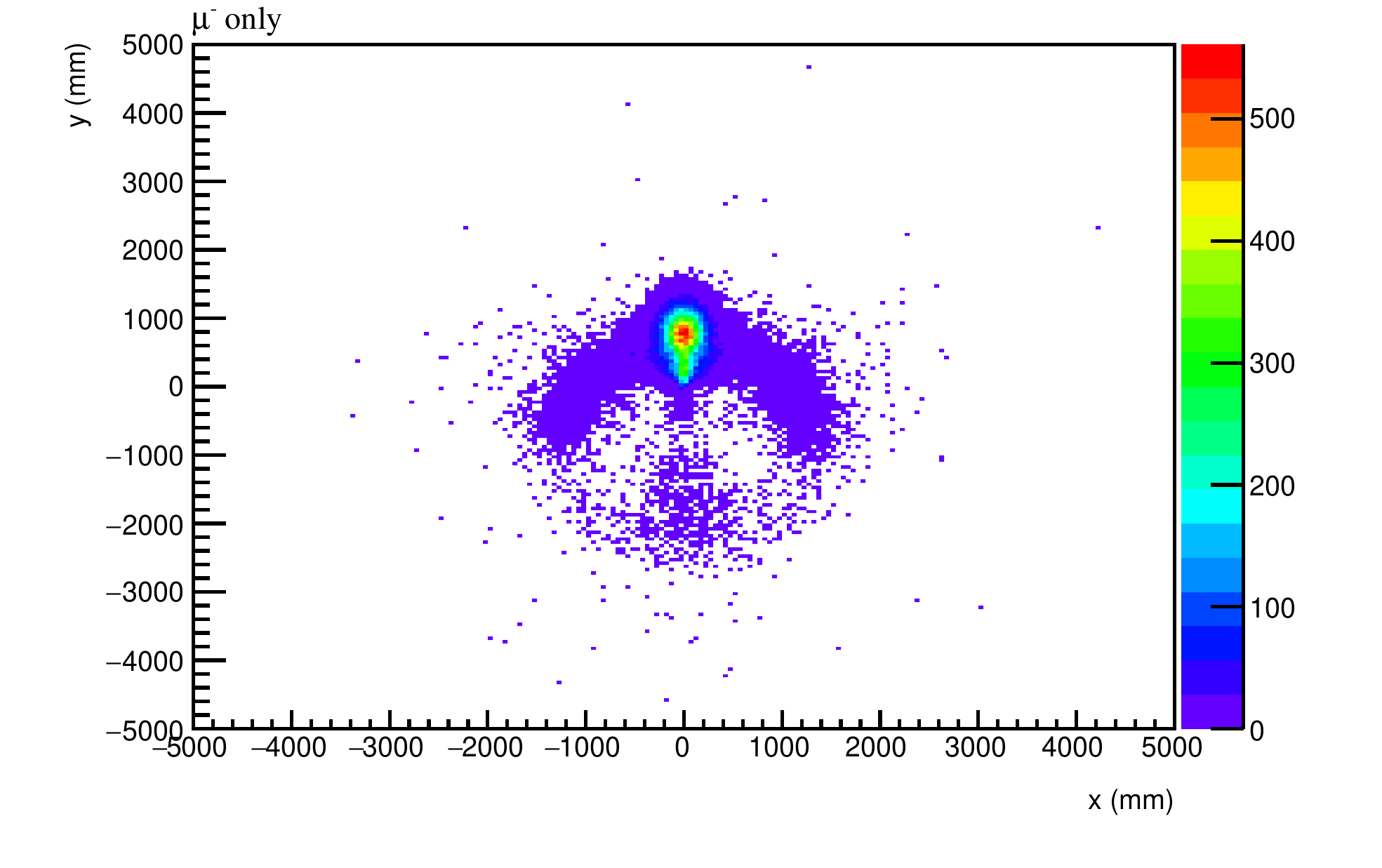}
\end{center}
\caption{\small Top: Muon illumination at the plane z=34.4~m.
Bottom: same but divided per charge (left: positive muons, right: negative muons).}
\label{fig:muon_ill}
\end{figure}

\begin{figure}[h]
\begin{center}
\includegraphics[width=0.7\textwidth]{figs/mu_ill_z34.pdf}
\includegraphics[width=0.7\textwidth]{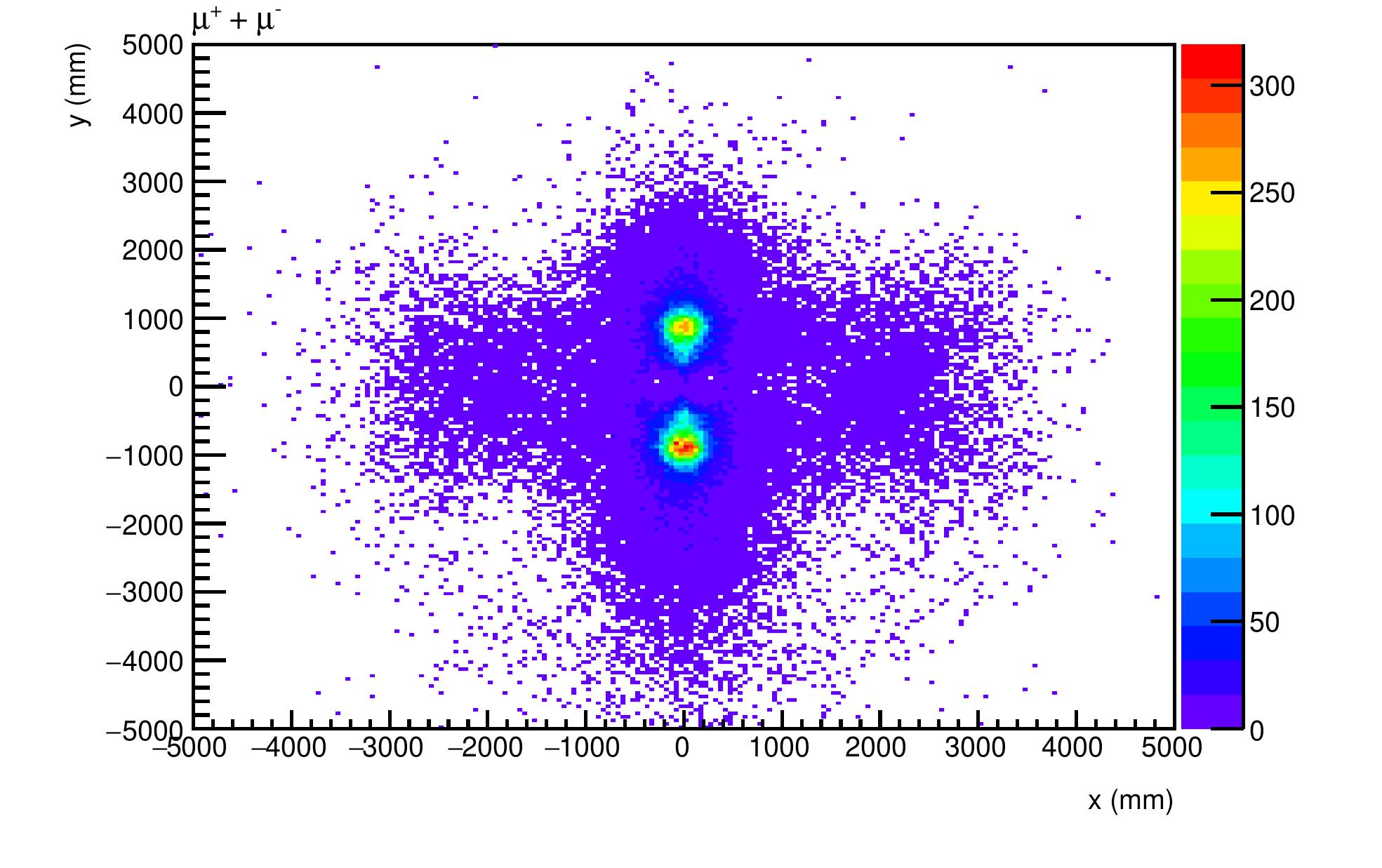}
\includegraphics[width=0.7\textwidth]{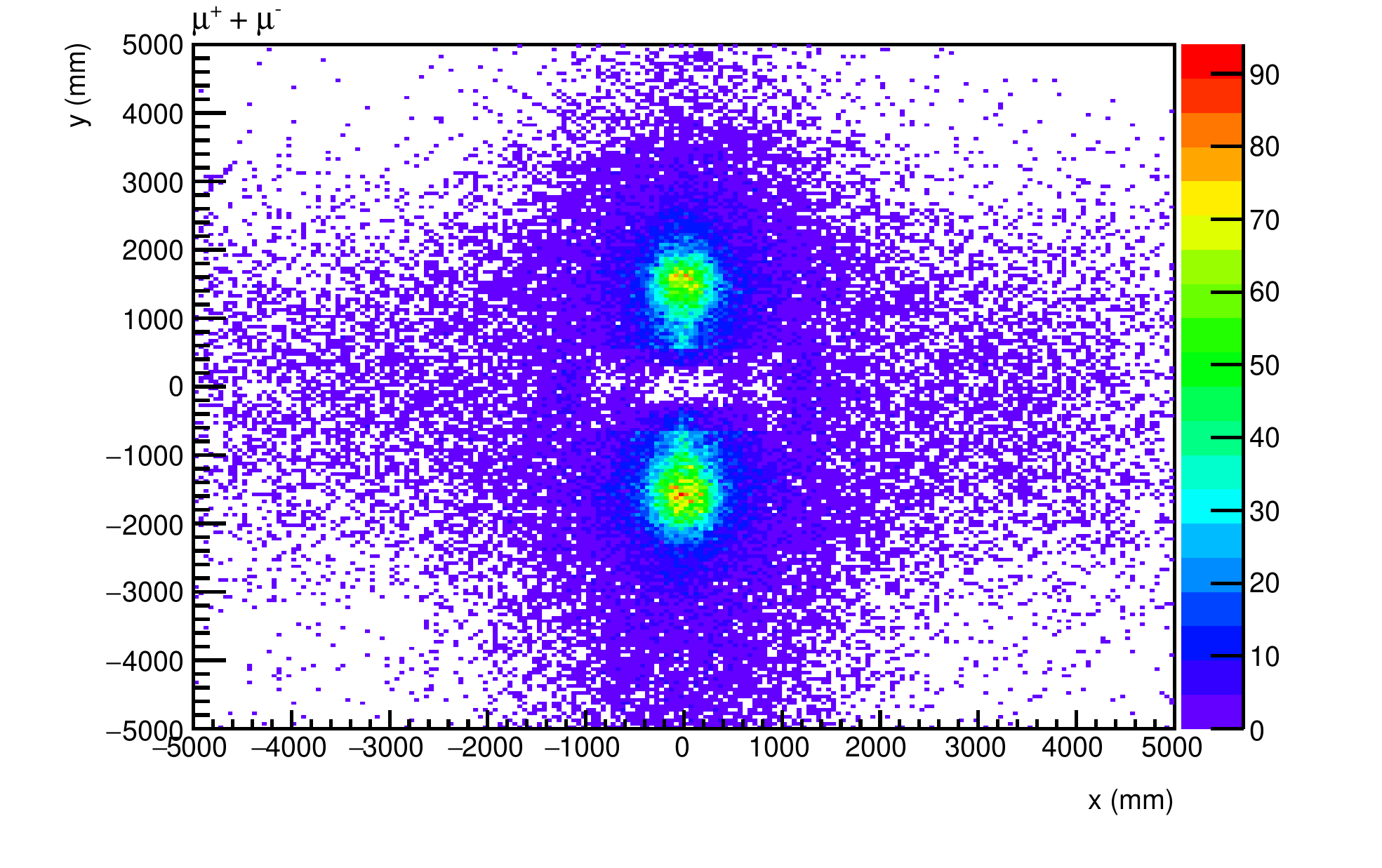}
\end{center}
\caption{\small Evolution of the muon illumination emerging from TAXes. Top: illumination at the plane z=34.4~m. Center: illumination at z=43.3~m (middle of the SHADOWS decay volume). Bottom: illumination at z=55~m (position of the first station of the SHADOWS tracker).}
\label{fig:muon_ill_evolution}
\end{figure}

\vskip 2mm
The illumination of the muon background immediately after the 'achromat' is shown in Figure~\ref{fig:muon_ill}, for both muon charges together and for each charge separately.
Muons emerging from the dump and the achromat have a non-negligible divergence: Figure~\ref{fig:muon_ill_evolution} shows the evolution of the muon flux as a function of the distance from the dump. At z~=~55~m (which corresponds roughly to the position of the first tracking station of the first SHADOWS spectrometer) the muon flux extends up to several meters with respect to the beam axis. The rate expected within the tracker acceptance is $\sim$~10-15~MHz at the NA62 nominal intensity.

\clearpage
\subsection{Reduction of the muon background: The magnetic muon sweeping system}
\begin{figure}[h]
\begin{center}
\includegraphics[width=0.49\textwidth]{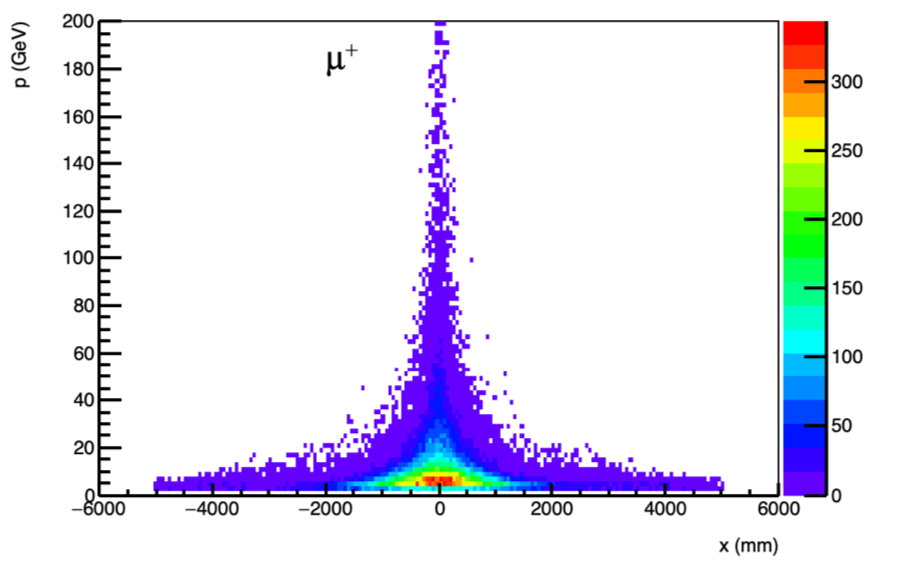}
\includegraphics[width=0.49\textwidth]{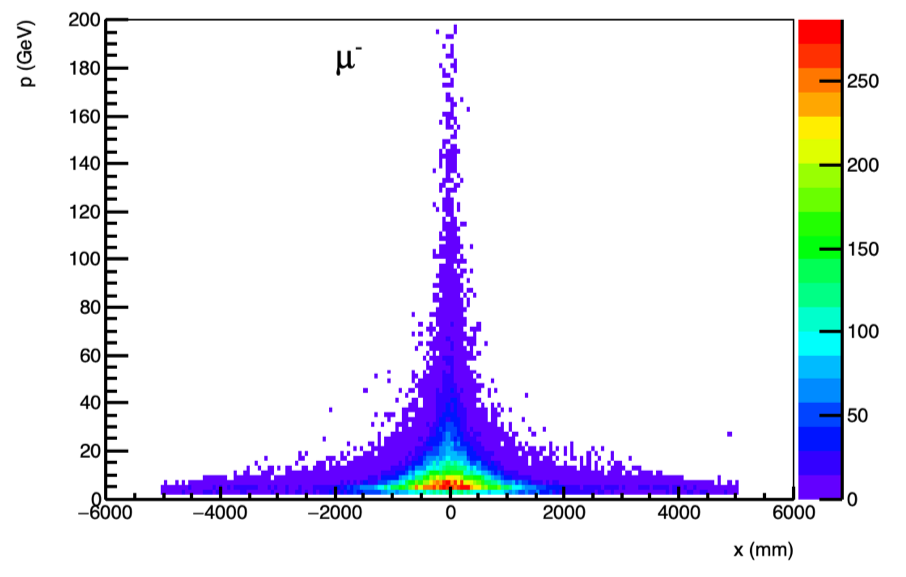}
\end{center}
\caption{\small Muons at z=34.4~m in the plane momentum versus lateral displacement: muons with a lateral displacement of more than 50~cm have momentum below 20 GeV. }
\label{fig:muons_p_vs_x}
\end{figure}

The muon flux evolution detailed in Figure~\ref{fig:muon_ill_evolution} shows that is paramount to get rid of off-axis muons immediately after the second dipole, where they are still concentrated in a small transverse area. Luckily enough, an analysis of their momentum distribution shows that off-axis muons are typically made of low-momentum ($< 20 $~GeV) tracks. This is shown in Figure~\ref{fig:muons_p_vs_x} where muons emerging from the second dipole at z=34.4~m are shown in the plane $p$ versus $x$: for $|x|>500$~mm the muons have $p<20$~GeV.
This observation suggests implicitly also the solution: low-p muons can be swept away using a magnetic field in a magnetised iron block.

\begin{figure}[h]
\begin{center}
\includegraphics[width=0.6\textwidth]{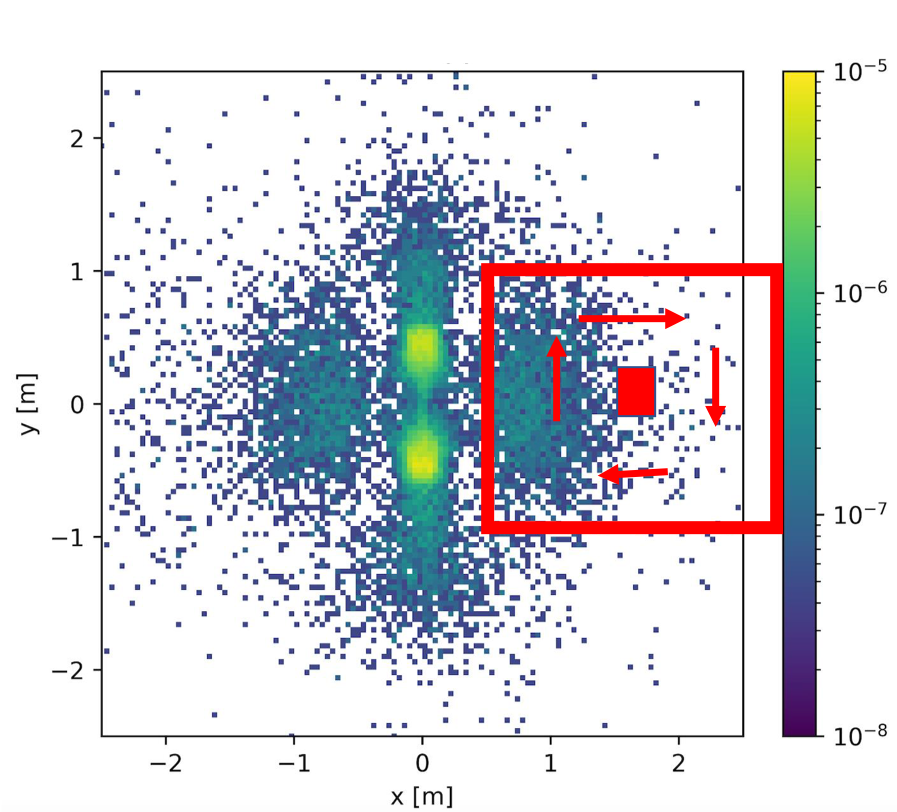}
\end{center}
\caption{\small Schematic view of the position of a magnetised iron block to sweep out the muons emerging from the BEND2 dipole with a lateral displacement in the SHADOWS direction. }
\label{fig:scraper_muon_ill}
\end{figure}

\vskip 2mm
The SHADOWS muon sweeping system based on magnetic elements is currently being conceptually designed by the BE-EA-LE group.
A preliminary sketch of the magnetized iron block (MIB) is shown in Figure~\ref{fig:scraper_muon_ill}. Assuming a maximum field value of 1.7~T/m, a MIB 5~m long produces an effective field of 8.5 Tm. The sweeping iron block will be placed immediately after the BEND2 dipole and will be part of the shielding structure of the TAXes area. Preliminary results show that the muon flux hitting the SHADOWS spectrometer can be reduced already by a factor of $\sim$20 without detailed optimization, bringing the muon flux below $\sim$1 MHz total rate in the SHADOWS acceptance.  When normalized to the detector active area, this flux is only about a factor of three higher than what has been measured in the NA62 acceptance when NA62 is operated in beam-dump mode. A measurement of the muon flux in the SHADOWS area will be performed during the current NA62 run, when NA62 will be operated in beam-dump mode to collect $o(10^{18})$~pot as planned in Ref.~\cite{NA62_Addendum}.

\begin{figure}[h]
\begin{center}
\includegraphics[width=0.49\textwidth, height=6cm]{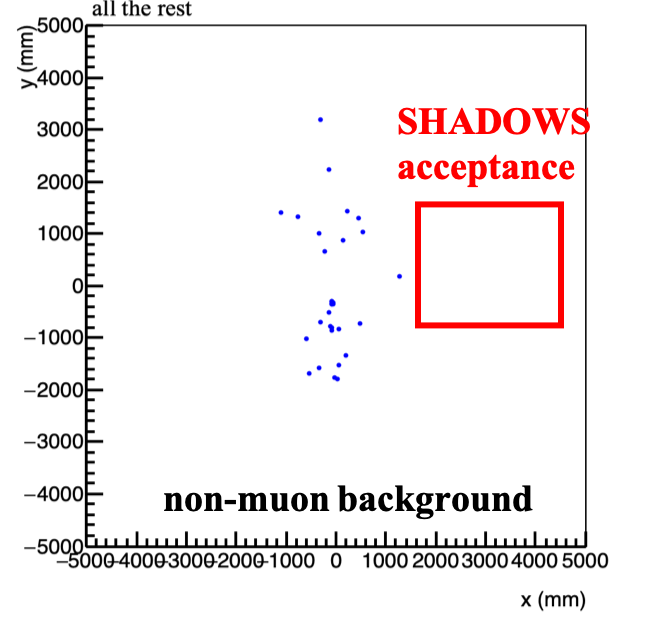}
\includegraphics[width=0.49\textwidth, height=6cm]{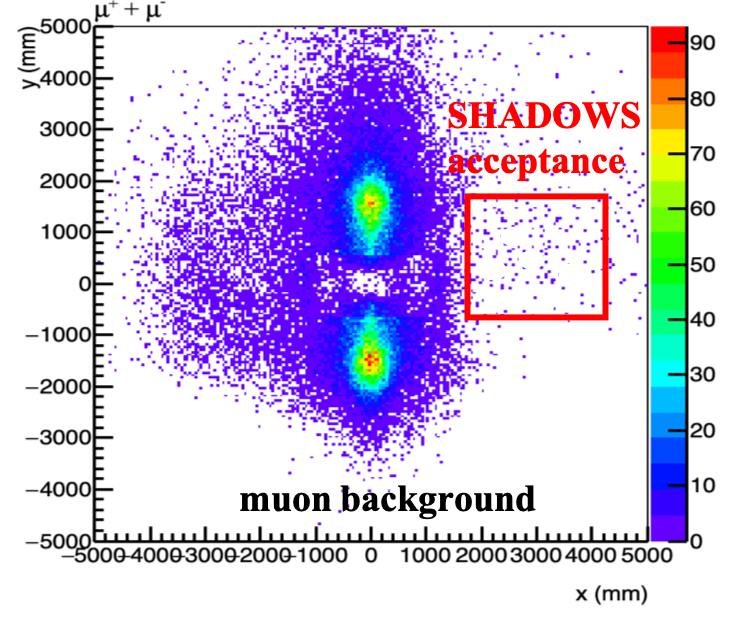}
\end{center}
\caption{\small Residual background at z=55.3~m (position of the first tracking station). Left: non-muon background; Right: muon background. In the SHADOWS acceptance the non muon background is fully negligible and the muon one is reduced by more than an order of magnitude with respect to the one without sweeping system.}
\label{fig:mu-after-scraper}
\end{figure}

\begin{figure}[h]
\begin{center}
\includegraphics[width=0.49\textwidth, height=6cm]{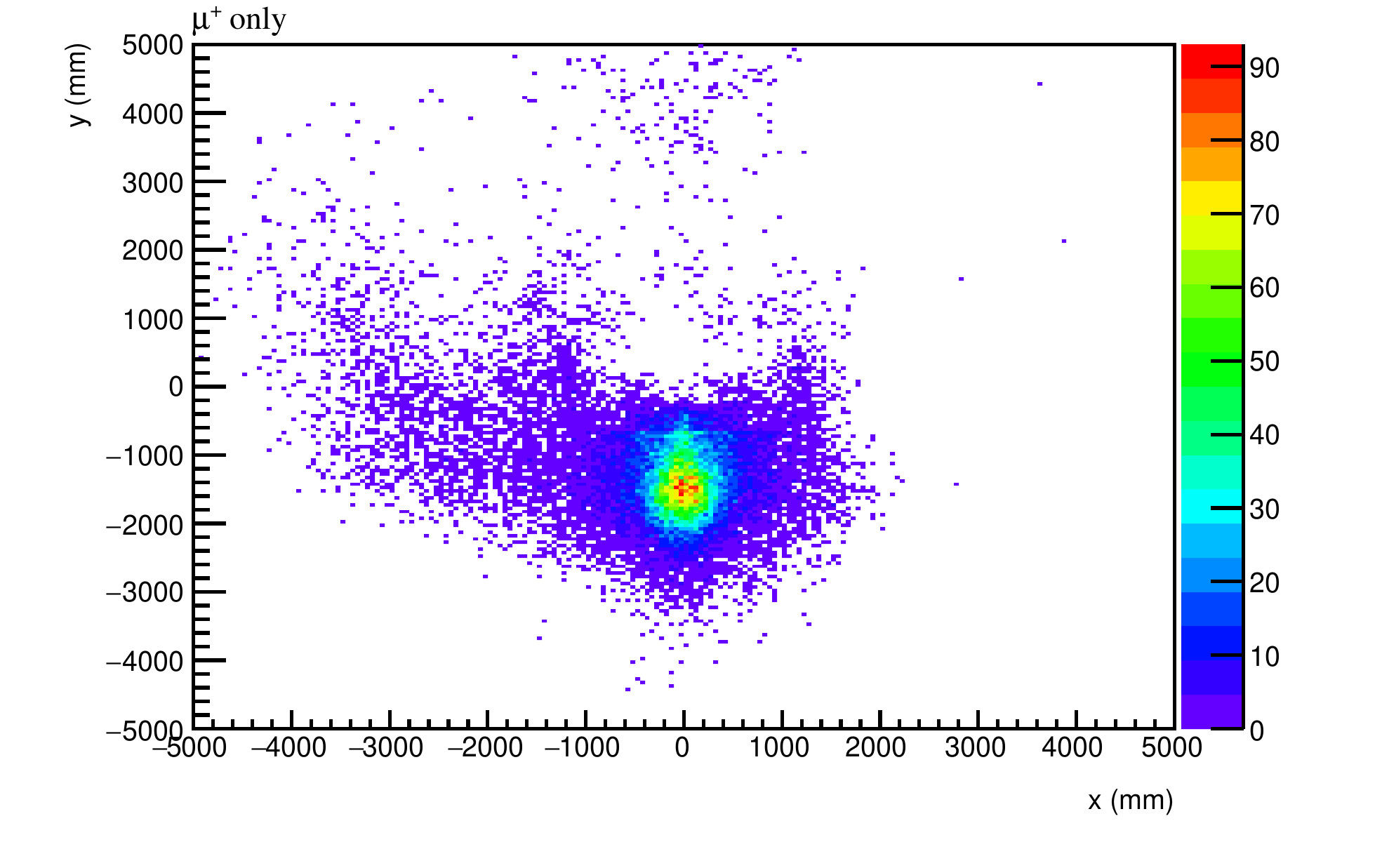}
\includegraphics[width=0.49\textwidth, height=6cm]{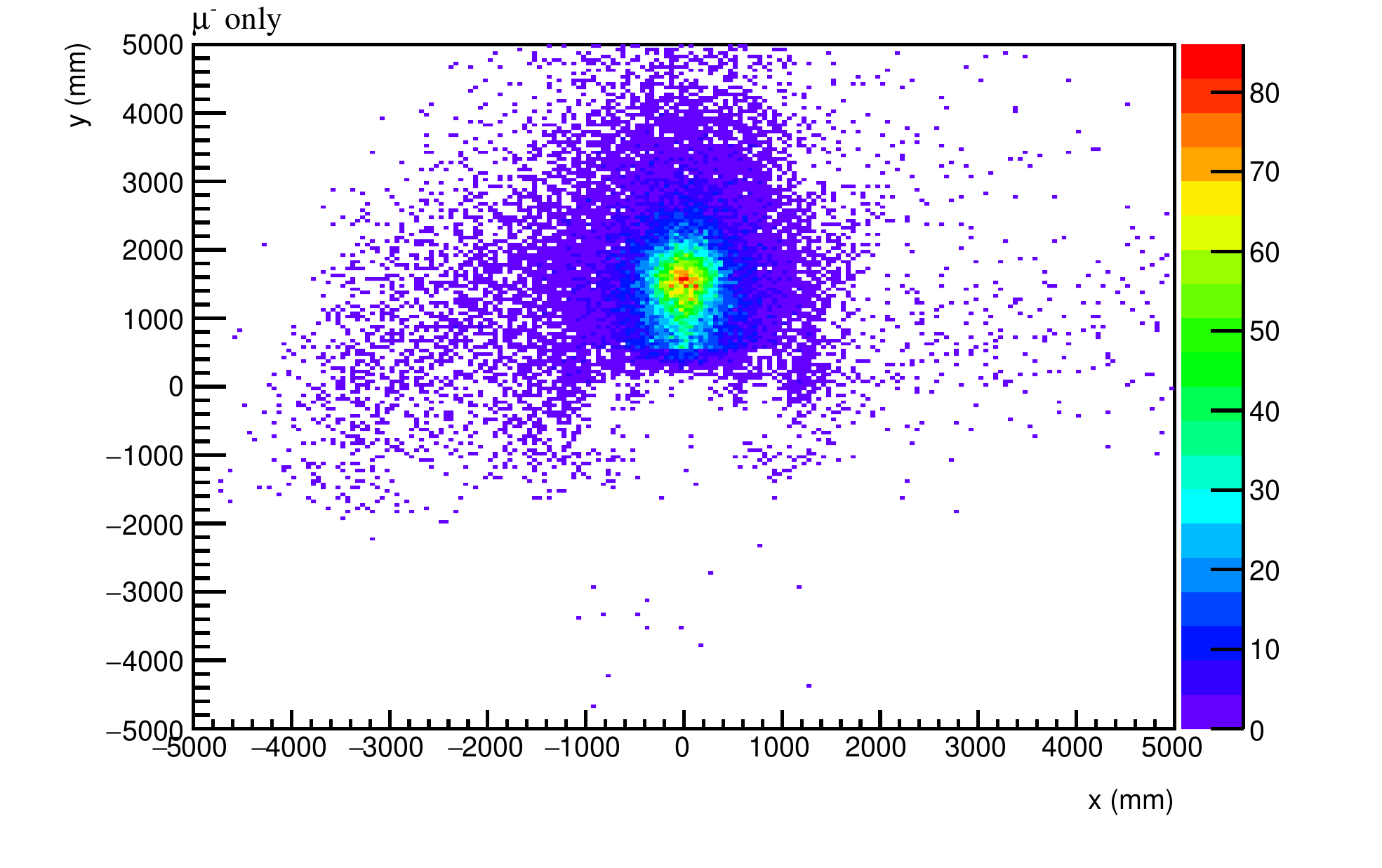}
\end{center}
\caption{\small Residual background at z=55.3~m (position of the first tracking station). Left: muon background with positive charge. Right: muon background with negative charge. The residual background is due to the bending of tracks back into the SHADOWS acceptance due to the opposite field in the other side of the magnetized block.}
\label{fig:mu-after-scraper-2charges}
\end{figure}

\vskip 2mm
Figure~\ref{fig:mu-after-scraper} shows the simulated muon illumination of the first SHADOWS tracking station with the MIB added to the setup. The non-muon background is fully negligible and the muon one is reduced by more than one order of magnitude already at the first attempt. A closer look at the residual flux shows that the residual muon flux within the SHADOWS acceptance is due to muons of the same charge, see Figure~\ref{fig:mu-after-scraper-2charges}. This is due to the fact that whilst muons bent towards the beam line do not encounter any other magnetic fields, those bent in  the opposite direction towards the cavern wall are back-scattered by the field of opposite polarity present in the external part of the MIB (see Figure~\ref{fig:scraper_muon_ill}). An improved design of the MIB is currently ongoing within the BE-EA-LE group.

\begin{figure}[h]
\begin{center}
\includegraphics[width=0.45\textwidth]{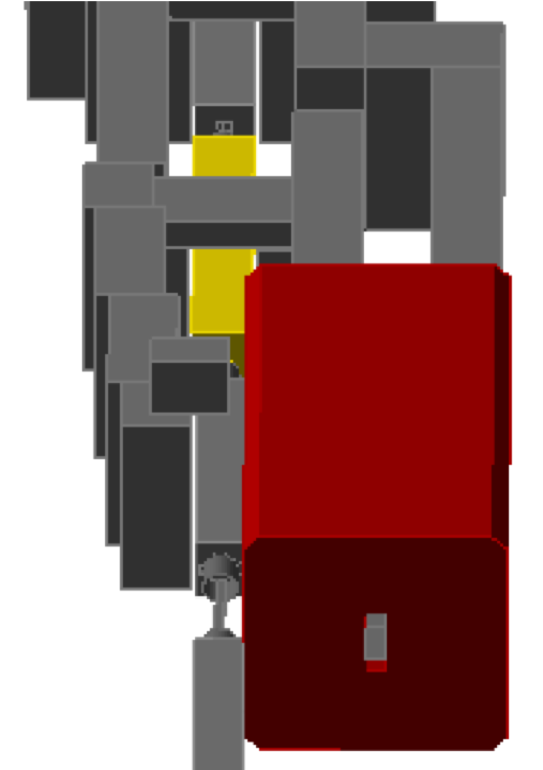}
\includegraphics[width=0.8\textwidth]{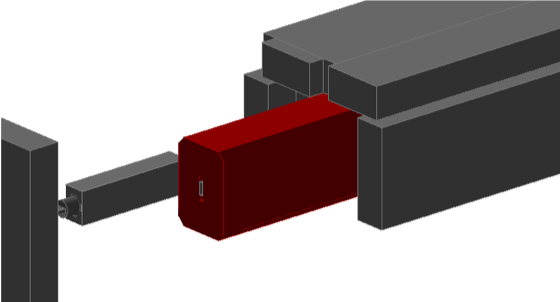}
\includegraphics[width=0.8\textwidth]{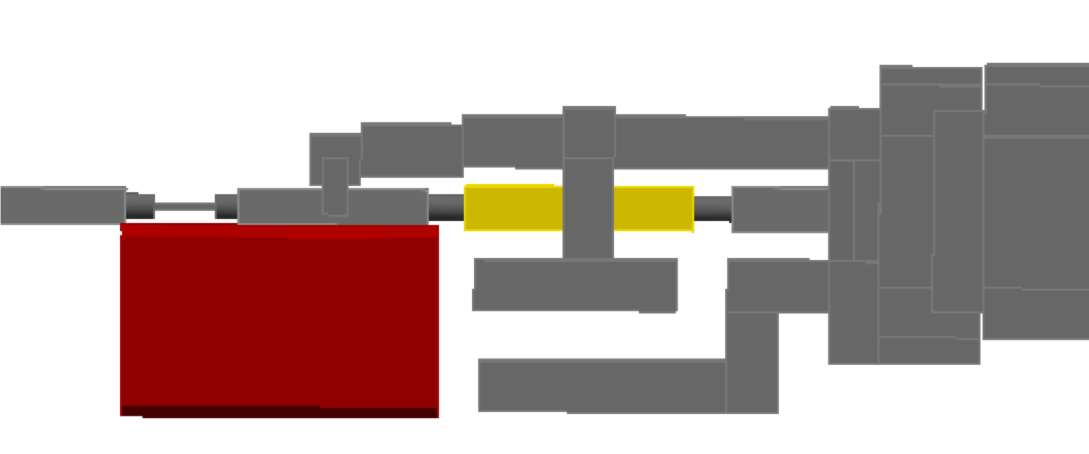}
\end{center}
\caption{\small Magnetized iron block for muon background sweeping placed right after the end of the last dipole of the achromat. The grey structure defines the concrete blocks used to shield the TAX area. Top and centre plot: 3D view; Bottom plot: top view. The yellow block defines the last dipole.}
\label{fig:scraper}
\end{figure}

\subsection{Reduction of the muon background: detector-based methods}
In addition to the muon sweeping system, an active veto in front of the decay vessel can tag charged tracks coming from the dump: an active veto with an efficiency of $\sim 99.5\%$ would reduce the untagged flux to a few permille of the total flux.

\vskip 2mm
Beyond the muon sweeping system and the active veto, additional handles are available for further reducing the background.
The backgrounds arising from inelastic interactions of the residual muon flux hitting SHADOWS detector material and from combinatorial mixing of two (or more) muon tracks are expected to have the following characteristics:

\begin{enumerate}
\item The total momentum of selected tracks when extrapolated backward to the dump does not point to the position of the proton beam impinging point;
\item  combinatorial tracks form rarely a fake vertex in the 20~m long fiducial volume;
\item combinatorial tracks are mostly non coincident in-time. This is particularly true for the case of the combinatorial muon background, 
because muons generated in the interactions of the primary proton beam with the dump remember the timing structure of the proton spill and therefore are almost uniformly spread along the 3.3-sec effective spill duration.
\end{enumerate}

These characteristics allow us to draw also the guidelines for reducing these backgrounds efficiently. 
The spectrometer must have:
\begin{itemize}
\item[a)] a tracking system able to reconstruct the vertex of 2-opposite charged tracks with $o(1)$~cm resolution in the transverse view over 20~m long decay volume and an impact parameter with respect to the impinging point of the proton beam onto TAXes with a resolution of $\sim$~1~cm.
\item[b)] A timing detector able to measure the track arrival times  with a resolution $\sigma_t \sim$~o(200)~ps will reduce out-of-time combinatorial tracks by a factor about $6 \cdot \sigma_t / 3.3$~sec.  
\item[c)] An electromagnetic calorimeter with a good time resolution and the capability also of reconstructing the direction of the photons.

\item[d)] A muon detector with a time resolution comparable to the one of the timing detector in order to identify and reject combinatorial muon tracks.
\end{itemize}

These characteristics can be easily satisfied by the detector options outlined in the next Section.

\clearpage
\section{The SHADOWS detector}
\label{sec:spectrometer}

\vskip 2mm
The SHADOWS detector must be able to reconstruct and identify most of the visible final states of FIPs decays. These are listed in Table~\ref{tab:decays} for the main portals used in the Physics Beyond Colliders exercise~\cite{Beacham:2019nyx}. 

\begin{table}[htbp]
\caption{Main decay modes for FIPs represented in the four portals of the PBC exercise. $\ell = e,\mu,\tau $.}
\label{tab:decays}
\vspace{.1cm}
\begin{center}
\begin{small}
\begin{tabular}{ll}
\hline \hline
Scalar portal &  $\ell^+ \ell^-$, $\pi^+ \pi^-$, $K^+ K^-$ \\
Pseudo-scalar portal & $\ell^+ \ell^-$, $\gamma \gamma$,  $\pi^+ \pi^-$, $K^+ K^-$\\
Vector portal  & $\ell^+ \ell^-, \pi^+ \pi^-, K^+ K^-$ \\
Fermion (neutrino) portal & $\ell^{\pm} \pi^{\mp}, \ell^{\pm} K^{\mp}, \ell^{\pm} \rho^{\mp} (\rho^{\mp} \to \pi^{\pm} \pi^0), \ell^+ \ell^- \nu$\\ 
\hline \hline
\end{tabular}
\end{small}
\end{center}
\end{table}

To this aim a standard spectrometer with excellent tracking and timing performance and some particle identification capability is required. The spectrometer should be made of:

\begin{itemize}
    \item[-] {\it a tracking system} able to reconstruct with high accuracy the mass, the decay vertex and the impact parameter with respect to the impact point of the beam on the dump for FIP decays with at least two charged tracks in the final state. We require: i) a mass resolution of a few MeV for masses of the order of a few GeV; ii) a vertex resolution of $\sim 1$~cm in the transverse plane over a volume length of $\sim$~20~m;  iii) an impact parameter resolution  of $\sim 1$~cm for FIP decays into two charged tracks when the total momentum is extrapolated backward at the impact point of the beam onto the dump. All these requirements are important to separate the signal from the background;

   \item[-] {\it a timing detector} with $\sim (100-200)$~ps time resolution in order to reduce any combinatorial background (and in particular the muon one) by requiring the tracks to be coincident in time. Combinatorial tracks are in fact intrinsically respectively out-of-time as they have a origin time spread over the 3.3~sec effective duration of a typical P42 proton spill;
    \item[-] {\it an electromagnetic calorimeter} able to reconstruct the energy, the position and (possibly) the direction of photons coming from FIP decays. 
    \item[-] {\it a hadron calorimeter} for efficient $\mu$/hadron separation;
    \item[-] {\it a muon detector} to positively identify muons and with timing capabilities to reinforce the rejection of the combinatorial muon background in combination with the timing detector.
    
\end{itemize}

\begin{figure}[h]
\begin{center}
\includegraphics[width=\textwidth]{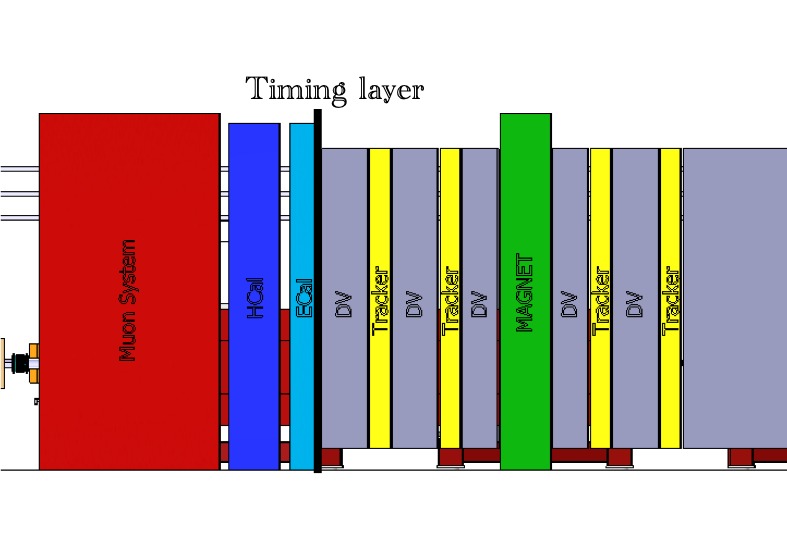}
\end{center}
\caption{\small Schematic layout of the spectrometer. The beam runs from right to left.}
\label{fig:spectrometer-layout}
\end{figure}

A schematic layout of the spectrometer is shown in Figure~\ref{fig:spectrometer-layout}.

The front face of the decay volume should be equipped with a high-efficient veto system able to tag beam-induced charged tracks emerging from the dump. Since the detector is off-axis the inelastic interactions of neutrinos with the air of the decay volume are not a concern. However charged tracks not tagged by the veto system could still enter the decay volume, make inelastic interactions with the air contained in it, and mimic a FIP decay. In order to reduce such a background the decay volume could be either placed in a mild ($\sim 10^{-2}-10^{-3}$ mbar) vacuum or filled with $He$.

\vskip 2mm
In the following Sections we describe possible technologies suitable for the SHADOWS sub-detectors. The list should be considered only indicative as other options could be equally viable. The final choice of the technologies will be driven by the overall detector optimisation and the interest of experimental groups.
While the R\&D of new technologies is certainly encouraged, this is not mandatory: in fact the relatively low background rate at the spectrometer level (see Section~\ref{sec:background}) allows the SHADOWS detectors to be built using existing technologies already available on the market.

\subsection{Tracking system}
The design of the tracking system and the dipole magnet is driven by the requirement of achieving $\sim 1$~cm (or less) resolution on the transverse position of the decay vertex of a feebly-interacting particle in two charged tracks over a decay volume length of about 20~m. 
This is in fact one of most effective handles to reduce the combinatorial background. A preliminary study shows that this resolution can be achieved by a tracking system with a dipole magnet with a bending power of about 1.5~T$\cdot$m and four tracking stations with a single station resolution of about 250~$\mu$m. The four stations should be placed two before and two after the magnet, with a distance between the two stations placed at the same side of the magnet of $\sim 500 $~mm and a gap of 1500~mm between the two stations adjacent to the dipole. 

Table~\ref{tab:tracking} shows the dependence of the momentum resolution as a function of the track momentum for a tracking system able to reconstruct with 1~cm accuracy in the transverse plane the decay vertex in any position along the 20~m long decay volume. 

\begin{table}[htbp]
\caption{Dependence of the momentum resolution and the curvature angle $\rho$ inside the magnetic field as a function of the momentum of the tracks for a fixed resolution in the decay vertex over 20~m long decay volume.}
\label{tab:tracking}
\vspace{.1cm}
\begin{center}
\begin{tabular}{ccccc}
\hline \hline
 p [GeV] & $\sigma_{xy}$ [mm] & $\sigma_z$ [mm] & $\rho$ [m] & $\sigma(p)/p$ (\%) \\ \hline
 1  & 10.6  & 500.0 & 3.33 & 0.2 \\
 2  & 10.6  & 500.0 & 6.67 & 0.4 \\
 5  & 10.6 & 500.0  & 16.67 & 1.1 \\
 10 & 10.6 & 500.0 & 33.33  & 2.2 \\
 20 & 10.6 & 500.0 & 66.67  & 4.4 \\
 50 & 10.6 & 500.0 & 166.67 & 11.1 \\
 100 & 10.6 & 500.0 & 333.33 & 22.2 \\
 200 & 10.6 & 500.0 & 666.67 & 44.4 \\ 
\hline \hline
\end{tabular}
\end{center}
\end{table}

Possible technologies already available on the market are:

\begin{enumerate}
\item {\it NA62 tracking system~\cite{Danielson:2010fta}:} Straw tubes in vacuum, 10~mm diameter, based on a gas mixture  Ar(70\%): CO2 (30\%). 
One straw chamber is composed of four views $(X, Y, U, V)$, with one double-layer per view. The hit resolution is better than 
$400 \mu$m over most of the straw diameter per single layer, and there are eight layers per tracking station. The warm dipole magnet provides 0.9 Tm bending power. This system achieves a mass resolution of 3-4 MeV for a long-lived particle of mass of 1~GeV decaying in two charged tracks. The decay vertex resolution is better than 1~cm over a 100~m long decay volume and the accuracy of the impact parameter with respect to the  beam impinging onto the TAXes is better than 1 cm for a distance between the decay vertex and the dump of (100-180)~m.

\item {\it LHCb Outer Tracker~\cite{LHCbOuterTrackerGroup:2013epe}}: gas-tight straw-tube modules based on  Ar(70\%): CO$_2$ (30\%) gas mixture.
Each module contains two staggered layers of drift-tubes with inner diameters of 4.9~mm.  The drift coordinate resolution per layer is $\sim 200$~$\mu$m. The mass resolution for a $B_s \to \mu^+ \mu^-$ decay at the LHC energies is $\sim $~20~MeV.

\item{\it  Fibre Tracker (LHCb upgrade phase 1)~\cite{LHCbSciFi}}: The LHCb SciFi tracker consists of three tracking stations (T1,T2,T3) each with four detection planes aligned with different stereo angles $(0^{\circ},\pm 5^{\circ})$ with respect to the vertical.  Each detection plane is made of 250~$\mu$m diameter, 2.5~m long scintillating fibres. The system is expected to provide 
a hit resolution per station of $< 80 \mu$m with 4~Tm bending magnet.

\end{enumerate}

\subsection{Timing detector}
The timing detector has to provide a time resolution of $o(100-150)$~ps to reject combinatorial backgrounds made of out-of-time charger tracks, mostly muons.
Several options are possible on the market, here we simply list some of them.

\begin{enumerate}
    \item{\it Scintillating bar timing detector:}
    scintillating bars read out by arrays of silicon photomultipliers. This option has been considered for the SHiP timing detector. The detector comprises more than 500 bars of EJ200 scintillating material with dimensions 168~cm$\times$6~cm$\times$1~cm broken. The end of each bar is read out by an array of eight SiPM attached to custom PCBs. A time resolution of $\sim$80~ps almost uniform along the bar length has been demonstrated on prototypes~\cite{SHiP:2019hap}.

\item{\it Glass Multigap Resistive Plate Chambers, glass-RPC}: This option is based on the RPC technology derives from a novel concept of Multigap Resistive Plate Chambers (MRPC) with a Sealed Glass Stack (SGS).
Also this option has been considered for the SHiP timing detector. A time resolution of $< 60$~ps has been measured on prototypes.

\item{\it Scintillating pad detector}: a double layer of the redesigned scintillating pads considered for the muon detector (see below) could provide a timing detector with
about 150~ps time resolution.

\end{enumerate}

\subsection{Electromagnetic calorimeter}
The electromagnetic calorimeter must be able to reconstruct the energy, the position and (possibly) the direction of photons coming from FIP decays, either directly or via intermediate resonances (eg. $\pi^0$s).

Given the relatively low rate and low energies involved, no radiation issues are expected. Several technologies fit SHADOWS requirements and are listed here below. Other technologies can be certainly considered. 

\begin{itemize}

    \item[-] {\it $PbWO_4$ crystals (CMS electromagnetic calorimeter)}: \\
The CMS ECAL~\cite{CMS_ECAL} is a homogeneous and hermetic calorimeter containing 61200 lead tungstate ($PbWO_4$) scintillating crystals mounted in the barrel, closed at each end by end-caps each containing 7324 crystals. 

\vskip 2mm
The high-density(8.28~g/cm$^3$), short radiation length ($X_0$=0.89~cm), and small Moli\`ere radius (R=2.2~cm) of $PbWO_4$ allow the construction of a compact calorimeter with fine granularity. The crystals emit blue-green scintillation light with a broad maximum at wavelengths 420–430~nm. They are read out by APD (barrel)~\cite{Baccaro:1999ns} and  vacuum phototriodes (endcaps)~\cite{Bell:2004eu}. The light yield is about 4.5 photo-electrons for 1~MeV of deposited energy in the crystals. The barrel is made of 23~cm long crystals with front face cross sections of around 
2.2~cm$\times$2.2~cm, whilst the endcaps comprise 22~cm long crystals with front face cross sections of 2.86~cm$\times$2.86~cm. The energy resolution is 
$\sigma(E)/E = 2.8\%/\sqrt{E(\rm GeV)} \oplus 12\% / E({\rm GeV}) \oplus 0.3\%$. 

\vskip 2mm
The CMS ECAL endcaps will be dismounted during long shutdown 3  and a fraction of the crystals can be reconditioned for their use in the SHADOWS experiment. This opportunity makes this choice particularly attractive for SHADOWS.

    \item[-] {\it Shashlik technology (LHCb ECAL)}: \\
    The LHCb electromagnetic calorimeter~\cite{Machikhiliyan:2009zz} structure employs the shashlik technology~\cite{Atoian:1992ze}
, when interleaving scintillator-absorber layers are pierced by optical fibers of light collection system. The light readout devices are photomultipliers (PMTs). The LHCb ECAL is made of 3312 separate modules of square section, of different sizes. The sampling stack comprises 66 lead plates and 67 scintillator (BASF-165H polystyrene basedplastic doped with 2.5\% p-terphenyl and 0.01\% POPOP) planes, separated by thin ($120\, \mu$m) TYVEC paper sheets. Wavelength-shifting (WLS) fibers (KURARAY Y-11(250) MSJ, 1.2~mm diameter) of the light collection system run parallel to the beam axis and penetrate the entire module body. Fibers serving the same cell are grouped together at the rear of the module and are read out by the same PM. The cells have transverse dimensions of $4.04\times4.04 /6.06\times 6.06 / 12.12\times 12.12$~cm$^2$ depending on the zone. This calorimeter has an energy resolution $\sigma(E)/E \sim 10\%/\sqrt{e} \oplus 1\%$ and a fast time response in order to cope with the LHC bunch spacing of 25~ns.
    
    \item[-] {\it The "SplitCal" calorimeter (SHiP proposal)~\cite{Bonivento:2018eqn}} \\
    The SplitCal calorimeter proposed for the SHiP experiment is a lead sampling calorimeter, with absorber plates orthogonal to the incoming beam direction, and two types of active layers: scintillator layers for the energy and time measurement, with coarse spatial segmentation and high resolution layers to determine the transverse position and direction of the shower.
    Preliminary studies show that a moderate energy resolution of $\sigma(E)/E \sim 10-15\%/\sqrt(E({\rm GeV})$, a time resolution of a few ns and photon direction measurement are achievable with this design.
    
\end{itemize}

\subsection{Hadronic calorimeter}
The hadron calorimeter has to positively identify hadrons and measure their energy. A detailed study of its requirements and related performance will be done in the future, but we expect that a traditional structure of  a sampling calorimeter made from alternate layers of iron and scintillator, as for example the MUV1 and MUV3 system of the NA62 experiment~\cite{NA62:2017rwk}, could be adequate for SHADOWS requirements.

\subsection{Muon system}
The muon system has to identify muons with high efficiency ($>$ 95\%), reduce the hadron contamination to less than 1\% in a momentum range between 5 - 100~GeV/c, and reject the combinatorial muon background pairs from the beam muon halo together with the Timing detector.
The combinatorial muon background from beam halo is spread along spills 3.3-sec long and can be controlled by requiring a tight ($<3 \cdot \sigma_t$, $\sigma_t \sim 150$~ps) time coincidence of the two tracks. 

\begin{figure}[hbt]
  \begin{center}
\includegraphics[width=0.6\textwidth, height=8cm]{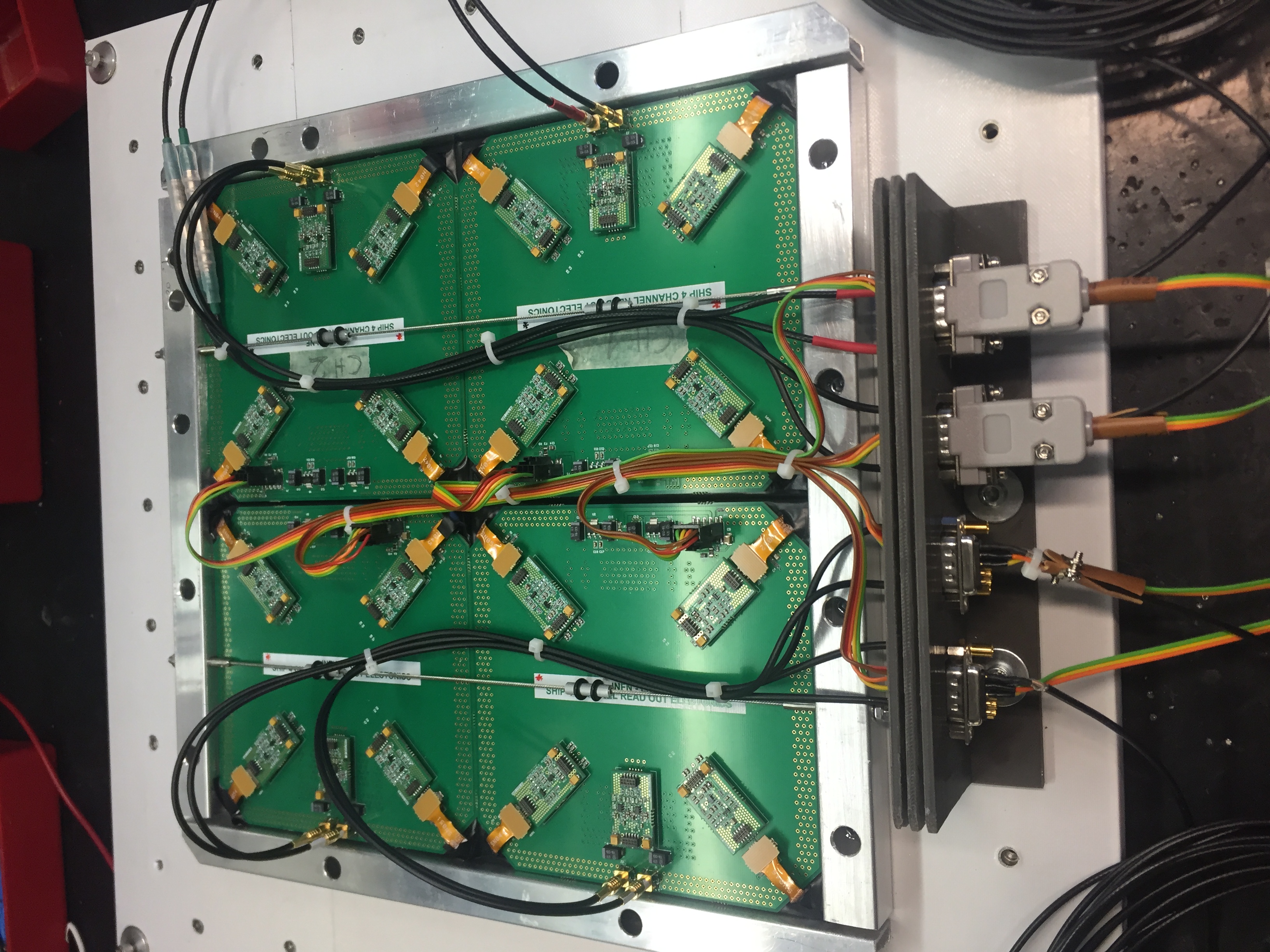}
    \caption{Prototype of the SHADOWS muon system: 4-scintillating tiles with direct SiPM readout equipped with a customized front-end electronics.}
\label{fig:tiles}
\end{center}
\end{figure}

\vskip 2mm
The detector could be made of four active stations, of $\sim(2.5\times 2.5)$~m$^2$ transverse area, interleaved by iron filters $\sim 3.0 \lambda_I$ thick. In order to achieve a time resolution of $o$(150)~ps over the four stations, a time resolution of $\sim$~300~ps per station is required.
The baseline technology consists of scintillating tiles with direct SiPM readout. The basic unit is a $\sim$225~cm$^2$ area (15$\times$ 15)~cm$^2$ and $\sim$~10~mm thick tile of organic scintillator, read out by four SiPMs. Tests performed at  the Laboratori Nazionali di Frascati of INFN and INFN Bologna~\cite{Balla:2021njd} show that a time resolution of 250-300~ps per tile can be achieved with an efficiency of $>99.8\%$ and a light yield of $\sim 230$ photo-electrons per minimum-ionizing particle. Figure~\ref{fig:tiles} shows the first 4-tile prototype.

In the SHADOWS muon detector the tiles will be grouped in {\it modules} of 16 (32) tiles each. Each stations will be equipped with 16 (8) modules.

\begin{figure}[hbt]
  \begin{center}
\includegraphics[width=\textwidth]{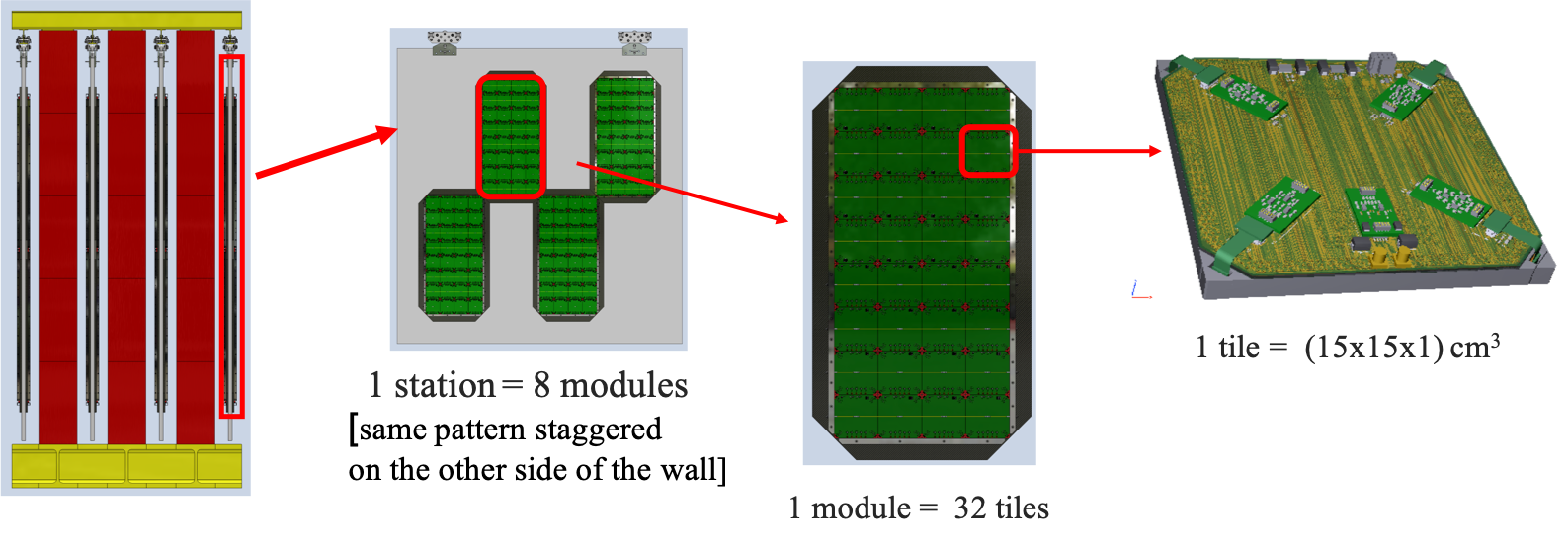}
    \caption{Schematic layout of the SHADOWS muon system. The system is made of four active stations interleaved by passive filters. Each station is made by 16 or 8 modules, 16- or 32-tile each, respectively. The modules are organised in a chess-like structure. Each tile has an area of $\sim 225$~cm$^{2}$.}
\label{fig:shadows-muon}
\end{center}
\end{figure}

\vskip 2mm
The modules are installed on both sides of a supporting structure (Al wall) in a staggered {\it chess} structure:
such a design allows both the dead zones to be minimised and the weight on the supporting wall to be equally shared between the two wall sides.
Each station can slide on rails to ease the access to the modules during installation/repair. A schematic layout of the SHADOWS muon system is shown in Figure~\ref{fig:shadows-muon}.

\subsection{Upstream Veto}
The Upstream Veto should be installed in the front face of the decay volume to tag the residual muon background emerging from the TAXes and not yet swept out by the magnetized iron block (see Section~\ref{sec:background}).
The veto must have high efficiency, excellent time resolution and a few cm granularity to stand the high rate expected in that region. A possibility could be to install a double layer of scintillating tiles similar to those developed for the muon system but of smaller area. Other technologies are clearly possible and can be considered.

\clearpage
\section{Tentative schedule}
\label{sec:schedule}

\begin{figure}[h]
\begin{center}
\includegraphics[width=0.9\textwidth]{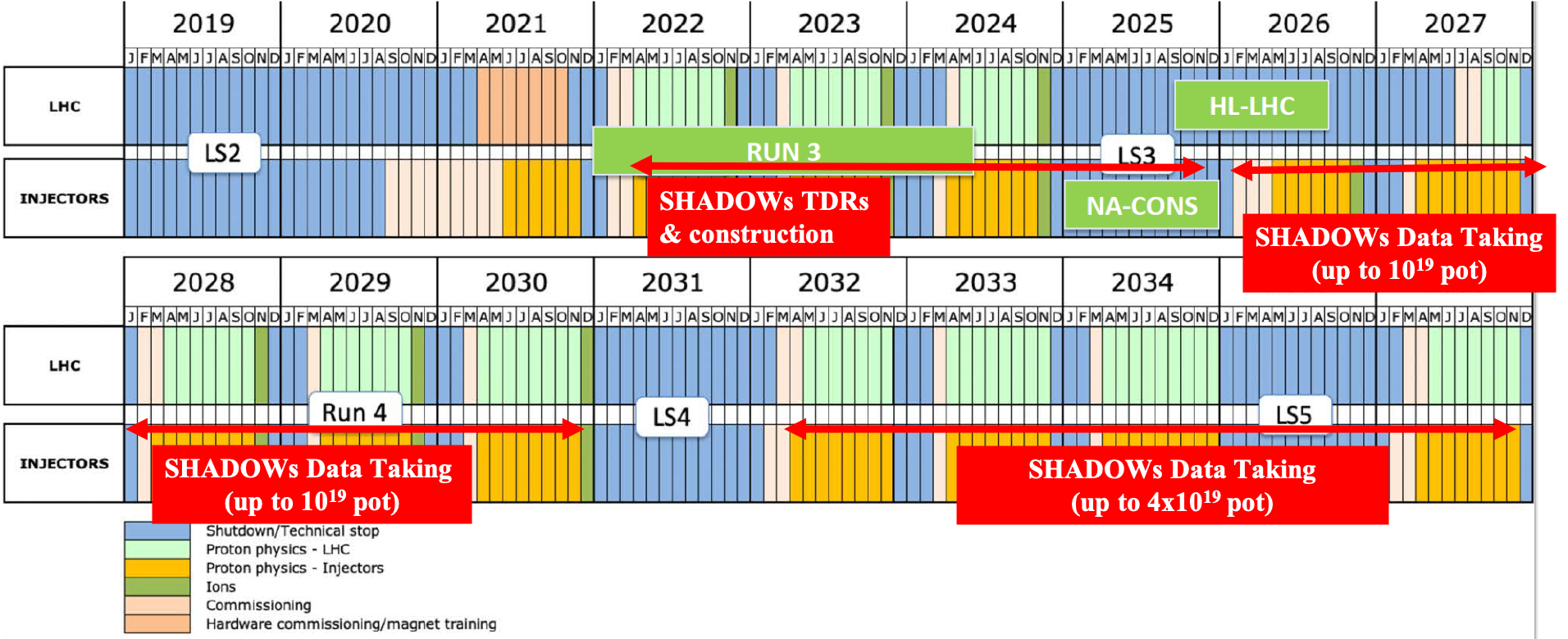}
\end{center}
\caption{\small SHADOWS tentative timescale for the installation of the first spectrometer in zone 1.}
\label{fig:shadows_schedule}
\end{figure}

Figure~\ref{fig:shadows_schedule} shows the tentative time schedule for the installation and operation of the SHADOWS experiment in the first phase (one spectrometer installed in zone 1). This assumes the current schedule for the CERN accelerator complex. Assuming the presentation of a Technical Proposal from the Collaboration within 2022 and a positive recommendation from the CERN SPS Committee and Research Board, we could aim at installing the detector during the Long Shutdown 3 and start with a pilot run at the beginning of Run 4.  The goal is to integrate  $5 \cdot 10^{19}$ pot in beam-dump mode within Run 5, with at least one spectrometer. 

\section{Conclusions and Outlook}
\label{sec:conclusions}
Feebly-interacting particles (FIPs) may provide an answer to many open questions inparticle physics, astrophysics, and cosmology. These include the origin of the neutrino masses and oscillations, the baryon asymmetry of the universe, the dark matter, the strong CP problem, the cosmological inflation, and the apparent fine tuning of the EW scale.

\vskip 2mm
In the coming years a wealth of experimental results for FIPs are expected from all the major laboratories in the world. These include accelerator based experiments (ATLAS,CMS, LHCb, NA64, NA62 at CERN; MEG-II and Mu3e at PSI; Belle II at KEK; BDX and HPS at JLab; MiniBooNE at FNAL; T2K ND280 and KOTO at J-Park) as well as dark matter direct detection experiments and dedicated experiments searching for axions/ALPs in Europe and US.

\vskip 2mm
In addition to established experimental efforts, a multitude of new initiatives has recently emerged both at CERN (NA62 in dump mode, SHiP, MATHUSLA,FASER, CODEX-b) and elsewhere (LDXM at SLAC, long baseline neutrino near detectors at Fermilab and in Japan, Beam Dump experiment at Mainz, and others) aiming at covering in the coming decade uncharted regions of FIPs parameter space inaccessible to traditional experiments at colliders.

\vskip 2mm
SHADOWS, profiting of the unique P42 beam line characteristics in terms of intensity and energy and of the profound expertise related to its operation of the CERN BE-EA-LE group and the NA62 collaboration, aims at becoming one of the leading experiments worldwide in the field during this decade.

\vskip 2mm
The experiment is currently being designed. Groups or individuals interested to contribute to the SHADOWS physics case and/or to the SHADOWS detector can contact: \\
{\it Gaia.Lanfranchi@lnf.infn.it, Augusto.Ceccucci@cern.ch} .

\clearpage

\bibliographystyle{JHEP}
\bibliography{biblio.bib}

\end{document}